\documentclass[12pt,twocolumn,amsmath,amssymb,aps,longbibliography]{revtex4-1}  
\usepackage{hyperref}
\usepackage{graphicx}
\usepackage[caption=false]{subfig}
\usepackage{xtab}
\usepackage{filecontents}

\newcommand{\tit}{\textit}
\newcommand{\beq}{\begin{equation}}
\newcommand{\eneq}{\end{equation}}

\begin{document}

\preprint{Open Science}

\title{Frequency patterns of semantic change: Corpus-based evidence of a near-critical dynamics in language change}

\author{Quentin Feltgen}
 \altaffiliation[Corresponding author]{}
 \email{quentin.feltgen@ens.fr}
\affiliation{\footnotesize{Laboratoire de Physique Statistique, \'Ecole Normale Sup\'erieure, PSL Research University; Universit\'e Paris Diderot, Sorbonne Paris-Cit\'e; Sorbonne Universit\'es, UPMC -- Univ. Paris 06; CNRS; Paris, France.}}

\author{Benjamin Fagard}
 \homepage{http://www.lattice.cnrs.fr/Benjamin-Fagard-98}
\affiliation{\footnotesize{Laboratoire Langues, Textes, Traitements informatique, Cognition (Lattice, CNRS, ENS \& Universit\'e Paris 3, PSL \& USPC), \'Ecole normale sup\'erieure, Paris, France.}}

\author{Jean-Pierre Nadal}
 \homepage{http://www.lps.ens.fr/~nadal/}
\affiliation{\footnotesize{Laboratoire de Physique Statistique, \'Ecole Normale Sup\'erieure, PSL Research University; Universit\'e Paris Diderot, Sorbonne Paris-Cit\'e; Sorbonne Universit\'es, UPMC -- Univ. Paris 06; CNRS; Paris, France.}}
\affiliation{\footnotesize{\'Ecole des Hautes \'Etudes en Sciences Sociales, PSL Research University, CNRS, Centre d'Analyse et de Math\'ematique Sociales, Paris, France.}}

\date{November 2017}

\begin{abstract}
It is generally believed that, when a linguistic item acquires a new meaning, its overall frequency of use rises with time with an S-shaped growth curve. Yet, this claim has only been supported by a limited number of case studies. In this paper, we provide the first corpus-based large-scale confirmation of the S-curve in language change. Moreover, we uncover another generic pattern, a latency phase preceding the S-growth, during which the frequency remains close to constant. We propose a usage-based model which predicts both phases, the latency and the S-growth. The driving mechanism is a random walk in the space of frequency of use. The underlying deterministic dynamics highlights the role of a control parameter which tunes the system at the vicinity of a saddle-node bifurcation. In the neighborhood of the critical point, the latency phase corresponds to the diffusion time over the critical region, and the S-growth to the fast convergence that follows. The durations of the two phases are computed as specific first passage times, leading to distributions that fit well the ones extracted from our dataset. We argue that our results are not specific to the studied corpus, but apply to semantic change in general.
\begin{description}
\small{\item[Keywords]language change; grammaticalization; language modeling; S-curve; \\ corpus-based
\item[Published 8 November 2017] R. Soc. Open Science. DOI: \href{https://doi.org/10.1098/rsos.170830}{10.1098/rsos.170830}}
\end{description}
\end{abstract}

\maketitle 

\tableofcontents

\section*{Introduction}

Language can be approached through three different, complementary perspectives. Ultimately, it exists in the mind of language users, so that it is a cognitive entity, rooted in a neuro-psychological basis. But language exists only because people interact with each other: It corresponds to a convention among a community of speakers, and answers to their communicative needs. Thirdly, language can be seen as something in itself: An autonomous, emergent entity, obeying its own inner logic. If it was not for this third Dasein of language, it would be less obvious to speak of language change as such. 

The social and cognitive nature of language informs and constrains this inner consistency. Zipf's law, for instance, may be seen as resulting from a trade-off between the ease of producing the utterance, and the ease of processing it~\cite{i2003least}. It relies thus both on the cognitive grounding of the language, and on its communicative nature. Those two external facets of language, cognitive and sociological, are similarly expected to channel the regularities of linguistic change. Modeling attempts (see~\cite{feltgen2017modeling} for an overview) have explored both how socio-linguistic factors can shape the process of this change~\cite{loreto2011statistical,ke2008language} and how this change arises through language learning by new generations of users~\cite{nowak2002computational,griffiths2007language}. Some models also consider mutations of language itself, without providing further details on the social or cognitive mechanisms of change~\cite{yanovich2016genetic}. In this paper, we adopt the view that language change is initiated by language use, which is the repeated call to one's linguistic resources in order to express oneself or to make sense of the linguistic productions of others. This approach is in line with exemplar models~\cite{pierrehumbert2001lenition} and related works, such as the Utterance Selection Model~\cite{baxter2006utterance} or the model proposed by Victorri~\cite{victorri1994continuity}, 
which describes an out-of-equilibrium shaping of semantic structure through repeated events of communication. 

Leaving aside socio-linguistic factors, we focus on a cognitive approach of linguistic change, more precisely of semantic expansion. Semantic expansion occurs when a new meaning is gained by a word or a construction (we will henceforth refer more vaguely to a linguistic `form', so as to remain as general as possible). For instance, \tit{way}, in the construction \tit{way too}, has come to serve as an intensifier (e.g. `The only other newspaper in the history of Neopia is the Ugga Ugg Times, which, of course, is way too prehistoric to read.'~\cite{neopia}). The fact that polysemy is pervasive in any language~\cite{ploux2010semantic} suggests that semantic expansion is a common process of language change and happens constantly throughout the history of a language. Grammaticalization~\cite{hopper2003grammaticalization} ---~a process by which forms acquire a (more) grammatical status, like the example of \tit{way too} above~--- and other interesting phenomena of language change~\cite{erman1993pragmaticalization,brinton2005lexicalization}, fall within the scope of semantic expansion.

Semantic change is known to be associated with an increase of frequency of use of the form whose meaning expands. This increase is expected indeed: As the form comes to carry more meanings, it is used in a broader number of contexts, hence more often. This implies that any instance of semantic change should have its empirical counterpart in the frequency rise of the use of the form. This rise is furthermore believed to follow an S-curve. The main reference on this phenomenon remains undisputedly the work of Kroch~\cite{kroch1989reflexes}, which unfortunately grounds his claim on a handful of examples only. It has nonetheless became an established fact in the literature of language change~\cite{aitchison2013language}. The origin of this pattern largely remained undiscussed, until recently: Blythe \& Croft~\cite{blythe2012s}, in addition to an up-to-date aggregate survey of attested S-curves patterns in the literature (totalizing about forty cases of language change), proposed a modeling account of the S-curve. However, they show that, in their framework, the novelty can rise only if it is deemed better than the old variant, a claim which clearly does not hold in all instances of language change. Their attempt also suffers, as most modeling works on the S-curve, from what is known as the Threshold Problem, the fact that a novelty will fail to take over an entire community of speakers, because of the isolated status of an exceptional deviation~\cite{nettle1999using}, unless a significant fraction of spontaneous adopters support it initially.

On the other hand, the S-curve is not a universal pattern of frequency change in language. From a recent survey of the frequency evolution of 14 words relating to climate science~\cite{bentley2012word}, it appears that the S-curve could not account for most of the frequency changes, and that a more general Bass curve would be appropriate instead. Along the the same line, Ghanbarnejad et al. \cite{ghanbarnejad2014extracting} investigated thirty instances of language change: 10 regarding the regularization of tense in English verbs (e.g. cleave, clove, cloven~$>$~cleave, cleaved, cleaved), 12 relating to the transliteration of Russian names in English (e.g. Stroganoff~$>$~Stroganov), and eight to spelling changes in German words (ss~$>$~{\ss}~$>$~ss) following two different ortographic reforms (in 1901 and 1996). They showed that the S-curve is not universal and that, in some cases, the trajectory of change rather obeys an exponential. This would be due to the preponderance of an external driving impetus over the other mechanisms of change, among which social imitation. The non-universality of the S-curve contrasts with the survey in~\cite{blythe2012s}, and is probably due to the specific nature of the investigated changes (which, for the spelling ones, relates mostly to academic conventions and affects very little the language system). This hypothesis would tend to be confirmed by the observation that, for the regularization of tense marking, an S-curve is observed most of the time (7 out of 10). It must also be stressed that none of these changes are semantic changes. 

In this paper, we provide a broad corpus-based investigation of the frequency patterns associated with about four hundred semantic expansions (about tenfold the aggregate survey of Blythe \& Croft~\cite{blythe2012s}). It turns out that the S-curve pattern is corroborated, but must be completed by a preceding latency part, in which the frequency of the form does not significantly increase, even if the new meaning is already present in the language. This statistical survey also allows to obtain statistical distributions for the relevant quantities describing the S-curve pattern (the rate, the width, and the length of the preceding latency part). 

Aside from this data foraging, we provide a usage-based model of the process of semantic expansion, implementing basic cognitive hypotheses regarding language use. By means of our model, we relate the micro-process of language use at the individual scale, to the observed macro-phenomenon of a recurring frequency pattern occurring in semantic expansion. The merit of this model is to provide a unified theoretical picture of both the latency and the S-curve, which are understood in relation with Cognitive Linguistics notions such as inference and semantic organization. It also predicts that the statistical distributions for the latency time and for the growth time should be of the same family as the Inverse Gaussian distribution, a claim which is in line with our data survey. 

\section*{Quantifying change from corpus data}

We worked on the French textual database {\em Frantext}~\cite{frantext}, to our knowledge the only textual database allowing for a reliable study covering several centuries (see Material and Methods and Appendix~\ref{A1}).  We studied changes in frequency of use for 408 forms which have undergone one or several semantic expansions, on a time range going from 1321 up to nowadays. We choose forms so as to focus on semantic expansions leading to a functional meaning ---~such as discursive, prepositional, or procedural meanings. Semantic expansions whose outcome remains in the lexical realm (as the one undergone by \tit{sentence}, whose meaning evolved from `verdict, judgment' to `meaningful string of words') have been left out. Functional meanings indeed present several advantages: They are often accompanied by a change of syntagmatic context, allowing to track the semantic expansion more accurately (e.g. \tit{way} in \tit{way too} + adj.); they are also less sensitive to socio-cultural and historical influences; finally, they are less dependent on the specific content of a text, be it literary or academic.

\begin{figure*}[!tbp]
\centering{\includegraphics[width=\linewidth]{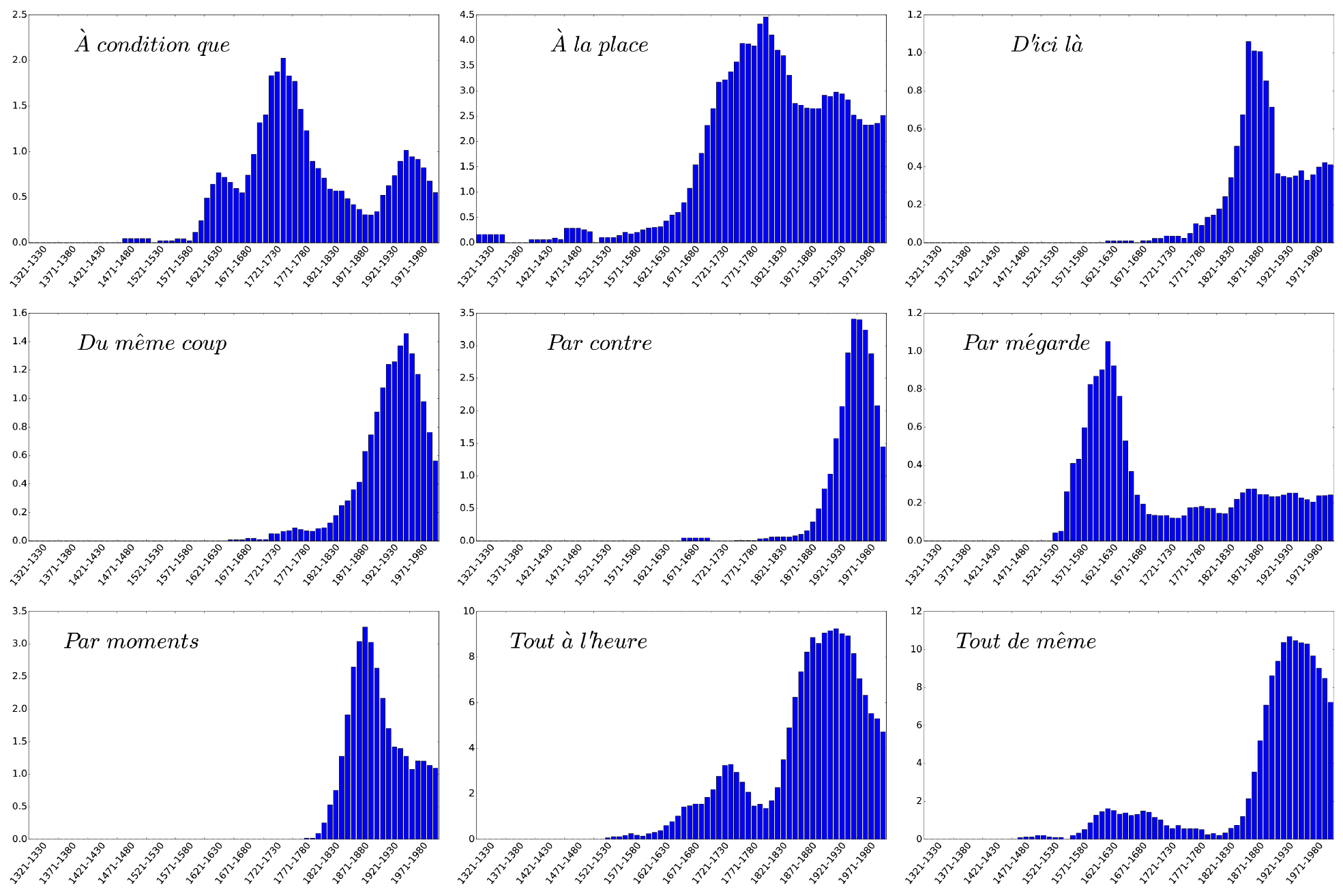}}
\caption{\small Frequency evolution on the whole time range (1321-2020) of nine different forms. Each blue bar shows the frequency associated to a decade. Frequency has been multiplied by a $10^5$ factor for an easier reading.}
\label{fig:frequency}
\end{figure*}

\begin{figure*}[!tbp]
\centering{\includegraphics[width=\linewidth]{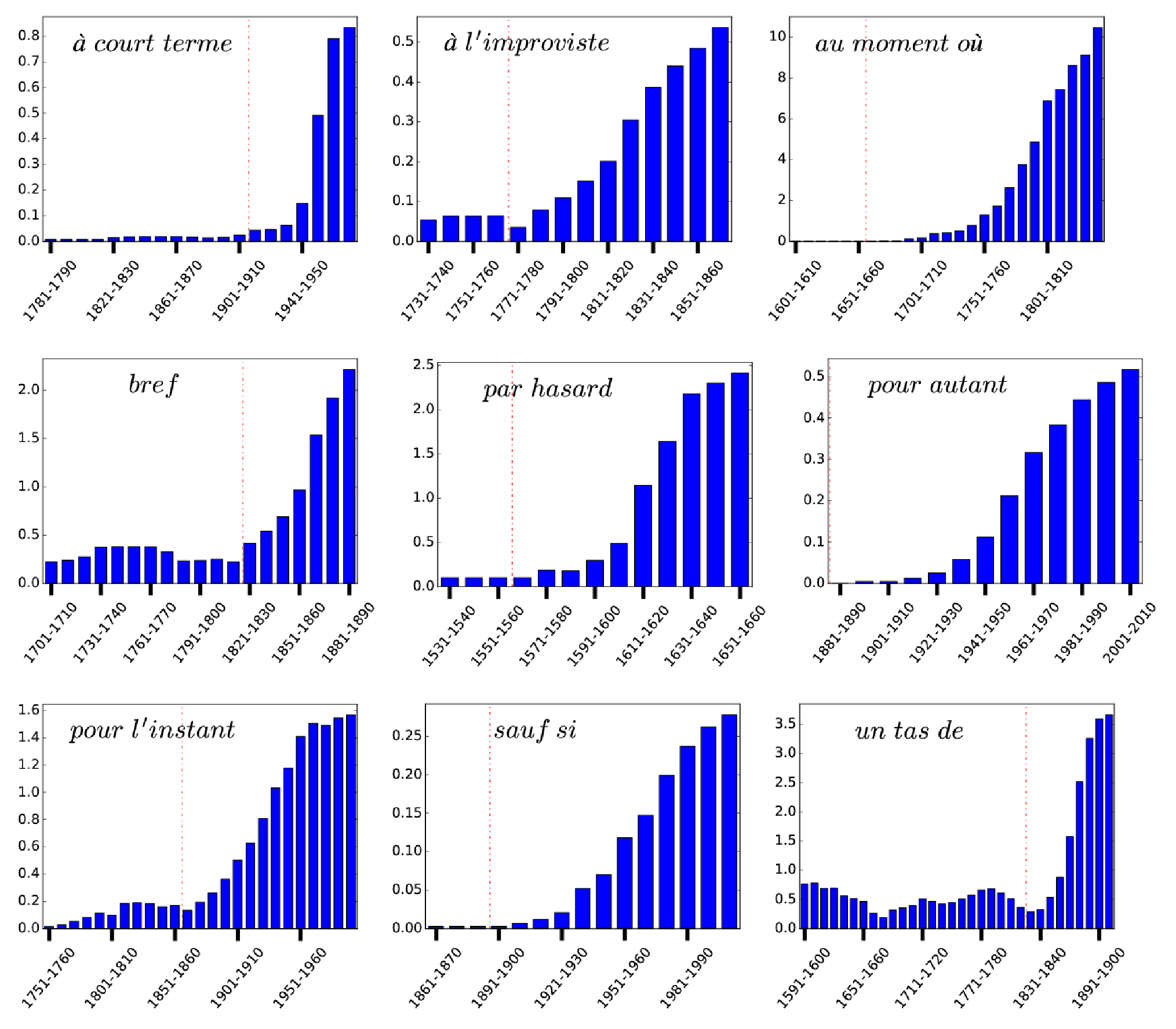}}
\caption{\small Extracted pattern of frequency rise for nine selected forms. The latency period and the S-growth are separated by a red vertical line.}
\label{fig:panel_sigmo}
\end{figure*}

\begin{figure*}[!tbp]
\centering{\includegraphics[width=\linewidth]{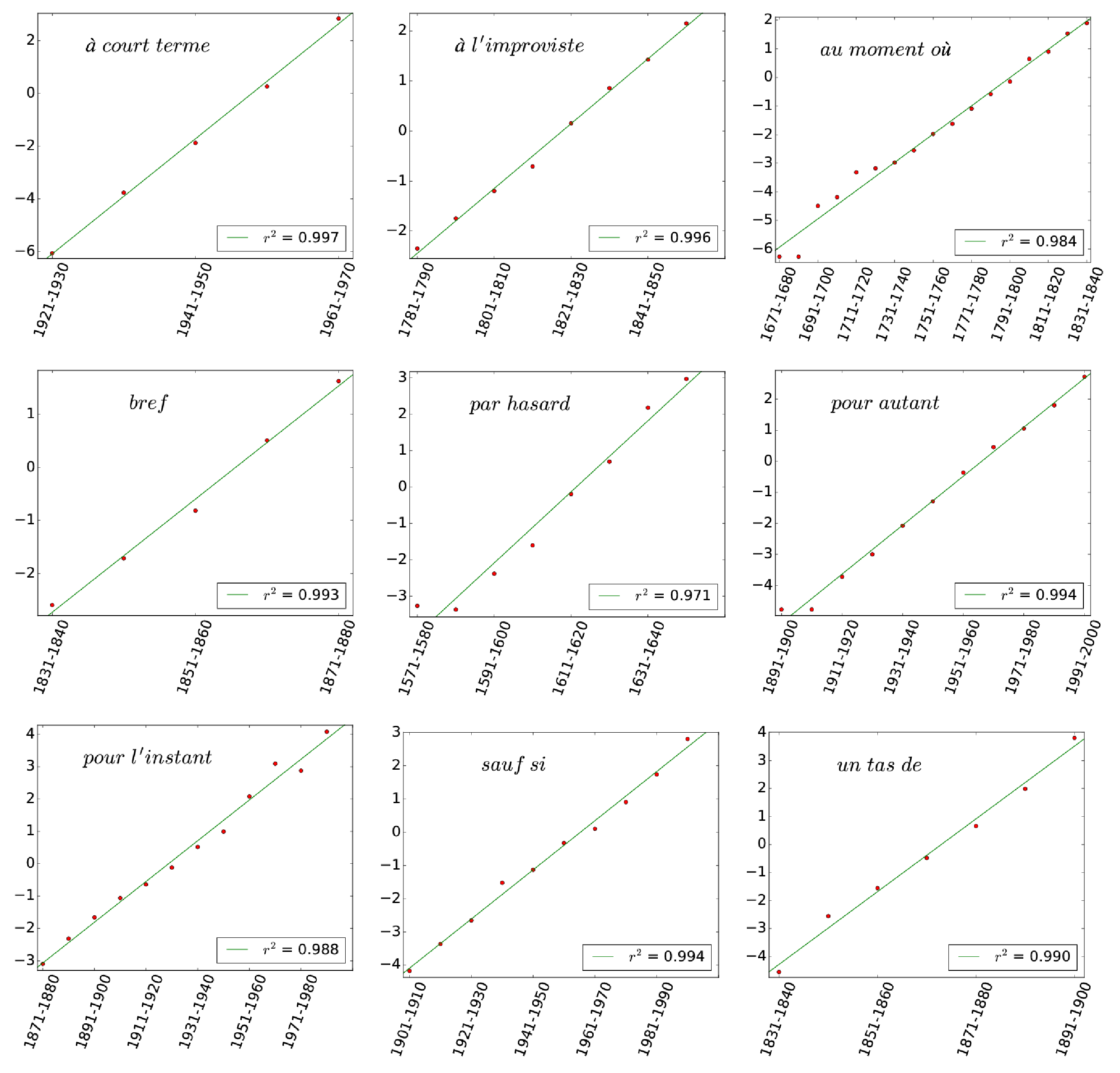}}
\caption{\small Logit transforms of the S-growth part of the preceding curves. Red dots correspond to data points and the green line to the linear fit of this set of points. The $r^2$ coefficient of the linear fit is displayed as well.}
\label{fig:panel_logit}
\end{figure*}

The profiles of frequency of use extracted from the database are illustrated on Fig.~\ref{fig:frequency} for nine forms. We find that 295 cases (which makes up more than 70\% of the total) display at least one sigmoidal increase of frequency in the course of their evolution, with a p-value significance of 0.05 compared to a random growth. We provide a small selection of the observed frequency patterns (Fig.~\ref{fig:panel_sigmo}), whose associated logit transforms (Fig.~\ref{fig:panel_logit}) follows a linear behavior, indicative of the sigmoidal nature of the growth (see Material and Methods). We thus find a robust statistical validation of the sigmoidal pattern, confirming the general claim made in the literature. 

Furthermore, we find two major phenomena besides this sigmoidal pattern. The first one is that, in most cases, the final plateau towards which the frequency is expected to stabilize after its sigmoidal rise is not to be found: The frequency immediately starts to decrease after having reached a maximum (Fig.~\ref{fig:frequency}). However, such a decrease process is not symmetrical with the increase, in contrast with other cases of fashion-driven evolution in language, e.g. first names distribution~\cite{coulmont2016diffusion}. Though this decrease may be, in a few handful of cases, imputable to the disappearance of a form (ex: \tit{apr\`es ce}, replaced in Modern French by \tit{apr\`es quoi}), in most cases it is more likely to be the sign of a narrowing of its uses (equivalent, then, to a semantic depletion). 

\begin{figure*}[!tbp]
\centering{\includegraphics[width=\linewidth]{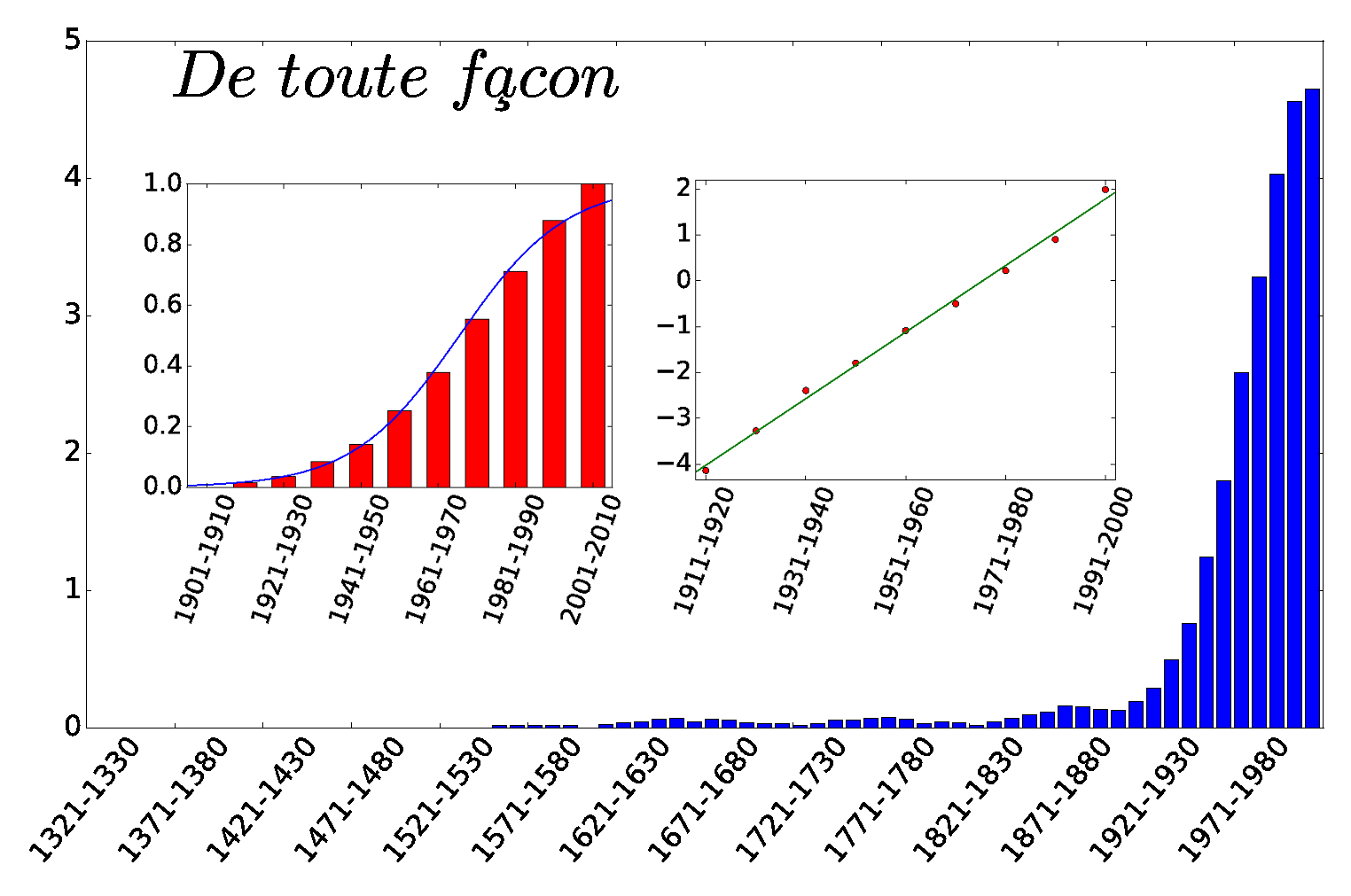}}
\caption{\small Overall evolution of the frequency of use of \tit{de toute fa\c con} (main panel), with focus on the S-shape increase (left inner panel), whose logit transformation follows a linear fit (right inner panel) with an $r^2$ of 0.996. Preceding the S-growth, one observes a long period of very low frequency (up to 35 decades).}
\label{fig:star}
\end{figure*}

The second feature is that the fast growth is most often (in 69 \% of cases) preceded by a long latency up to several centuries, during which the new form is used, but with a comparatively low and rather stable frequency (Fig.~\ref{fig:panel_sigmo}). How the latency time is extracted from data is explained in Materials \& Methods. One should note that the latency times may be underestimated: If the average frequency is very low during the latency part, the word may not show up at all in the corpus, especially in decades for which the available texts are sparse. The pattern of frequency increase is thus better conceived of as a latency followed by a growth, as exemplified by \tit{de toute fa\c{c}on} (Fig.~\ref{fig:star}) --- best translated by \tit{anyway} in English, since the present meanings of these two terms are very close, and remarkably, despite quite different origins, the two have followed parallel paths of change.

To our knowledge, this latency feature has not been documented before, even though a number of specific cases of sporadic use of the novelty before the fast growth has been noticed. For instance, it has been remarked in the case of \tit{just because} that the fast increase is only one stage in the evolution~\cite{hilpert2009assessing}. Other examples have been mentioned~\cite{denison2003log}, but it was described there as the slow start of the sigmoid. On the other hand, the absence of a stable plateau has been observed and theorized as a `reversible change'~\cite{best1990methodischer} or a `change reversal'~\cite{nevalainen2015descriptive}, and was seen as an occasional deviation from the usual S-curve, not as a pervasive phenomenal feature of the evolution. We rather interpret it as an effect of the constant interplay of forms in language, resulting in ever-changing boundaries for most of their respective semantic dominions. 

In the following, we propose a model describing both the latency and the S-growth periods. The study of this decrease of frequency following the S-growth is left for future work. 

\section*{Model}

\subsection*{A cognitive scenario}
To account for the specific frequency pattern evidenced by our data analysis, we propose a scenario focusing on cognitive aspects of language use, leaving all socio-linguistic effects back-grounded by making use of a representative agent, mean-field type, approach. We limit ourselves to the case of a competition between two linguistic variants, given that most cases of semantic expansion can be understood as such, even if the two competing variants cannot always be explicitly identified. Indeed, the variants need not be individual forms, and can be schematic constructions, paradigms of forms, or abstract patterns. Furthermore, the competition is more likely to be local, and to involve a specific and limited region of the semantic territory. If the invaded form occupies a large semantic dominion, then loosing a competition on its border will only affect its meaning marginally, so that the competition can fail to be perceptible from the point of view of the established form. 

The idealized picture is therefore as such: Initially, in some concept or context of use $C_1$, one of the two variants, henceforth noted $Y$, is systematically chosen, so that it conventionally expresses this concept. The question we address is thus how a new variant, say $X$, can be used in this context and eventually evict the old variant $Y$?

\ The main hypothesis we propose is that the new variant almost never is a brand new merging of phonemes whose meaning would pop out of nowhere. As Haspelmath highlights~\cite{haspelmath1999grammaticalization}, a new variant is almost always a periphrastic construction, i.e., actual parts of language, put together in a new, meaningful way. Furthermore, such a construction, though it may be exapted to a new use, may have showed up from time to time in the time course of the language history, in an entirely compositional way; this is the case for \tit{par ailleurs}, which incidentally appears as early as the \textsc{xiv}\textsuperscript{th} in our corpus, but arises as a construction in its own right during the first part of the \textsc{xix}\textsuperscript{th} century only. In other words, the use of a linguistic form $X$ in a context $C_1$ may be entirely new, but the form $X$ was most probably already there in another context of use $C_0$, or equivalently, with another meaning.

We make use of the well-grounded idea~\cite{hudson2007language} that there exists links between concepts due to the intrinsic polysemy of language: There are no isolated meanings, as each concept is interwoven with many others, in a complicated tapestry. These links between concepts are asymmetrical, and they can express both universal mappings between concepts~\cite{heine1997cognitive,dellert2016using} and cultural ones (e.g. entrenched metaphors~\cite{lakoff2008metaphors}). As the conceptual texture of language is a complex network of living relations rather than a collection of isolated and self-sufficient monads, semantic change is expected to happen as the natural course of language evolution and to occur repetitively throughout its history, so that at any point of time, there are always several parts of language which are undergoing changes. The simplest layout accounting for this network structure in a competitive situation consists then in two sites, such that one is influencing the other through a cognitive connexion of some sort. 

\subsection*{Model formalism} 

\ We now provide details on the modeling of a competition between two variants $X$ and $Y$ for a given context of use, or concept, $C_1$, also considering the effect exerted by the related context or concept $C_0$ on this evolution. 

$\bullet$ Each concept $C_i, i=0,1$, is represented by a set of exemplars of the different linguistic forms. We note $N_{\mu}^i(t)$ the number at time $t$ of encoded exemplars (or occurrences) of form $\mu \in \{X, Y\}$, in context $C_i$, in the memory, of the representative agent. 

$\bullet$ The memory capacity of an individual being finite, the population of exemplars attached to each concept $C_i$ has a finite size $M_i$. For simplicity we assume that all memory sizes are equal ($M_0 = M_1 = M$). As we consider only two forms $X$ and $Y$, for each $i$ the relation $N_{X}^i(t) + N_{Y}^i(t) = M$  always hold: We can focus on one of the two forms, here $X$, and drop out the form subscript, granted that all quantities refer to $X$.

$\bullet$ The absolute frequency $x^i_t$ of form $X$ at time $t$ in context $C_i$ --- the fraction of `balls' of type $X$ in the bag attached to $C_i$ --- is thus given by the ratio $N^i(t)/ M$. In the initial situation, $X$ and $Y$  are assumed to be established conventions for the expression of $C_0$ and $C_1$ respectively, so that we start with $N^0(t = 0) = M$ and $N^1(t = 0) = 0$.

$\bullet$ Finally, $C_0$ exerts an influence on context $C_1$, but this influence is assumed to be unilateral. Consequently, the content of $C_0$ will not change in the course of the evolution and we can focus on $C_1$. An absence of explicit indication of context is thus to be understood as referring to $C_1$. 

\ The dynamics of the system runs as follows. At each time $t$, one of the two linguistic forms is chosen to express concept $C_1$. The form $X$ is uttered with some probability $P(t)$, to be specified below, and $Y$ with probability $1-P(t)$. In order to keep constant the memory size of the population of occurrences in $C_1$,  a past occurrence is randomly chosen (with a uniform distribution) and the new occurrence takes its place. This dynamics is then repeated a large number of times. Note that this model focuses on a speaker perspective (for alternative variants, see Appendix~\ref{hearer_mech}). 

\begin{figure*}[!tbp]
\centering{\includegraphics[width=\linewidth]{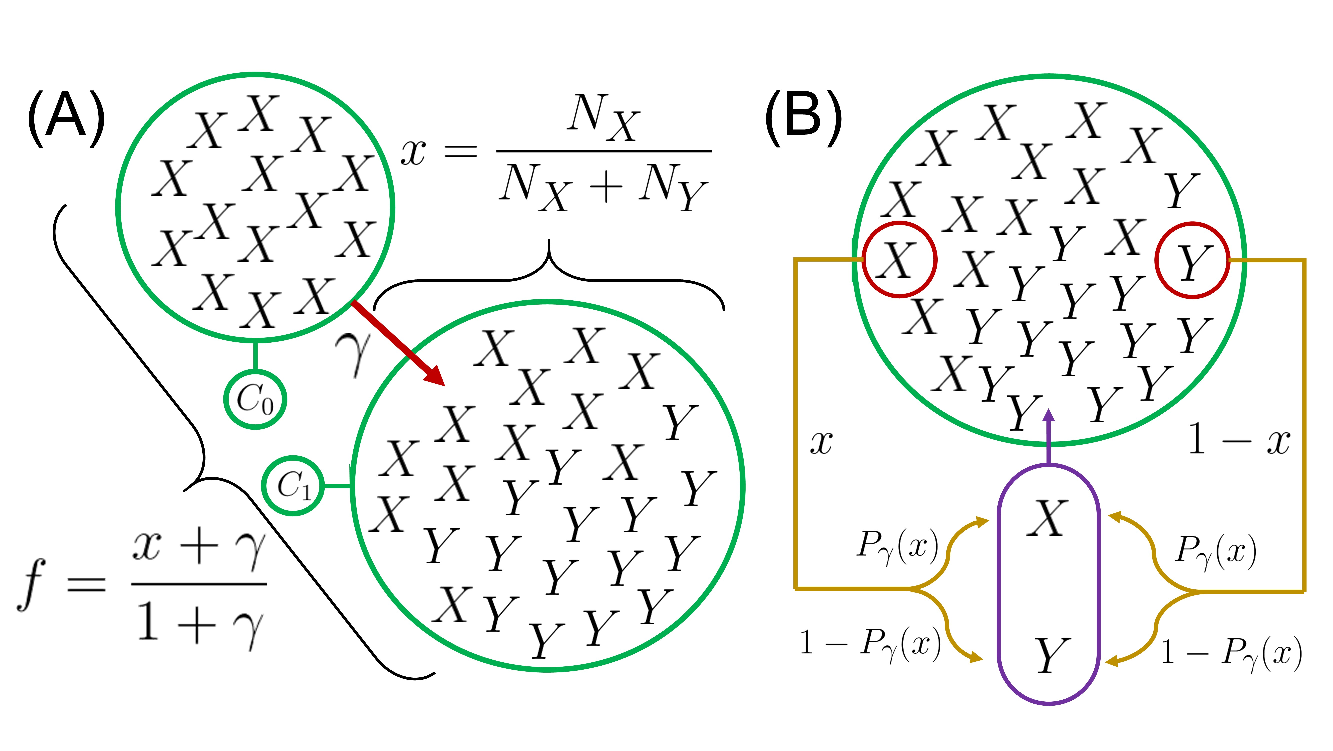}}
\caption{\small Schematic representation of model mechanisms. (A) Difference between absolute frequency $x$ and relative frequency $f$ in context $C_1$. Absolute frequency $x$ is given by the ratio of $X$ occurrences encoded in $C_1$. Effective frequency $f$ also takes into account the $M$ occurrences contained in the influential context $C_0$, with a weight $\gamma$ standing for the strength of this influence. (B) Schematic view of the process. At each iteration, either $X$ or $Y$ is chosen to be produced and thus encoded in memory, with respective probability $P_{\gamma}(x)$ and $1-P_{\gamma}(x)$; the produced occurrence is here represented in the purple capsule. Another occurrence, already encoded in the memory, is uniformly chosen to be erased (red circle) so as to keep the population size constant. Hence the number of $X$ occurrences, $N_X$, either increases by $1$ if $X$ is produced and $Y$ erased, decreases by $1$ if $Y$ is produced and $X$ erased, or remains constant if the erased occurrence is the same as the one produced.}
\label{fig:process}
\end{figure*}

We want to explicit the way $P(t)$ depends on $x(t)$, the absolute frequency of $X$ in this context at time $t$. The simplest choice would be $P(t)=x(t)$. However, we want to take into account several facts. As context $C_0$ exerts an influence on context $C_1$, denoting by $\gamma$ the strength of this influence (see Appendix~\ref{interpret_gamma} for an extended discussion on this parameter), we assume the probability $P$ to rather depend on an effective frequency $f(t)$ (Fig.~\ref{fig:process}A), 
{\small\beq 
f(t) = \frac{N^1(t) + \gamma N^0(t)}{M + \gamma M} = \frac{x(t)+\gamma }{1+\gamma} \,. \eneq}
We now specify the probability $P(f)$ to select $X$ at time $t$ as a function of $f=f(t)$. First, $P(f)$ must be nonlinear. Otherwise, the change would occur with certainty as soon as the effective frequency $f$ of the novelty is non-zero: That is, insofar two meanings are related, the form expressing the former will also be recruited to express the latter. This change would also start in too abrupt a way, while sudden, instantaneous takeovers are not known to happen in language change. Second, one should preserve the symmetry between the two forms, that is, $P(f) = 1 - P(1-f)$, as well as verify $P(0)=0$ and $P(1)=1$. Note that this symmetry is stated in terms of the effective frequency $f$ instead of the actual frequency $x$, as production in one context always accounts for the contents of neighboring ones. 

For the numerical simulations, we made the following specific choice which satisfies these constraints:
{\small\beq
P(f) = \frac{1}{2} \left\{ 1 + \tanh \left( \beta \, \frac{f - (1 - f)}{\sqrt{f(1 - f)}} \right) \right \} \,,
\eneq}
where $\beta$ is a parameter governing the non-linearity of the curve. Replacing $f$ in terms of $x$, the probability to choose $X$ is thus a function $P_{\gamma}(x)$ of the current absolute frequency $x$:
{\small\beq
P_{\gamma}(x) = \frac{1}{2} \left\{ 1 + \tanh \left( \beta \, \frac{2x- 1+\gamma}{\sqrt{(x+\gamma)(1-x)}} \right) \right \} 
\eneq
}

\subsection*{Analysis: Bifurcation and latency time}
\ The dynamics outlined above (Fig.~\ref{fig:process}B) is equivalent to a random walk on the segment $[0;1]$ with a reflecting boundary at $0$ and an absorbing one at $1$, and with steps of size $1/M$. The probability of going forward at site $x$ is equal to $(1-x)P_{\gamma}(x)$, and the probability of going backward to $x(1-P_{\gamma}(x))$. 

For large $M$, a continuous, deterministic approximation of this random walk leads, after a rescaling of the time $M\,t~\rightarrow~t$, to a first order differential equation for $x(t)$:
{\small\beq \label{speed}
\dot{x} = P_{\gamma}(x) -x \,.
\eneq}

\begin{figure*}[!tbp]
\centering{\includegraphics[width=\linewidth]{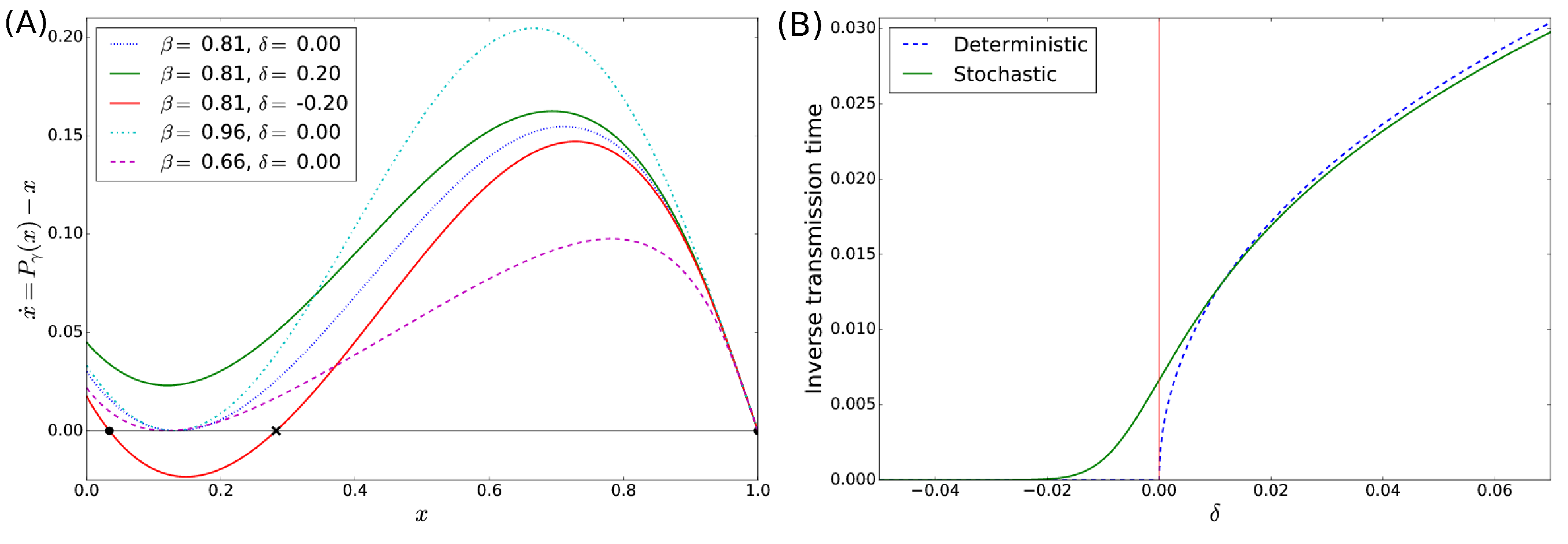}}
\caption{\small Evidence of a near-critical behavior. (A)~Speed $\dot{x}$ of the deterministic process for each of the sites, for different values of $\beta$ and $\delta = (\gamma - \gamma_c) / \gamma_c$, the distance to threshold.  Depending on the sign of $\delta$, there is either one or three fixed points. (B)~Inverse transmission time (time required for the system to go from $0$ to $1$), for the deterministic process (blue dotted line), and for the averaged stochastic process (green line), as a function of the control parameter $\delta$. Deterministic transmission time diverges at the transition while averaged stochastic transmission time remains finite.}
\label{fig:critical}
\end{figure*}

This dynamics admits either one or three fixed points (Fig.~\ref{fig:critical}A), $x = 1$ always being one. Below a threshold value $\gamma_c$, which depends on the non-linearity parameter $\beta$, a saddle-node bifurcation occurs and two other fixed points appear close to a critical frequency $x_c$. The system, starting from $x = 0$, 
is then stuck at the smallest stable fixed point. The transmission time, i.e. the time required for the system to go from $0$ to $1$, becomes therefore infinite (Fig.~\ref{fig:critical}B). Above the threshold value $\gamma_c$, only the fixed point $x = 1$ remains, so that the new variant eventually takes over the context for which it is competing. Our model thus describes how the strengthening of a cognitive link can trigger a semantic expansion process. 

Slightly above the transition, a stranglehold region appears where the speed almost vanishes. Accordingly, the time spent in this region diverges. The frequency of the new variant will stick to low values for a long time, in a way similar to the latent behavior evidenced by our dataset. This latency time in the process of change can thus be understood as a near-critical slowing down of the underlying dynamics. 

Past this deterministic approximation, there is no more clear-cut transition (Fig.~\ref{fig:critical}B) and the above explanation needs to be refined. The deterministic speed can be understood as a drift velocity of the Brownian motion on the $[0;1]$ segment, so that in the region where the speed vanishes, the system does not move in average. In this region of vanishing drift, the frequency fluctuates over a small set of values and does not evolve significantly over time. Once it escapes this region, the drift velocity drives the process again, and the replacement process takes off. Latency time can thus be understood as a first-passage time out of a trapping region. 

\section*{Numerical results}

\subsection*{Model simulations}

\ We ran $10,000$ numerical simulations of the process described above (Fig.~\ref{fig:process}B), with the following choice of parameters: $\beta = 0.808$, $\delta = 0.0$ and $M = 5000$, where $\delta = (\gamma - \gamma_c) / \gamma_c$ is the distance to the threshold. The specific value of $\beta$ has been chosen to maximize $x_c$. Since $x_c$ is the frequency at which the system gets stuck if $\gamma$ is slightly below the threshold, it corresponds to the assumption that, even if the convention is not replaced, there is room for synonymic variation and the new variant can be used marginally. We chose $\delta = 0.0$ in order for the system to be purely diffusive in the vicinity of $x_c$. The choice of $M$ is arbitrary. 

Even if this set of parameters remains the same throughout the different simulation runs, the quantities describing each of the $10,000$ S-curves generated that way, especially the rate and the width, will change. It becomes then possible to obtain the statistical distributions of these quantities. Thus, while there is no one-to-one comparison between a single outcome of the numerical process and a given instance of change, we can discuss whether their statistical properties are the same. 

\begin{figure*}[!tbp]
\centering{\includegraphics[width=\linewidth]{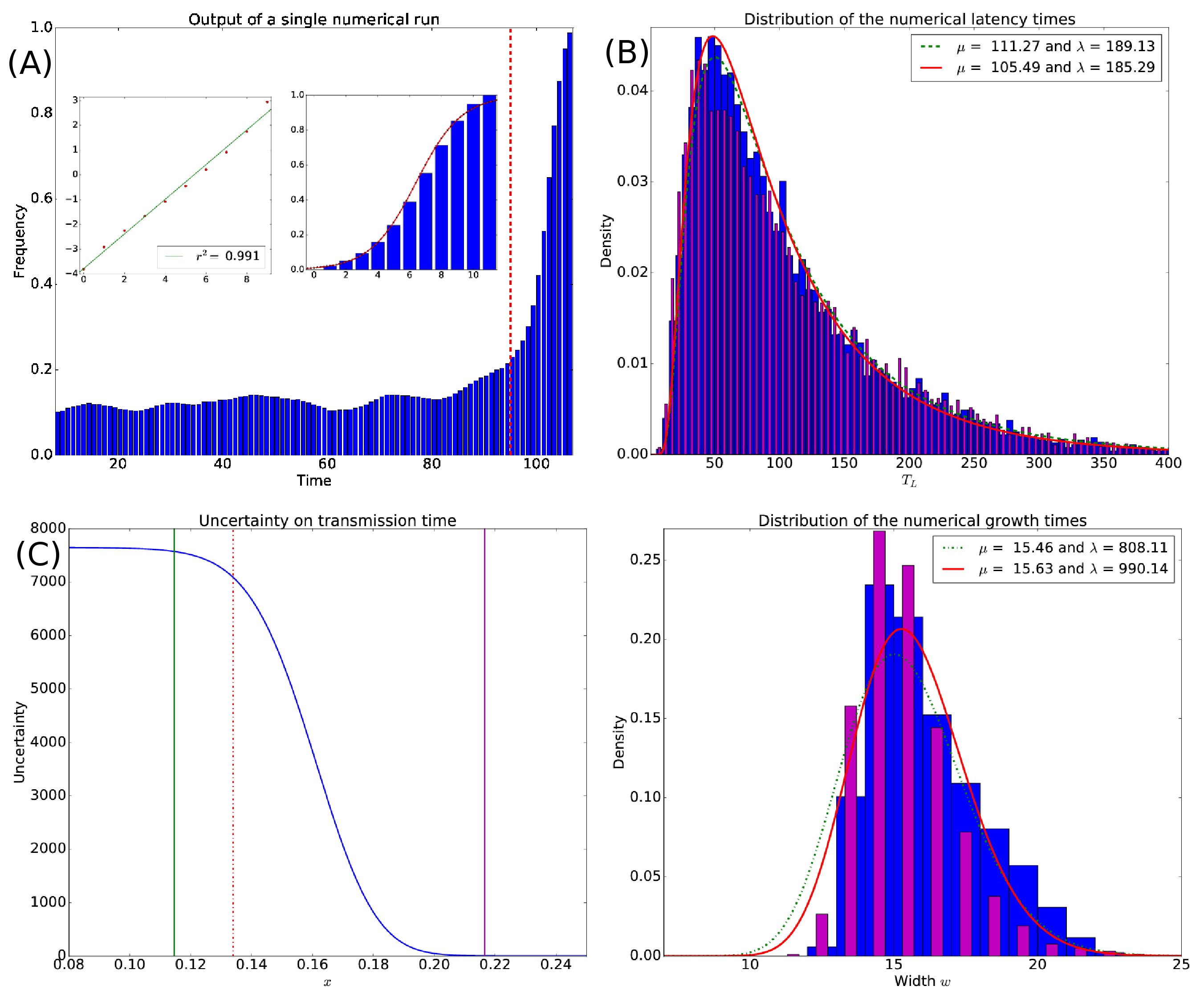}}
\caption{\small Numerical simulation of latency and growth times at the critical threshold. (A)~Time evolution of the frequency of produced occurrences (output of a single run). Growth part and latency part are separated by a red dotted line. The logit transform (with linear fit) of the growth is shown in the left inset, alongside with the sigmoidal fit of the rescaled frequency of the growth part (right inset). (B)~Distribution of latency times (top) and growth times (bottom) over $10k$ processes, extracted from an empirical approach (blue wide histogram) and a first-passage time one (magenta thin histogram), with their respective Inverse Gaussian fits (in red: Empirical approach; in green: First-passage time approach). (C)~Uncertainty on the transmission time given the position of the walker. The entrance and the exit of the trap are shown, respectively, by green and magenta lines. The red dotted line indicates the critical frequency $x_c$. The trap corresponds to the region where the uncertainty drops from a high value to a low value.}
\label{fig:numerical_pattern}
\end{figure*}

From the model simulations, data is extracted and analyzed in two parallel ways. On one side, simulations provide surrogate data: We can mimic the corpus data analysis and count how many tokens of the new variant are produced in a given timespan (set equal to $M$), to be compared with the total number of tokens produced in this timespan. We then extract 'empirical' latency and growth times (Fig.~\ref{fig:numerical_pattern}A), applying the same procedure as for the corpus data.

One the other side, for each run we track down the position of the walker, which is the frequency $x(t)$ achieved by the new variant at time $t$. This allows to compute first passage times. We then alternatively compute analytical latency and growth times (`analytical' to distinguish them from the former `empirical' times) as follows. Latency time is here defined as the difference between the first-passage times at the exit and the entrance of a `trap' region (see Appendix~\ref{IB} for additional details). Analytical growth time is defined as the remaining time of the process once this exit has been reached. Their distribution over $10,000$ runs of the process are fitted with an Inverse Gaussian distribution, which would be the expected distribution if the jump probabilities were homogeneous over the corresponding regions (an approximation then better suited for latency time than for growth time). Figure \ref{fig:numerical_pattern}B shows the remarkable agreement between  the `empirical' and `analytical' approaches, together with their fits by an Inverse Gaussian distribution.

Crucially, those two macroscopic phenomena, latency and growth, are thus to be understood as of the same nature, which explains why their statistical distribution must be of the same kind. Furthermore, the boundaries of the trap region leading to the best correspondence between first passage times and empirically determined latency and growth times are meaningful, as they correspond to the region where the uncertainty on the transmission time significantly decreases (Fig.~\ref{fig:numerical_pattern}C). 

\subsection*{Confrontation with corpus data}
\begin{figure*}[!tbp]
\centering{\includegraphics[width=\linewidth]{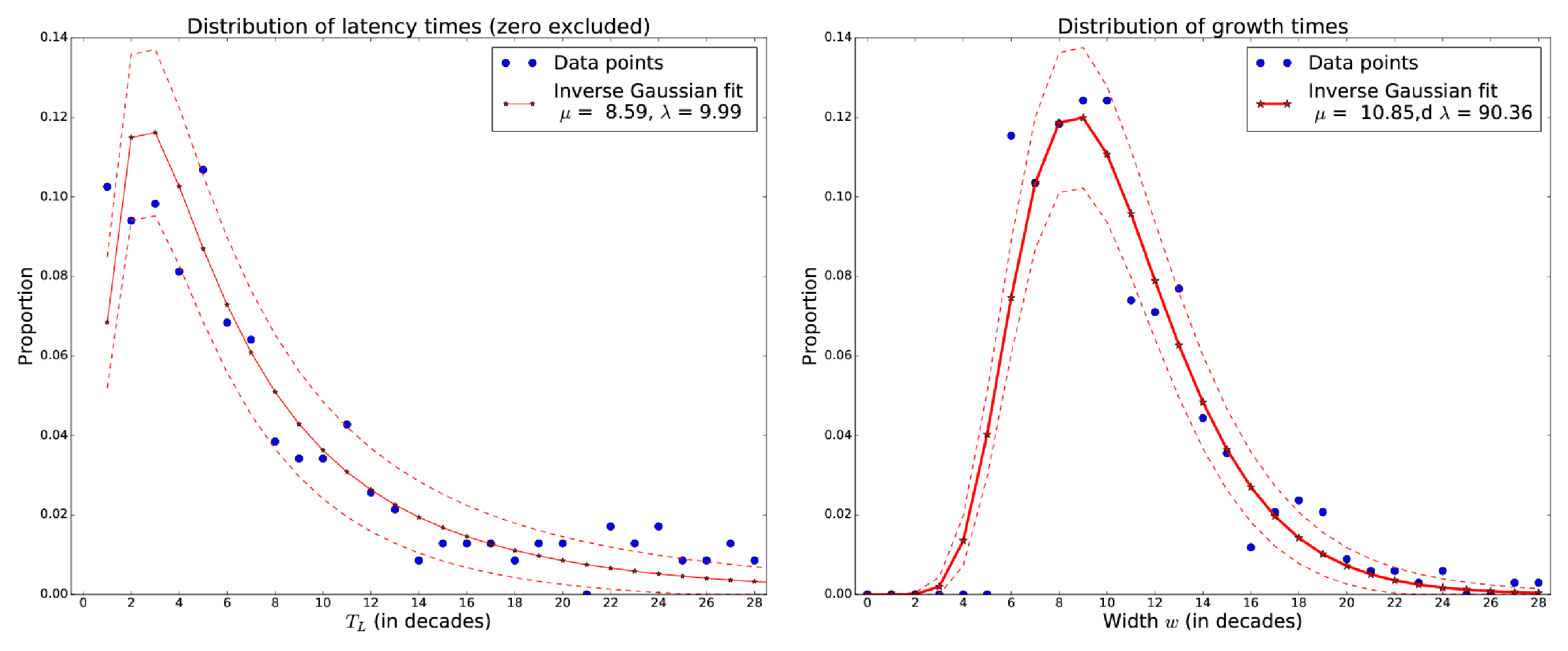}}
\caption{\small Inverse Gaussian fit of the latency times $T_L$ (left) and the growth times $w$ (right), extracted from corpus data. Data points are shown by blue dots, the Inverse Gaussian fit being represented as a full red curve with star-shaped marks. The dashed red lines represent the standard deviation from the model. We detail in Materials \& Methods how we extracted these growth times and latency times from corpus data.}
\label{fig:data_IG}
\end{figure*}

\ Our model predicts that both latency and growth times should be governed by the same kind of statistics, Inverse Gaussian being a suited approximation of those. Inverse Gaussian distribution is governed by two parameters, its mean $\mu$ and a parameter $\lambda$ given by the ratio $\mu^3 / \sigma^2$, $\sigma^2$ being the variance. We thus try to fit the corpus data with an Inverse Gaussian distribution (Fig.~\ref{fig:data_IG}). In both cases, the Kullback-Leibler divergence between the data distribution and the Inverse Gaussian fit is equal to 0.10. The rate $h$ (slope of the logit) also follows a non-trivial distribution, as shown in Appendix~\ref{A4}.

Although there are short growth times in the frequency patterns of the forms we studied, below six decades they are not described by enough data points to assess reliably the specificity of the sigmoid fit. On Fig.~\ref{fig:data_IG} there are therefore no data for these growth times. The Inverse Gaussian fit is not perfect, and is not expected to be: The model only predicts the distribution to be of the same family as the Inverse Gaussian. Satisfyingly, among a set of usual distributions (exponential, Poisson, Gaussian, Maxwellian), the Inverse Gaussian proves to be the most adequate for both the growth and the latency (see Appendix~\ref{A4} for additional details).

The main quantitative features extracted from the dataset are thus correctly mirrored by the behavior of our model. We confronted the model with the data on other quantities, such as the correlation between growth time and latency time, two quantities which our model predicts to be independent. There again, the model proves to match appropriately these quantitative aspects of semantic expansion processes (see Appendix~\ref{A5}).

\section*{Discussion} 

Based on a corpus-based analysis of frequency of use, we have established two robust stylized facts of semantic change: An S-curve of frequency growth, already evidenced in the literature, and a preceding latency period during which the frequency remains more or less constant, typically at a low value.  We have proposed a model predicting that these two features, albeit qualitatively quite different, are two aspects of one and the same phenomenon.

Our analysis is based on the a priori assumption that a frequency rise is caused by a semantic expansion. An alternative would be the reverse mechanism, that semantic expansion is induced by an increase in the frequency of use. Actually, it is not infrequent to find unambiguous traces of the semantic expansion throughout and even before the latency phase. Also, we often looked for forms in a syntactic context compatible only with the new meaning --- e.g. for \tit{j'imagine} we searched specific intransitive patterns, like ``il y a de quoi, j'imagine, les faire \'etrangler'' (1783) (``There's good reason to have them strangled, I suppose'') --- so that, in such cases, it leaves no doubt that the latency phase and the frequency rise are posterior to the semantic expansion. The model, however, does not exclude that both mechanisms are at work, as discussed in Appendix~\ref{interpret_gamma}.

The detailed hypotheses on which our model lies are well-grounded on claims from Cognitive Linguistics: Language is resilient to change (non-linearity of the $P$ function); language users have cognitive limitations; the semantic territory is organized as a network whose neighboring sites are asymmetrically influencing each other. The overall agreement with empirical data tends to suggest that language change may indeed be cognitively driven by semantic bridges of different kinds between the concepts of the mind, and constrained by the mnemonic limitations of this very same mind. 

According to our model, the onset of change depends on the strength of the conceptual link between the source context and the target context: If the link is strong enough, that is, above a given threshold, it serves as a channel so that a form can `invade' the target context and then oust the previously established form. In a sense, the sole existence of this cognitive mapping is already a semantic expansion of some sort, yet not necessarily translated into linguistic use. Latency is specifically understood as resulting from a near-critical behavior: If the link is barely strong enough for the change to take off, then the channel becomes extremely tight and the invasion process slows down drastically. These narrow channels are likely to be found between lexical and grammatical meanings~\cite{heine2002role,diewald2006context}. This would explain why the latency-growth pattern is much more prominent in the processes of grammaticalization, positing latency as a phenomenological hint of this latter category.

As acknowledged by a few authors~\cite{ogura1996snowball,croft2000explaining}, it is interesting to note that, in the literature, the S-growth is given two very different interpretations. According to the first one, an S-curve describes the spread of the novelty in a community of speakers~\cite{osgood1954psycholinguistics,weinreich1968empirical,haspelmath2004directionality,ke2008language}, as for the second one, it reflects the spread in language itself, the new variant being used in an increasing number of contexts~\cite{mcmahon1994understanding,levin2006collective,aitchison2013language,burridge2016understanding}. According to the interpretation we give to our model, the diffusion chiefly happens over the linguistic memory of the whole speech community. It does not involve some binary conversion of individuals towards the new variant; it is a spread within the individuals rather than a spread among them. On the other hand, the S-curve arises in the taking over a single context, and does not rely on a further diffusion over additional contexts to appear. Though the latter spread needs thus not be responsible for the S-shape, it may nonetheless influence the evolution in other ways (e.g. the total duration). The interplay between the specific features of an S-curve and the structure of the conceptual network remains to be investigated. 

We note, however, that our model may be given a different, purely socio-linguistic interpretation, as discussed in Appendix~\ref{socio_interpretation}. Nevertheless, several arguments argue against this interpretation. First, the semantic evolution involves very long timescales, up to several centuries~\cite{levin2006collective}; second, societal diffusion, of a new technological device for instance, is associated to a specific scaling law between the steep and duration of the S-curve of -2/3~\cite{michard2005theory}, which is very different from the behavior of the forms in our dataset, where no scaling law is to be found (the two parameters are related by a trivial -1.0 exponent; see Appendix~\ref{A5}). 

Recently, the nature of linguistic change has been investigated through different case studies, separating internal (imitation between members of a community) and external  (e.g. linguistic reforms from language academies) factors of change~\cite{ghanbarnejad2014extracting}. While internal factors give rise to an S-curve, external factors lead to an exponential growth of frequency; hence, the S-curve is not the only dynamics by which language change can occur.  However, in this work, agents choose between the two variants on a binary basis, and language-based mechanisms, such as the network asymmetric links at the core of our own model, would count as an external mechanism. These strong differences make it difficult to quantitatively compare their approach and ours, albeit it is to be agreed that S-curves contain crucial information on language change and need to be investigated and quantified further on. Moreover, as semantic change is seldom driven by external forces such as linguistic reforms, the exponential pattern is not to be expected in this case, and indeed we have not found it in our dataset. 

Finally, we argue that our results, though grounded on instances of semantic expansion in French, apply to semantic expansion in general. The time period covered is long enough (700 years) to exclude the possibility that our results be ascribable to a specific historical, sociological, or cultural context. The French language itself has evolved, so that Middle French and contemporary French could be considered as two different languages, yet our analysis apply to both indistinctly. Besides, the latency-growth pattern is to be found in other languages; for instance, although Google Ngram cannot be used here for a systematic quantitative study, specific queries for constructions such as \tit{way too}, \tit{save for}, \tit{no matter what}, yield qualitative frequency profiles consistent with our claims. Our model also tends to confirm the genericity of this pattern, as it relies on cognitive mechanisms whose universality has been well evidenced~\cite{heine2002world,lapolla2015logical}.

\section*{Materials and methods}

\subsection*{Corpus data}
We worked on the \tit{Frantext} corpus~\cite{frantext}, which in 2016 contained 4674 texts and 232 millions of words for the chosen time range. More details are given in Appendix~\ref{A1}. It would have been tempting to make use of the large database Google Ngram, yet it was not deemed appropriate for our study, as we explain in Appendix~\ref{A6}.

We studied changes in frequency of use for about 400 instances of semantic expansion processes in French, on a time range going from 1321 up to nowadays. See Appendix~\ref{A7} for a complete list of the studied forms. 

\subsection*{Extracting patterns from corpus data} 

\paragraph{Measuring frequencies} 
We divided our corpus into 70 decades. Then, for each form, we recorded the number of occurrences per decade, dividing this number by the total number of occurrences in the database for that decade. The output number is called here the \tit{frequency} of the form for the decade, and is noted $x_i$ for decade $i$. In order to smooth the obtained data, we replaced $x_i$ by a moving average, that is, for $i \geq i_0 + 4$, $i_0$ being the first decade of our corpus:
$x_i \leftarrow \frac{1}{5} \sum_{k=i-4}^i x_k\,.$ 
 
\paragraph{Sigmoids}
We looked for major increases of frequency. When such a major shift is encountered, we automatically (see below) identify frequencies $x_{min}$ and $x_{max}$, respectively at the beginning and the end of the increasing period. If we respectively note $i_{start}$ and $i_{end}$ the decades for which $x_{min}$ and $x_{max}$ are reached, then we define the width (or growth time) $w$ of the increasing period as $w = i_{end}-i_{start} +1$. To quantify the sigmoidal nature of this growth pattern, we apply the logit transformation to the frequency points between $x_{min}$ and $x_{max}$:
{\small\beq
y_i =  \log \left( \frac{x_i-x_{min}}{x_{max}-x_i} \right) \,.
\label{eq:yi}
\eneq}
If the process follows a sigmoid $\tilde{x}_i$ of equation:
{\small\beq
\tilde{x}_i = x_{min}+\frac{x_{max}-x_{min}}{1+e^{-hi-b}}\,,
\eneq}
then the logit transform of this sigmoid satisfies:
$\tilde{y}_i =h\, i + b\,.$
We thus fit the $y_i$'s given by (\ref{eq:yi}) with a linear function, which gives the slope (or rate) $h$ associated with it, the residual $r^2$ quantifying the quality of the fit. The boundaries $i_{start}$ and $i_{end}$ have been chosen so as to maximize $w$, with the constraint that the $r^2$ of the linear fit should be at least equal to a value depending on the number of points, in order to insure that the criterion has a p-value significance of less than 0.05 according to a null model of frequency growth. Further explanations are provided in Appendix~\ref{pvalue}. 

\paragraph{Latency period}
In most cases (69\% of sigmoidal growths), one observes that the fast increasing part is preceded by a phase during which the frequency remains constant or nearly constant. The duration of this part, denoted by $T_L$ (latency time) in this paper, is identified automatically as follows. Starting from the decade $i_{start}$, previous decades $j$ are included in the latency period as long as they verify $ \lvert x_j - x_{min} \rvert < 0.15 * (x_{max} - x_{min})$ and $x_j > 0$, and cease to be included either as soon as the first condition is not verified, or if the second condition does not hold for a period longer than 5 decades. Then the start $i_{lat}$ of the latency point is defined as the lowest $j$ verifying both conditions, so that $T_L$ is given by $T_L = i_{start} - i_{lat}$. 

\section*{Data Availability}

The datasets supporting this article have been uploaded as part of the supplementary material (see Appendix~\ref{data}).

\section*{Acknowledgements}

We thank B. Derrida for a useful discussion on random walks and L. Bonnasse-Gahot for his useful suggestions. We also thank the two anonymous reviewers who provided relevant and constructive feedback on this paper.

\section*{Funding}

QF acknowledges a fellowship from PSL Research University. BF is a CNRS member. JPN is senior researcher at CNRS and director of studies at the EHESS.

\newpage

\appendix 

\onecolumngrid

\newpage

\section{Further data analysis}

\subsection{Null model of frequency growth and significance of the sigmoidal fit \label{pvalue}}

To evaluate the significance of the sigmoidal fit, we need to compare it with a null model of frequency growth. However, what would be the null hypothesis in this case is far from obvious. Given that the frequency has risen from $x_{min}$ to $x_{max}$ in a time $w$, which model of growth would be the closest to an assumption-free one? As the frequency can be rescaled using the following formula:
\begin{equation}
x \leftarrow \frac{x - x_{min}}{x_{max} - x_{min}} \, ,
\end{equation}
the matter can be simplified by considering a growth from $0$ to $1$. 

\subsubsection{Stochastic null model}

A simple choice is to consider the following random walk, with Gaussian jumps at each time step:
\begin{equation}
x_{t+1} = x_t + \frac{1}{w} (1 + \eta_t) \, ,
\end{equation}
where $\eta_t$ is a random term drawn from a normal distribution of mean $0$ and variance $1$, with the initial condition $x_0 = 0$. The mean process would be a linear growth from $0$ to $1$ with $w$ steps of size $1/w$. 

In the main text, we extracted the S-curve according to the following procedure:
\begin{itemize}
\item search for all pairs $t_{min}$ and $t_{max}$, with $t_{max} - t_{min} > 5$, so that the logit transform of the data points $x_t$ in-between is associated with a linear fit of sufficiently good quality
\item retain only the pairs associated with the greatest possible width $w$ ($w = t_{max} - t_{min} +1$);
\item select among those ones the pair with the best $r^2$ coefficient of the linear fit of the logit.
\end{itemize}
The question is then: what is a linear fit of sufficiently good quality? Now that we have a null model, we can devise a criterion $r^2_{min}(w)$ so that the fit is associated with a $p$-value below $0.05$: if the $r^2$ of the fit is higher than this criterion, then the sigmoidal fit is deemed significant. 

\begin{figure*}[!bp]
  \centering
  \includegraphics[width=\linewidth]{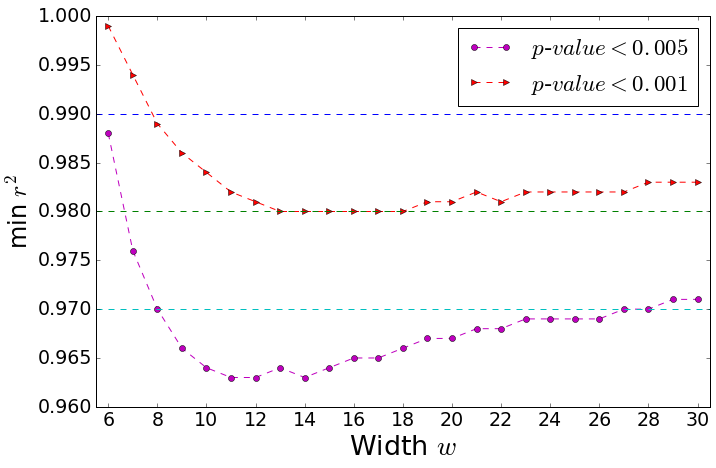}
  \caption{Minimal quality of the linear fit of the logit transform so as to insure the significance of a sigmoidal fit of the data as compared to a random null model of frequency growth.}
  \label{fig:r2_criterions}
\end{figure*}

To do so, for a given value of $w$, we generated $50,000$ growth processes and computed the ratio $p$ of processes obeying the criterion. This ratio gives thus the $p$-value associated with the criterion. The criterion was then increased so as to pass below the threshold $p<0.05$  (Fig.~\ref{fig:r2_criterions}). The same can be done for any threshold of significance (e.g. $p<0.001$). As can be seen from Fig.~\ref{fig:r2_criterions}, a very high criterion must be set to insure significance for low number of points (a width of $T$ is associated with $T-2$ points). The criterion is non-monotonic and increases for large number of points. Indeed, in these cases, the noise $1/w$ becomes weak and the process tends to a linear curve, which can be easily compatible with a sigmoid. In our data survey, we used the criterions associated with the $p < 0.05$ threshold of significance.

\subsubsection{Alternative null models}

We could have used other null models. A possibility we investigated is the following. We posit a saturating growth function $G(x)$ given by :
\begin{equation}
G(x) = \left[ 1 - (x - 1) ^{2n} \right]^{1/2n} \, ,
\end{equation}
where $n$ is a positive integer. This insures an infinite derivative at $x = 0$ and a null derivative at $x =1$: the process can start as quickly and end as slowly as one wishes. The outcome weakly depends on this parameter $n$, which can be set to $1$. Then, the null process of growth would be as follows:
\begin{equation}
x_{t+1} = x_t + \eta_t \, ,
\end{equation}
where $\eta_t$ is drawn from the distribution :
\begin{equation}
P(\eta_t,x_t, t)) = \frac{1}{Z} \exp \left\{ -  \frac{\lambda (\eta_t -x_t)}{G(t /w) - x_t}  \right\} \, ,
\end{equation}
with $\lambda$ a parameter that we set to $5$. 

This model allows for a wider diversity of processes (there can be sudden jumps), but can hardly be qualified as a null hypothesis. Also, it enforces a strict monotony, which is frequent in the data, but not necessary. Nonetheless, it gave rise to criterions close to those found in the preceding null model. As a conclusion, we can only stress that a null model of growth is already an assumption of some sort, and it is unclear how much theoretical a priori is feeding the null hypothesis.

\subsubsection{Robustness of the sigmoidal fit \label{section:robustness}}

We can alternatively address the statistical robustness of the sigmoidal fit. To do so, we compute, for each point, the expected fluctuation that the sigmoidal model would predict for a finite sample size associated with the number of occurrences $\tilde{N}_t$ characterizing decade. $t$ We make use of the standard confidence interval of 95\% probability:
\begin{equation} \label{eq:fluctu}
\tilde{n}_t =  \tilde{x}_t \tilde{N}_t \pm 1.96 \sqrt{\displaystyle \tilde{x}_t ( 1- \tilde{x}_t) \tilde{N}_t} \, ,
\end{equation}
where $\tilde{n}_t$ is the expected number of occurrences, and $\tilde{x}_t$ the probability of the form to be produced, according to the sigmoidal fit:
\begin{equation}
\tilde{x}_t (h,b,x_{min},x_{max}) = x_{min} + \frac{x_{max}-x_{min}}{1 + e^{-ht -b}} \, .
\end{equation}

Therefore, the actual number $x_t$ of occurrences must obey, for the sigmoidal fit $\tilde{x}_t$ to be consistent with the data:
\begin{equation}\label{eq:check}
 \tilde{x}_t - 1.96 \sqrt{\frac{\displaystyle \tilde{x}_t ( 1- \tilde{x}_t)}{\displaystyle \tilde{N}_t}} < x_t < \tilde{x}_t + 1.96 \sqrt{\frac{\displaystyle \tilde{x}_t ( 1- \tilde{x}_t)}{\displaystyle \tilde{N}_t}} \, , \ \forall t \in [t_{start}:t_{end}] ,
\end{equation}
where $t_{start}$ and $t_{end}$ are the time boundaries of the extracted pattern, respectively associated with frequencies $x_{min}$ and $x_{max}$. 

Note that, as the data $x_t$ is a gliding average of the frequency, the number of occurrences of decade $t$ is not straightforwardly given by the number $N_t$ of occurrences in the corpus. This is why we made use in the above formulae of an `effective' number of occurrences associated with decade $t$, $\tilde{N}_t$, given by:
\begin{equation}
\tilde{N}_t = \frac{1}{W}\sum \limits{k = t - W +1}^{t} N_k \, .
\end{equation}

Another remark to be made is that these expected fluctuations are due to the finite size of the sample. Other sources of fluctuations are nonetheless to be expected, such as inhomogeneities in the sample (e.g. if the linguistic data in the corpus is dominated by a handful of authors). Therefore, fluctuations in equation~\eqref{eq:fluctu} should be considered as lower bounds for the true fluctuations, which we cannot know precisely.

\begin{figure}[!tbp]
  \centering
  \subfloat[]{\includegraphics[scale=0.37]{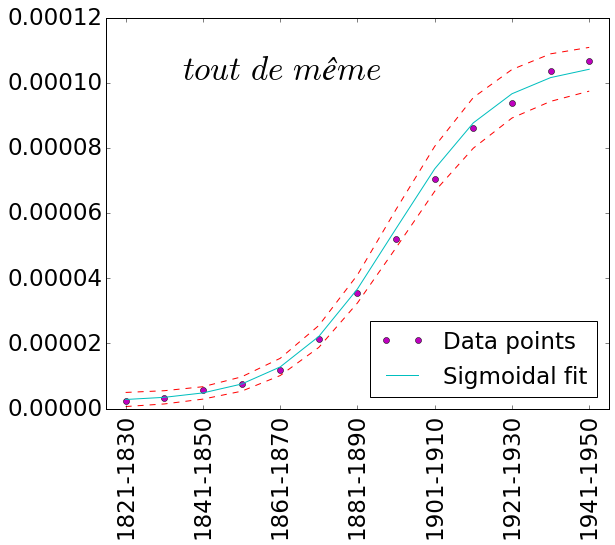}\label{fig:check_pass}}
  \hfill
  \subfloat[]{\includegraphics[scale=0.37]{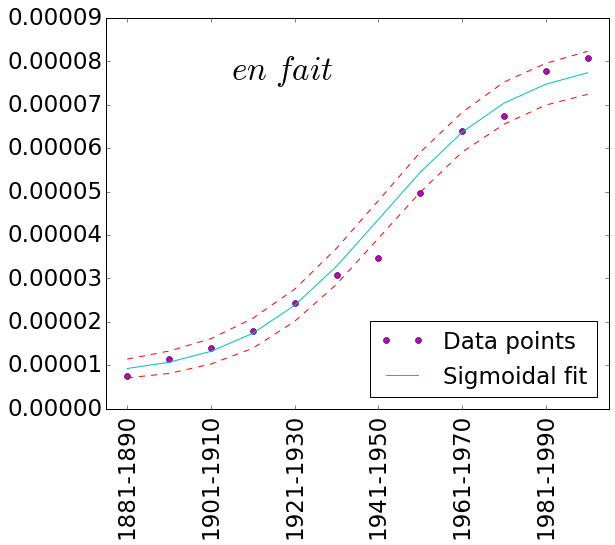}\label{fig:check_fail}}
  \caption{Frequency data (magenta dots) associated with the extracted pattern from (a) \textit{tout de m\^eme} and (b) \textit{en fait}. The sigmoidal model is shown in cyan and the associated confidence interval is shown in red dashed lines. All datapoints lie within this confidence interval in the first case (\textit{tout de m\^eme}). In the second case (\textit{en fait}), datapoints associated with decades 1941-1950 and 1951-1960 lie outside the interval. Therefore, the S-curve may not be a reliable fit of the data.}
\end{figure}

The robustness of all sigmoidal patterns extracted from our data have therefore been checked through equation~\eqref{eq:check}. The result of this check, for each pattern, has been reported on the Table of all studied forms (section~\ref{A7}). For 292 patterns out of the 338 extracted (approximately 86\% of the total), all data points lie within the confidence interval (Fig.~\ref{fig:check_pass}), which proves that the data is consistent with the sigmoidal fit. For the remaining 46 patterns, one or several datapoints lied outside the confidence interval (Fig.~\ref{fig:check_fail}). As the fluctuations are underestimated, we did not withdraw these patterns from the computation of the statistical patterns. This test serves only to assess the consistency of at least 86 \% of the sigmoidal patterns, supporting our claim that the present statistical analysis confirms the robustness of the S-curve agreed on in the literature. 

\subsection{Boundaries of the trap region \label{IB}}

The analytical definitions, used to compute  the latency and growth times in the model, are based on first passage times. In this section we outline the procedure we followed to compute these times.

\subsubsection{Analytical computation of mean first passage times}

Let us note $T_{n\rightarrow m}$ the first passage time at site $m$, starting at site $n$, $0 \leq n ,\; m \leq M$. This is a random variable for which one can write down a recursion equation for its generatrix function:
\beq
\left < e^{\lambda T_{n\rightarrow m}}\right > = R_n \left < e^{\lambda (T_{n+ 1\rightarrow m}+1)}\right> + L_n \left < e^{\lambda (T_{n-1\rightarrow m}+1)}\right> + (1 - L_n - R_n) \left < e^{\lambda( T_{n\rightarrow m}+1)}\right> \, ,
\label{eq:rec}
\eneq
where $R_n$ and $L_n$ are, respectively, the forward and backward jump probabilities, and $\left < .\right>$ denotes the average. We recall that $n=0$ is a reflecting boundary ($L_0=0, R_0 >0$), and $n=M$ an absorbing boundary ($R_M=L_M=0$). We have $T_{n\rightarrow n}=0$, and for the left boundary condition, that is for $n=0$:
\beq 
\left < e^{\lambda T_{0\rightarrow m}}\right > = R_0 \left < e^{\lambda (T_{1\rightarrow m}+1)}\right> + (1 - R_0) \left < e^{\lambda (T_{0\rightarrow m}+1)}\right>. 
\eneq

The first and second derivatives of equation (\ref{eq:rec}) with respect to $\lambda$ leads for $\lambda=0$ to recurrence relations for the first and second moment of  $T_{n\rightarrow m}$, respectively.

More specifically, we can compute the first two moments of the first passage time between one site and its immediate successor, $T_{i \rightarrow i +1}$:
\beq
\left < T_{i \rightarrow i +1 } \right > = t_i
\eneq
And:
\beq
  \left < T_{i \rightarrow i +1} ^ 2 \right > = u_i \, ,
\eneq
Where the $t_i$'s and $u_i$'s are iteratively computed from:
\beq
\begin{cases}\displaystyle t_0 = \frac{1}{R_0} \\ \\ \displaystyle u_0 = \frac{2 t_0 - 1}{R_0} \end{cases}
\eneq
And:
\beq
\begin{cases} \displaystyle t_{i} = \frac{1}{R_i}  + \frac{L_i}{R_i} t_{i-1}  \\ \\ \displaystyle u_i = 2 t_i^2 + \frac{L_i}{R_i} u_{i-1} \end{cases} \, . 
\eneq
From this, we can easily compute the first two moments for any $T_{n \rightarrow m}$:
\beq
\mu \left(T_{n \rightarrow m} \right) = \sum \limits _{k=n} ^ {m-1} t_k
\eneq
And:
\beq
\sigma ^ 2 \left(T_{n \rightarrow m} \right) = \sum \limits _{k=n} ^ {m-1} \left( u_k - t_k ^ 2 \right)
\eneq

\subsubsection{Trap boundaries}

In the main text, we explain latency time and growth time as first passage times. However, these two quantities are both empirically extracted from the macroscopic pattern obtained at the end of a run, in a procedure exactly transposed from the corpus data treatment. The question is then: Which trap boundaries $n_{in}$ and $n_{out}$ should we set in order for the properly defined time $T_{n_{in} \rightarrow n_{out}}$ to correspond statistically to the empirically defined latency time? 

Besides, growth time can be seen as well as a first passage time between two sites. Though the exit site should be $M$, it is more appropriate to define a cut-off $n_{last}$. Indeed, there is a discrepancy between the fact that, close to the absorbing point, the walk gets slowed down again, and that, in this region, the new variant is almost always produced anyway. In other terms, growth time, as extracted from the time evolution of the ratio of produced new variant occurrences, is not sensitive whether the end of the walk is reached or not. 

Let us note $\mu_g$ and $\sigma^2_g$, and $\mu_{lat}$ and $\sigma^2_{lat}$, respectively the mean and the variance of the growth and latency times (obtained from the distributions of those empirically extracted quantities from ten thousand runs). Then, over a reasonable range of $n$, we look for $m$ so that $\mu \left(T_{n \rightarrow m} \right)$ is as close as possible to $\mu_g$; we then choose the pair $(n;m)$ such that $\sigma ^ 2 \left(T_{n \rightarrow m} \right)$ is as close as possible to $\sigma^2_g$. This pair defines thus the region of growth, $(n_{out};n_{last})$. We then choose $n_{in}$ so as to fit the mode of the empirical latency distribution, assuming that first passage time is distributed according to an Inverse Gaussian (which entails that the mode is a known function of $\mu$ and $\sigma^2$). 

\subsection{Statistical distributions \label{A4}}

In the main paper, we presented the statistical distributions of both the latency times and the growth times obtained from corpus data, and proposed an Inverse Gaussian fit of the result, following the theoretical prediction that the distribution should be of the same family as the Inverse Gaussian. We can now consider whether other usual statistical distributions could be suited as well to account for the statistical features of our dataset.

\subsubsection{Growth time}

\begin{figure*}[!tp]
  \centering
  \includegraphics[width=\linewidth]{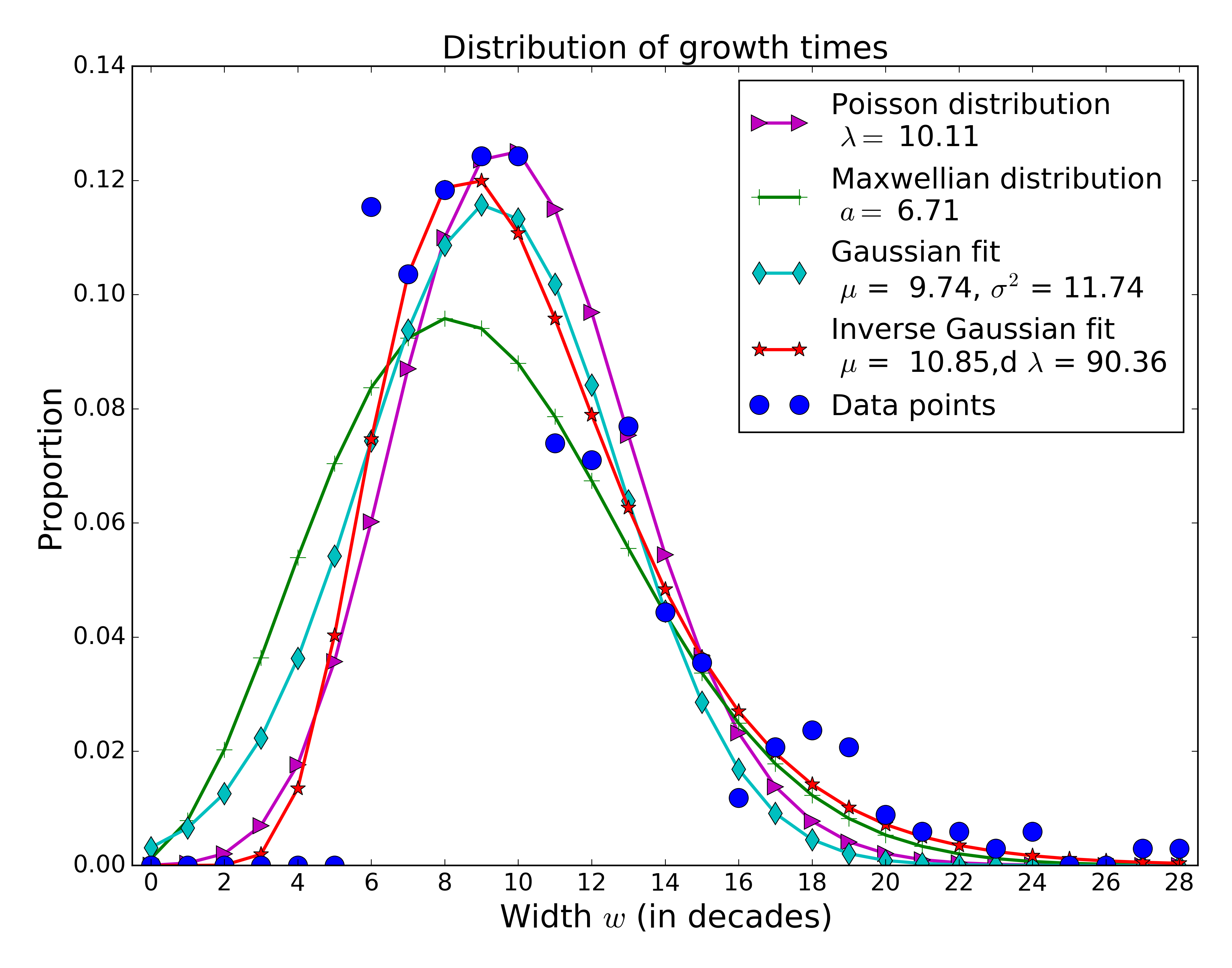}
  \caption{Several fits of the distribution of growth times as extracted from corpus data.}
  \label{fig:growth_distrib}
\end{figure*}

We tried to fit the distribution of growth times with three different usual statistical distributions: Poisson, Maxwellian, and Gaussian (Fig~\ref{fig:growth_distrib}). Aside from the Poisson distribution, the fit is qualitatively inadequate compared to an Inverse Gaussian fit. 

We can further assess which of these four trials is to be favored by computing the Kullback-Leibler divergence between these theoretical proposals and the corpus data. We remind that the Kullback-Leibler divergence is closely related to the likelihood, and maximizing the likelihood is strictly equivalent to minimizing the Kullback-Leibler divergence. We obtained Kullback-Leibler divergences of 0.21, 0.26, 0.35 and 0.10 for the Poisson, Maxwellian, Gaussian and Inverse Gaussian distributions, respectively. Other statistical tests have been performed to account for the difference in the number of parameters between these distributions (1 for Poisson vs. 2 for the three others) and reported on Table~\ref{table:growth_test}.  Therefore, even if the Poisson distribution seems adequate, it does not perform much better than the Maxwellian. This failure is imputable to the tail of the distribution, which is thicker than what a Poisson distribution would predict. This tail is captured by the Maxwellian, but the latter distribution fails to reproduce the peak of the distribution.

\begin{table}
\centering
\caption{Output of three statistical tests (Kullback-Leibler divergence ($D_{KL}$), Akaike Information Criterion (AIC) and Bayesian information Criterion (BIC), to compare different fits of the growth times distribution.}
\label{table:growth_test}
\begin{tabular}{|c|c|c|c|c|}
\hline
Test & Poisson & Maxwellian & Gaussian & Inverse Gaussian \\ \hline
$D_{KL}$ & 0.21 & 0.26 & 0.35 & 0.10 \\ \hline
AIC & 227  & 259 & 323 & 152 \\ \hline
BIC & 231 & 263 & 331 & 160 \\ \hline
\end{tabular}
\end{table}

Comparatively, the Inverse Gaussian fit is significantly better than the other three. It is adequate for both the peak and the tail. Therefore, albeit the data is not perfectly fit by the Inverse Gaussian, this distribution displays the right behavior, as we predicted from our model.

\subsubsection{Latency time}

\begin{figure*}[!tp]
  \centering
  \includegraphics[width=\linewidth]{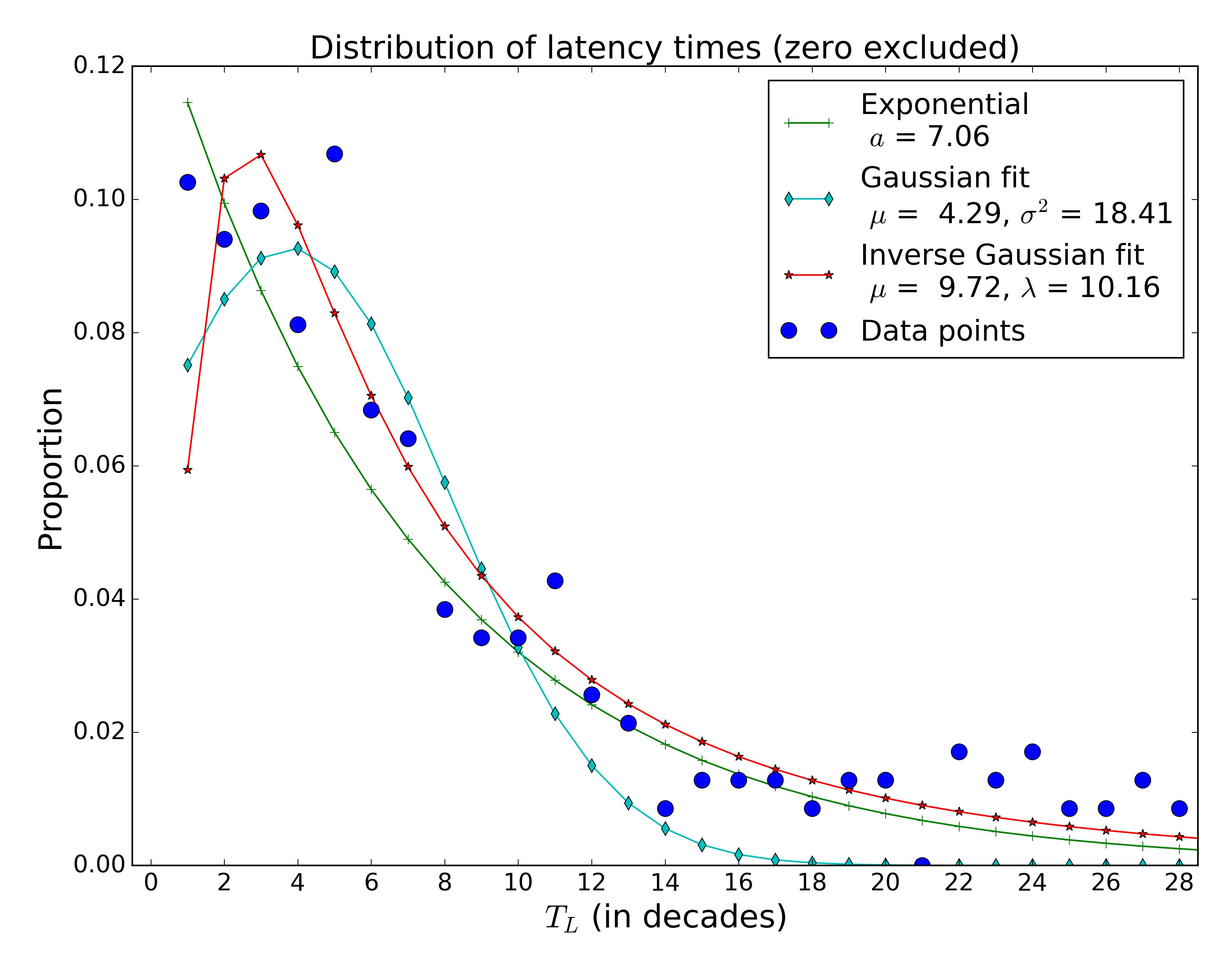}
  \caption{Several fits of the distribution of latency times as extracted from corpus data.}
  \label{fig:latency_distrib}
\end{figure*}

We can do the same for the distribution of latency times. We tried, besides the Inverse Gaussian, the exponential and the Gaussian distributions, as the Maxwellian and the Poisson distributions were largely inadequate (Fig.~\ref{fig:latency_distrib}).

The same statistical tests as before have been performed to select the best distribution (Table~\ref{table:latency_test}). Once more the Inverse Gaussian proves to be superior, even though the exponential also displays the right qualitative behavior.  Also, we can compare the parameters obtained from an optimization fit with the actual mean of the data, which is $8.59$. The mean should be given by the parameter $a$ of the exponential and the parameter $\mu$ of both the Gaussian and the Inverse Gaussian. In this regard, it is clear that the Gaussian can be ruled out (it predicts a mean of 4.29) while the exponential and the Inverse Gaussian are consistent with the data (they respectively predict a mean of 7.06 and 9.72). An interesting difference between the exponential distribution and the Inverse Gaussian one would be that the mode of the distribution is zero in the former case, and non-zero in the latter. This feature could be further investigated with a larger amount of data regarding the latency, so as to clarify the behavior of the distribution in the region of lower values of the latency time. A finer timescale would also allow to zoom in this region of low latency times, so as to investigate whether the behavior of the distribution is non-monotonic in this domain, as would predict the Inverse Gaussian.

\begin{table}
\centering
\caption{Output of three statistical tests (Kullback-Leibler divergence ($D_{KL}$), Akaike Information Criterion (AIC) and Bayesian information Criterion (BIC), to compare different fits of the latency times distribution.}
\label{table:latency_test}
\begin{tabular}{|c|c|c|c|}
\hline
Test & Exponential & Gaussian & Inverse Gaussian \\ \hline
$D_{KL}$ & 0.24  & 1.71 & 0.10 \\ \hline
AIC & 834  & 1393  & 717 \\ \hline
BIC & 837  & 1400  & 725  \\ \hline
\end{tabular}
\end{table}

Here again, the Inverse Gaussian appears to capture more closely the corpus data than the other usual statistical distributions, as predicted from the model. It is also worth noticing that the model predicts that the Inverse Gaussian would be suited for both the growth and the latency, while the other candidates are appropriate for only one of these quantities (the growth time for the Poisson distribution, the latency time for the exponential).

\subsubsection{Statistical distribution of the slopes}

\begin{figure*}[!bp]
  \centering
  \includegraphics[width=\linewidth]{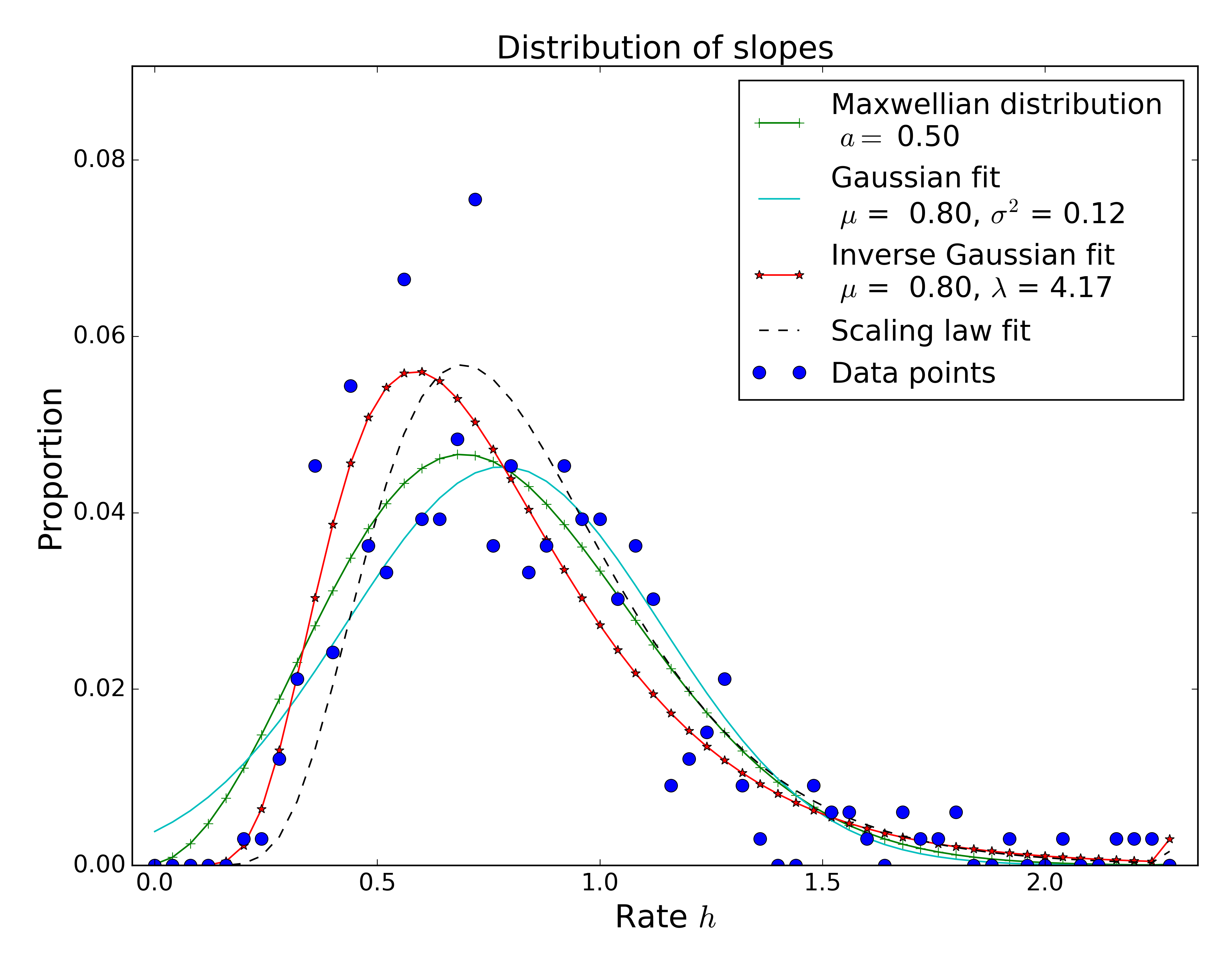}
  \caption{Several fits of the distribution of the slopes as extracted from corpus data.}
  \label{fig:slopes_distrib}
\end{figure*}

From the empirical procedure, we can also extract, for both the corpus and numerical datasets, the statistical distributions of the slopes of the logit transform of the sigmoidal part. Corpus data (Fig.~\ref{fig:slopes_distrib}) is best fitted by the Inverse Gaussian, by comparison with a Maxwellian and a Gaussian. Statistical tests favor consistently the Inverse Gaussian (Table~\ref{table:slopes_test}). All these three fits have been done without optimization, using the mean and the variance of the data to compute the parameters accordingly. 

\begin{table}
\centering
\caption{Output of three statistical tests (Kullback-Leibler divergence ($D_{KL}$), Akaike Information Criterion (AIC) and Bayesian information Criterion (BIC), to compare different fits of the slopes distribution.}
\label{table:slopes_test}
\begin{tabular}{|c|c|c|c|c|}
\hline
Test & Maxwellian & Gaussian & Inverse Gaussian & Scaling law fit \\ \hline
$D_{KL}$ & 0.14  & 0.20 & 0.10 & 0.14 \\ \hline
AIC & 243  & 293  & 222 & 252 \\ \hline
BIC & 240  & 285  & 215 & 244  \\ \hline
\end{tabular}
\end{table}

Why the distribution of the slopes would follow an Inverse Gaussian is unclear though. From the scaling relation between the slope and the width (see section~\ref{A5}), we can derive that the slopes $h$ must be distributed according to the density $\rho_h$ given by:
\begin{equation}
\rho_h(h) = \frac{e^{2.10}}{h^2} \rho_w\left( \frac{e^{2.10}}{h} \right) \, .
\end{equation}
Assuming for $\rho_w$ an Inverse Gaussian with parameters obtained from the Inverse Gaussian fit of the corpus data for the growth time, we can propose an estimate of the statistical distribution for the slopes. As can be seen on Fig.~\ref{fig:slopes_distrib}, this curve is qualitatively appropriate, hinting therefore at the consistency between our different results. The associated Kullback-Leibler divergence is equal to $0.14$, the same as for the Maxwellian, not far from an Inverse Gaussian fit. 

There is another prediction that we can make regarding this matter. If we assume that the growth is Inverse Gaussian, then according to the scaling law relating the width $w$ and the slope $h$ (see section~\ref{A5}):
\begin{equation}
h \approx \frac{e^{2,10}}{w} \, ,
\end{equation}
we can predict that, under the assumption that the width is Inverse Gaussian distributed:
\begin{equation}
\begin{split}
\left< h \right > &\approx e^{2.10}\left( \frac{1}{\mu_{w}} + \frac{1}{\lambda_w} \right) \\
&\approx e^{2,10}\left( \frac{1}{10.85} + \frac{1}{90.36} \right) \\
& \approx 0.84 
\end{split}
\end{equation}
which is close to what we find in the data ($\left< h \right > = 0.80$).

\begin{figure*}[!tp]
  \centering
  \includegraphics[width=\linewidth]{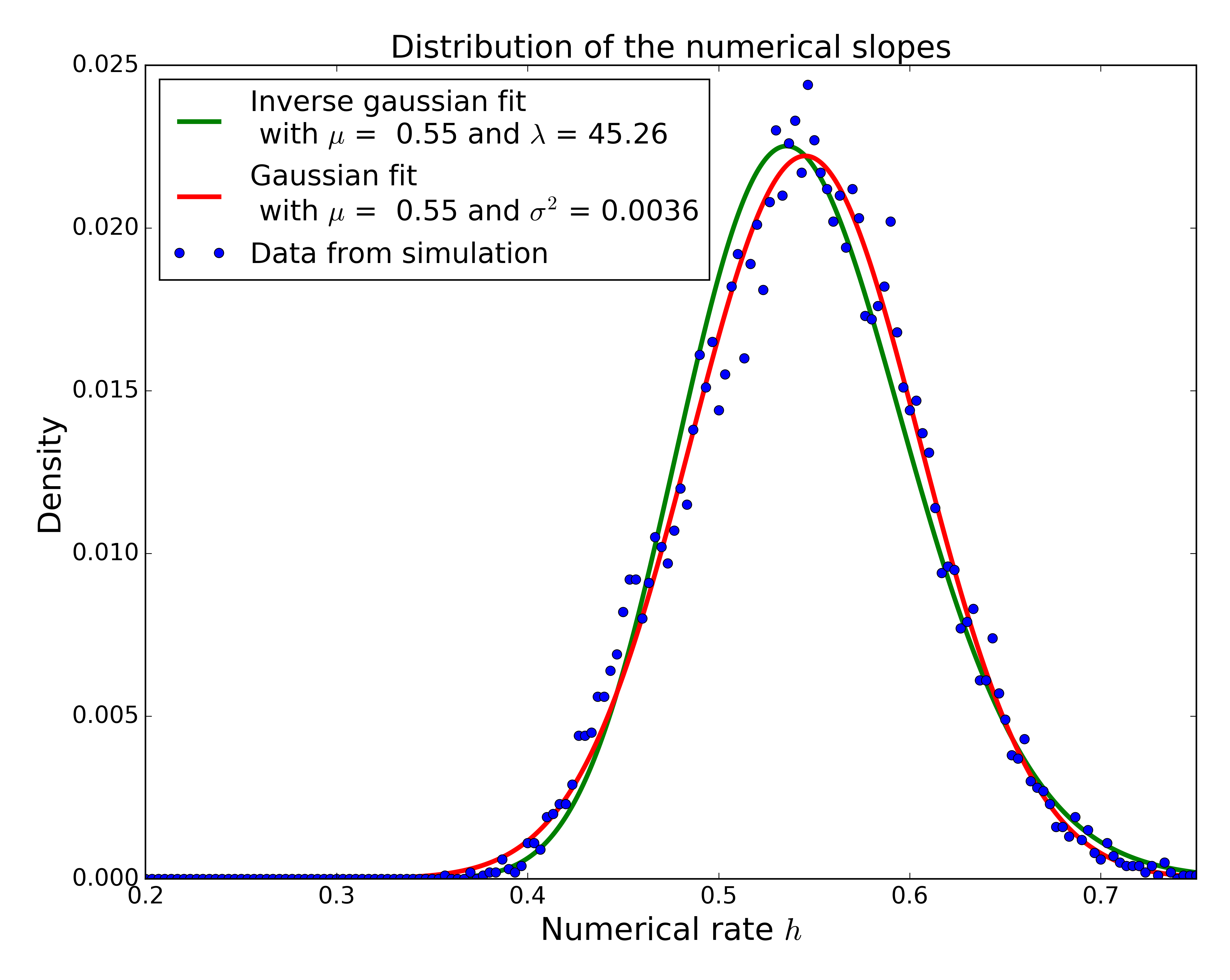}
  \caption{Inverse Gaussian and Gaussian fits of the distribution of slopes as extracted from numerical data.}
  \label{fig:slopes_numerical_distrib}
\end{figure*}

On the other hand, the distribution of the slopes generated from numerical data is best fitted by a Gaussian (Fig.~\ref{fig:slopes_numerical_distrib}), with a Kullback-Leibler divergence of 0.009 compared to 0.013 for the Inverse Gaussian.

This may be explained by the fact that an Inverse Gaussian distribution tends to a Gaussian one whenever parameter $\lambda$ tends to infinity. The fact that $\lambda$ is much bigger compared to $\mu$ in numerical data than in corpus data implies that there are more sources of variation for the growth part of the process in the data than what we considered in the model. We discuss this issue in the next subsection.

\subsection{Further comparisons with corpus data \label{A5}}

In our paper, we show that an Inverse Gaussian distribution is adequate to capture both latency time and growth time distributions, indicating that these two quantities are of the same nature, and result from the same mechanism of change. However, the agreement between our model and the corpus data goes further, as we show in this section. 

\subsubsection{P\'eclet number}

The parameters $\mu$ and $\lambda$ of the Inverse Gaussian distribution scale with the time length in the same way, so that is is relevant to consider their ratio, which is called the P\'eclet number~\cite{redner2001guide}. Note that, because the relation $\lambda~=~\mu^3/\sigma^2$ holds, the P\'eclet number is but the ratio between the squared mean and the variance. 

The P\'eclet number for latency times from corpus data is equal to $2.3$ while the model gives back a P\'eclet number of $1.4$, so they both are of the same order of magnitude. However, for growth times, we get $10.4$ for corpus data, and $63$ in the model, so that there is no agreement between the two.

Actually, this discrepancy is rather expected. Given the definition of the P\'eclet number, it means that the variance of the growth time is comparatively greater in the data than it is in our model. Yet, this can be understood in terms of the latter: Indeed, it has been stressed that the conceptual network of language is organized as a small-world network~\cite{gaume2008semantic}, 
and we have proposed that major semantic changes, characterized by the latency-growth pattern, would correspond to a leap from a cluster to another. It means that latency involves only one bridge, so that the set-up we explored should be enough to cover it. Growth, on the other hand, depends on the cluster size, and on the inner organization of the cluster. It thus involves a varying number of contexts, which explains why the variance of the growth would be greater in actual data, leading to a smaller P\'eclet number. 

Concerning the scale of the process, it could be tempting to compare mean latency between model and data to find the value of $M$ (size of the memory) which would correspond to the data. However, the scale entangles both $M$ and the size of the counting window. It also depends on the total number of involved contexts. There is thus no obvious way to compare the scales involved in the model and in the data. 

\subsubsection{Growth-Slope correlation and scaling law}

\begin{figure}[!tbp]
  \centering
  \subfloat[]{\includegraphics[scale=0.45]{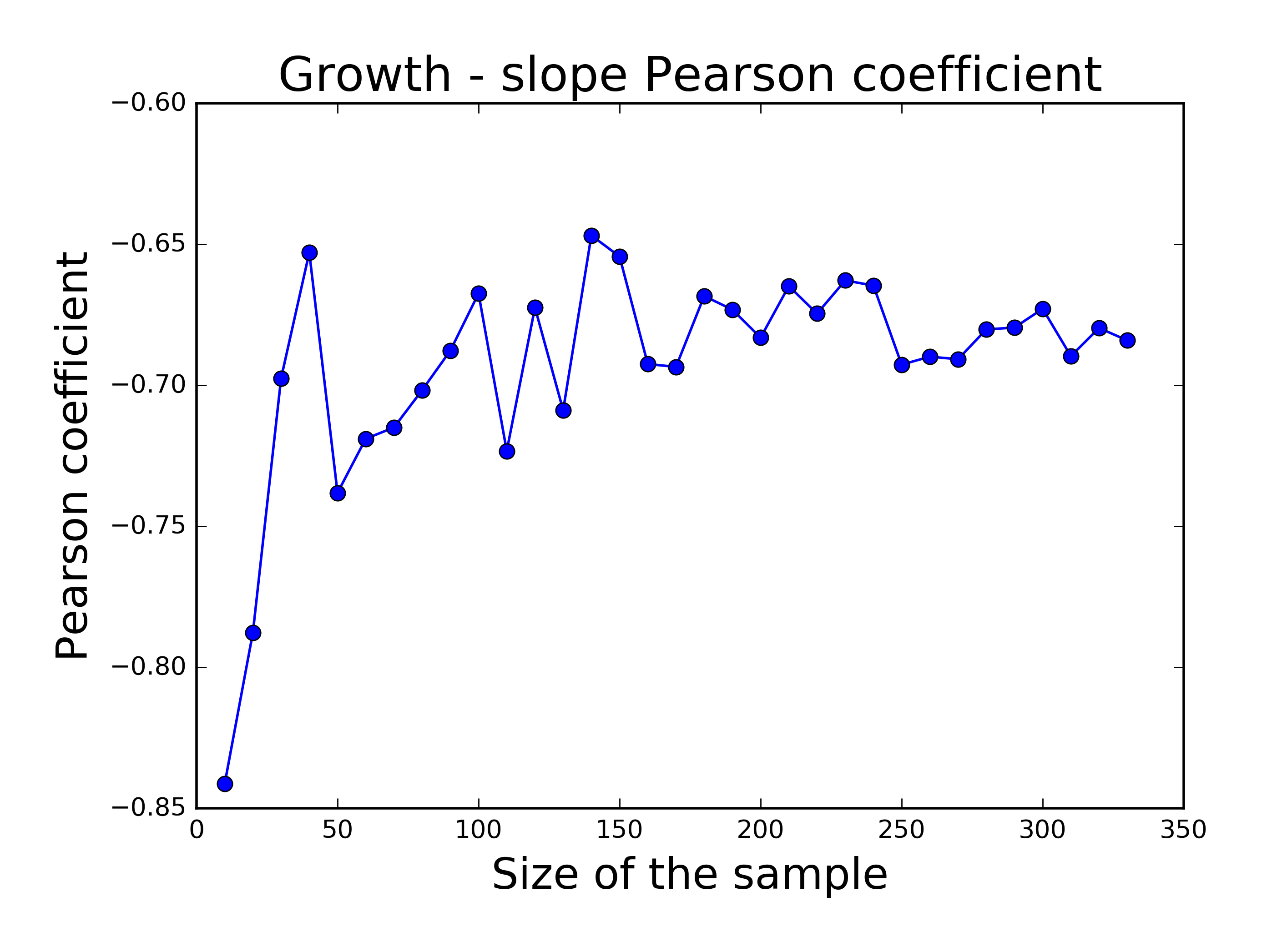}\label{fig:slorel_corpus}}
  \hfill
  \subfloat[]{\includegraphics[scale=0.45]{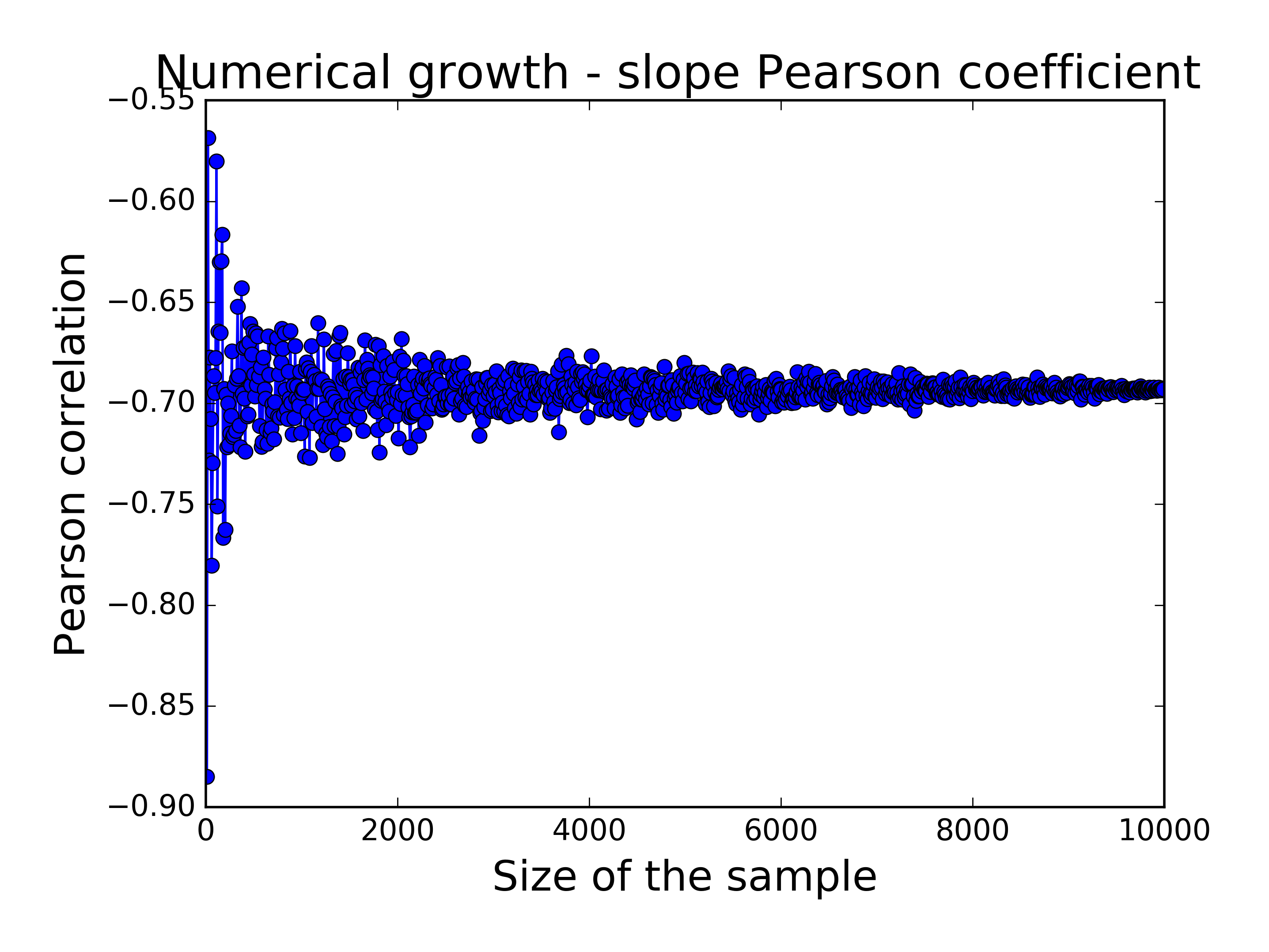}\label{fig:slorel_numerical}}
  \caption{Pearson coefficient for the correlation between growth time and slope obtained from (a) corpus data and (b) numerical simulations}
\end{figure}

Growth and slope are expected to be correlated. The two quantities are convincingly negatively correlated, both in corpus data (Pearson coefficient of $-0.69$, Fig.\ref{fig:slorel_corpus}) and in our model (Pearson coefficient also equal to $-0.69$, Fig.\ref{fig:slorel_numerical}).

\begin{figure*}[!tp]
  \centering
  \includegraphics[width=\linewidth]{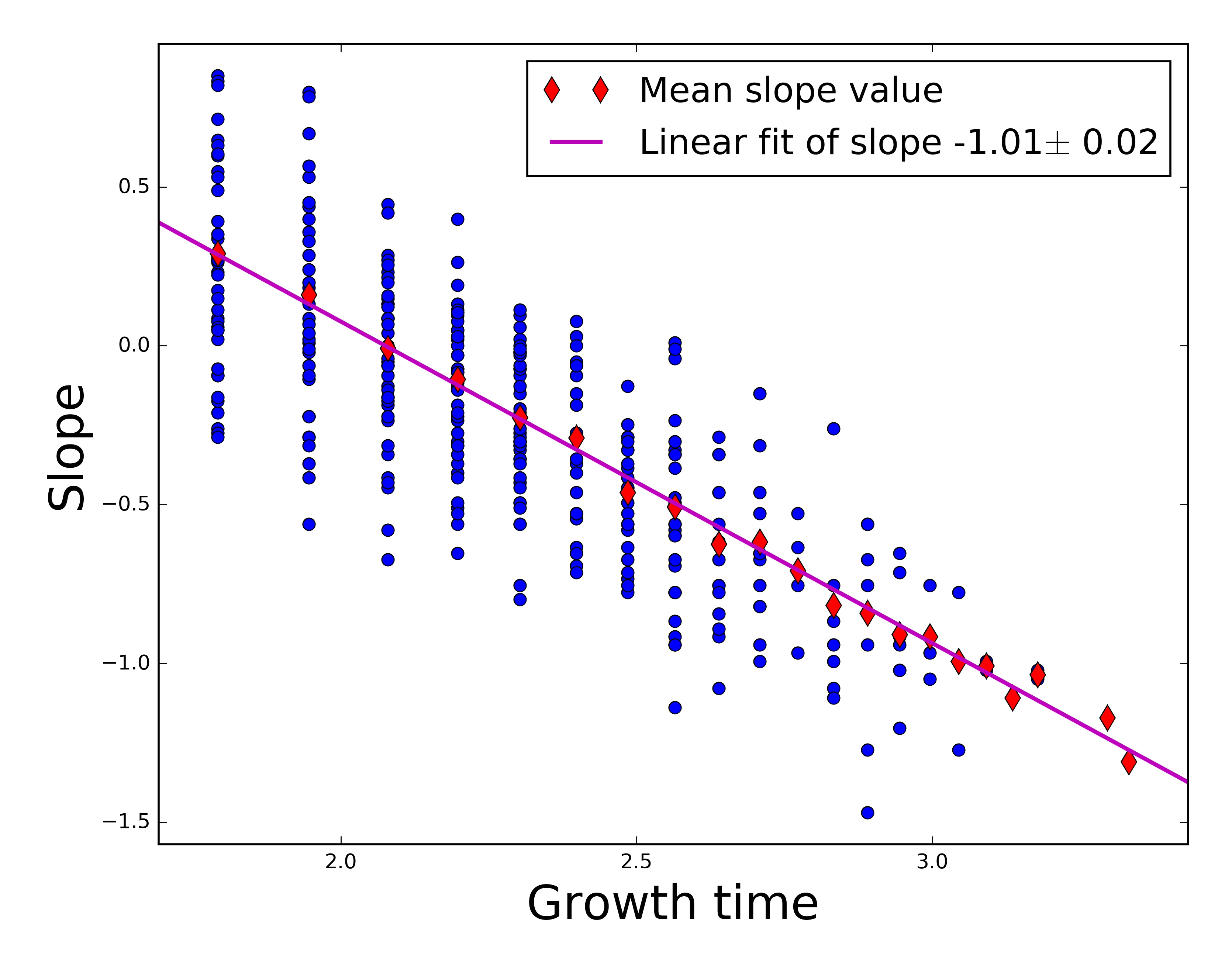}
  \caption{Scaling law between the slope $h$ and the growth time $w$. The fit is performed on the average values of the slope $h$ for all different $w$. The associated $r^2$ is equal to 0.994.}
  \label{fig:scaling_law}
\end{figure*}

It is also worthy to consider the possibility of a scaling law between these two quantities, in line with what has been evidenced for other socio-cultural changes~\cite{michard2005theory}, where an exponent $\alpha = -2/3$ is found between the rate $h$ and the width $w$ (slope and growth time, respectively). This exponent differs from the expected $-1$ exponent which would be expected for pure sigmoids. Our model also predicts such a scaling behavior with an exponent of $-2/3$. However, the corpus data is not characterized by any specific scaling law: The rate $h$ and the width $w$ are related through a trivial $-1$ exponent (Fig.~\ref{fig:scaling_law}):
\begin{equation}
\log h  = -1.01 \log w + 2.10 \, .
\end{equation}

The discrepancy between the scaling behavior of corpus data and that of numerical data is yet to be explained. Once more, it could be due to the difference between the model set-up (one site competition) and the whole process of a semantic expansion (pervasion of a cluster of the semantic network), but this is purely conjectural. 

\subsubsection{Latency-Growth correlation}

\begin{figure}[!tbp]
  \centering
  \subfloat[]{\includegraphics[scale=0.45]{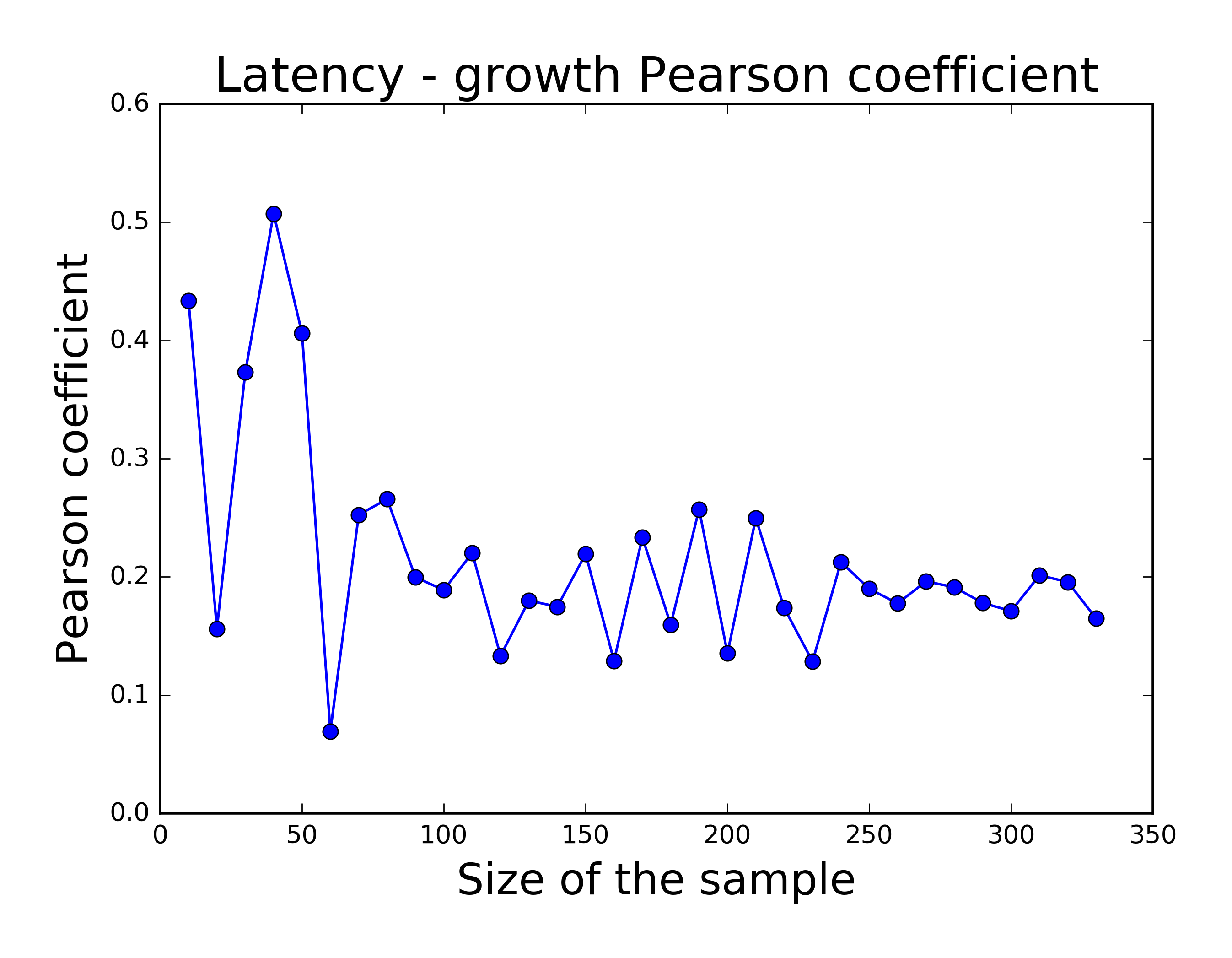}\label{fig:latrel_corpus}}
  \hfill
  \subfloat[]{\includegraphics[scale=0.45]{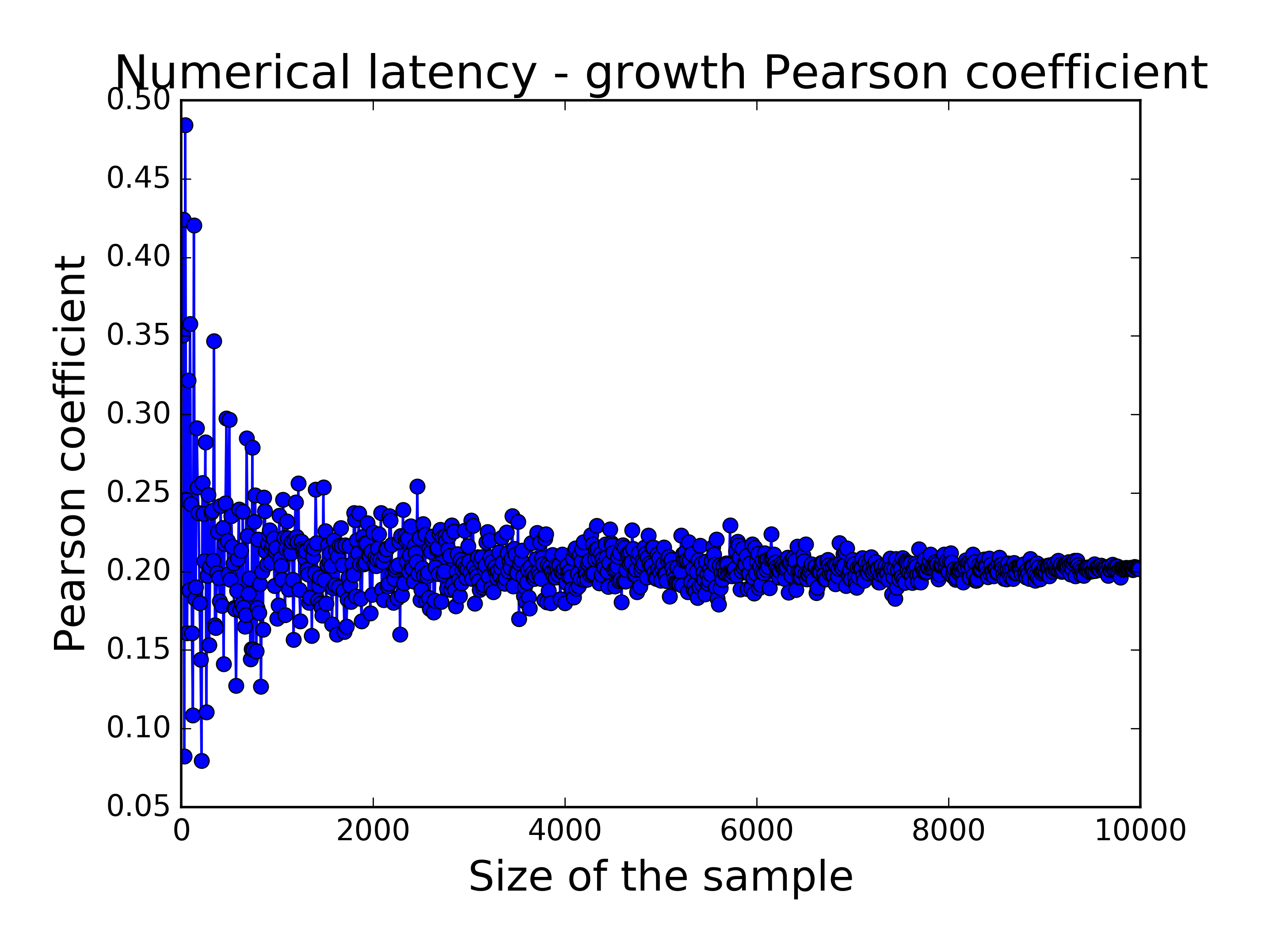}\label{fig:latrel_numerical}}
  \caption{Pearson coefficient for the correlation between growth time and latency time obtained from (a) corpus data and (b) numerical simulations}
\end{figure}

It may be intuitively expected for latency and growth times to be correlated: The longer the wait, the more momentum is gained. Yet, according to our model, there is no such correlation: Latency and growth times, as seen as first passage times in different parts of a Markov chain, are strictly independent quantities. However, in the empirical procedure, these two parameters become correlated, for the latency is defined as the time spent in a region comprised between $x_{t_{out}} \pm a ( 1 - x_{t_{out}})$, where $x_{t_{out}}$ is the frequency attained at the beginning of the growth process and $a$ is set to $0.17$ (and $0.15$ for corpus data). Thus, the higher this $x_{t_{out}}$, the smaller the margin, so that a high $x_{t_{out}}$ will be correlated with a short latency, as well as a shorter growth on average. These two quantities are thus weakly positively correlated, with a Pearson coefficient of $0.20$ (Fig. \ref{fig:latrel_numerical}). 

If we now turn to corpus data, we find a Pearson coefficient of $0.19$ (Fig.\ref{fig:latrel_corpus}). The correlation between latency and growth is weak, and can be entirely imputed to the details of the empirical procedure, as we have just seen for the numerical data. It thus means that growth time and latency time are two independent quantities, so that positing a Markovian nature of language change is in line with findings from corpus data. 

The latency and the slope are expected to be weakly negatively correlated as well, as a result from the scaling relation between the slope and the width. In the data, we find a Pearson coefficient of -0.16, to be compared with -0.23 in the model.

\section{Model variants}

\subsection{Hearer mechanism \label{hearer_mech}}

The model we propose in the paper describes a mechanism associated with language production: It is solely based on a speaker perspective. Yet, language change may not come only from innovation in producing language, but also in understanding it. Actually, these two aspects cannot be separated: If an innovation is possible in a speaker perspective, it must also be accessible from a hearer perspective. Be it a speaker or a hearer, a language user relies on the same cognitive entity. It seems thus necessary to consider model variants where the novelty can come from this complementary perspective, as well as from a combination of the two.

\subsubsection{Hearer variant}

Let us consider the same situation as for the listener model: There are two meanings, $C_0$ and $C_1$, to which are attached a pool of memories of linguistic tokens. Initially, $C_0$ is populated by $X$ tokens only, while $C_1$ is populated by $Y$ tokens only. Just as context $C_1$ is fed by the memory of $C_0$ when it came to express $C_1$, if a linguistic occurrence yields meaning $C_0$, it can elicit meaning $C_1$ as well. Occurrences of $X$ thus have a chance to populate context $C_1$, so that we will note $x$ the proportion of $X$ tokens in $C_1$, just as we did in the speaker-based model. If we ascribe to the inference $C_0 \Rightarrow C_1$ a probability equal to $\gamma$, then we can describe the dynamics as follows:
\begin{enumerate}
\item Either $C_0$ or $C_1$ are chosen to be expressed, with equal probabilities.
\item If $C_0$ has been chosen, $X$ is produced. If $C_1$ has been chosen, $X$ is produced with probability $P_0(x)$, otherwise $Y$ is produced. $P_0(x)$ is the same function as $P_{\gamma}(x)$, except that $\gamma$ is now set to 0 (there is no such thing as an effective frequency in this framework). 
\item The produced occurrence is recorded in the chosen context. If $C_0$ has been chosen, an additional occurrence of the same kind as the previous one is recorded in $C_1$ with probability $\gamma$ ($C_0$ has elicited the meaning $C_1$). 
\item A past occurrence is deleted whenever needed, so as to keep both memory sizes constant. 
\end{enumerate}

These dynamics correspond once more to a random walk where the jump probabilities, forward and backward,  respectively $R^H(x)$ and $L^H(x)$ (where $H$ stand for `hearer'), are given by: 
\beq
     \begin{cases}
         R^H(x) &= \displaystyle \frac{1}{2}  \left[ \gamma + P_0(x) \right] (1 -x) \\
         \\
         L^H(x) &= \displaystyle \frac{1}{2}  (1 - P_0(x) ) x
     \end{cases} \, ,
\eneq
to be compared with the jump probabilities in the speaker perspective (respectively $L^S(x)$ and $R^S(x)$ for the forward and backward jump probabilities):
\beq
     \begin{cases}
         R^S(x) &= \displaystyle  P_{\gamma}(x) (1 -x)\\
         \\
         L^S(x) &= \displaystyle(1 - P_{\gamma}(x) ) x 
     \end{cases} \, .
\eneq

These modified jump probabilities lead to a new expression for the drift velocity:
\beq \label{hearer_velocity}
\dot{x} = \frac{1}{2} \left[ P_0(x) - x + \gamma (1-x) \right ] \, .
\eneq
A change of variable $y = (1 + \gamma) x - \gamma$ leads to the same equation as equation 4 of the paper, with a slightly different timescale accounting for the fact that two contexts are now being called:
\beq
\frac{2}{1+\gamma}\dot{y}  = \left[ P_0\left( \frac{y + \gamma}{1 + \gamma} \right) - y \right ] \, .
\eneq 
Indeed, $P_0\left( \frac{y + \gamma}{1 + \gamma} \right)$ is exactly $P_{\gamma}(y)$, so that the fixed point in the hearer perspective $x_c ^{H}$ will be given, as a function of the fixed point $x_c^S$ of the speaker perspective, as:
\beq
x_c ^ H = \frac{x_c ^ S + \gamma} {1+\gamma} \, ,
\eneq
which is higher than $x_c^S$. This means that, in the hearer pespective, the latency frequency will also be higher. However, it does not entail that the change will be more or less likely to happen, since what triggers the change is the fact that $\gamma$ is equal to $\gamma_c$ or higher, and this parameter $\gamma_c$ remains the same throughout the perspective shift. 

\subsubsection{Combined model}

We can now combine the Listener and Hearer perspectives, by taking into account the effective frequency $f$ instead of the actual frequency $x$ in step $2$ of the dynamics outlined in the previous subsection. Then, in the above formulae, all $P_0(x)$ become $P_{\gamma}(x)$ (or equivalently, $P_0(f)$). The velocity is now set to:
\beq \label{total_velocity}
\dot{x} = \frac{1}{2} \left[ P_{\gamma}(x) - x + \gamma (1-x) \right ] \, .
\eneq
Setting $X = (x + \gamma) / (1  +\gamma)$, we get:
\beq
2 (1 + \gamma) \dot{x} = P_0(X) - X  + (1 - X) \gamma (2 + \gamma) \, .
\eneq
We can now define a renormalized parameter $\tilde{\gamma} = \gamma ( 2 +\gamma) $ to make this velocity similar to the one given by \eqref{hearer_velocity}. Setting $Y  = (1 + \tilde{\gamma}) X - \tilde{\gamma}$, we finally get:
\beq
2 \frac{1+\gamma}{1+\tilde{\gamma}} \dot{Y} = P_{\tilde{\gamma}}(Y) - Y \, .
\eneq
This implies that $(Y_C, \tilde{\gamma}_c) = (x_c^S, \gamma_c^S)$, so that the critical point $(x_c^T, \gamma_c^T)$ in this combined perspective is equal to:
\beq
(x_c^T, \gamma_c^T) = \left( \frac{x_c^S + \gamma_c^S}{1 + \gamma_c^S}  , \sqrt{1 + \gamma_c^S} - 1 \right) \, .
\eneq
In this case $\gamma_c^T$ is lower than its hearer and speaker perspectives counterparts. It entails that the change would happen more easily. $x_c^T$ is somewhere in between $x_c^S$ and $x_c^H$.

\subsubsection{Summary}

All three variants of the model give rise to the same picture of sigmoidal growth preceded by a period of latency. The data does not allow to discriminate between either one of these three possibilities. Yet, the hypothesis that the change is driven by both hearer and speaker mechanisms is the most probable, as all language users adopt the role of hearer and speaker alternatively. An enthralling perspective of research would be to devise a quantitative criterion so as to see which of the three mechanisms best account for real language data. One could also investigate which features of language change speaker and hearer perspectives are respectively able to account for independently, and if some features need the conjunction of both to appear. Obviously, all those questions hinge upon available data and the finding of relevant observable quantities to look at. 

\subsection{Interpretation of the cognitive strength $\gamma$ \label{interpret_gamma}}

In the proposed model, we make the assumption that all memory sizes are equal in the speaker perspective, and that all meanings $C_i$ are expressed with equal probability in the hearer perspective. Here we consider the alternative that the links in the network are not weighted: They are either 1 or 0. The asymmetric structure between the two contexts $C_0$ and $C_1$ is however maintained (i.e. the graph is a directed graph and the link between sites $C_0$ and $C_1$ is $1$ while the link between sites $C_1$ and $C_0$ is $0$). 

\subsubsection{Heterogeneous memory sizes} 

Now let us assume different memory sizes for the two concepts, denoting by $m$ and $M$ the memory sizes of $C_0$ and $C_1$, respectively. Then the effective frequency of $X$ in $C_1$ is given by:
\beq
f = \frac{N + m}{M+m} = \frac{x + m/M}{1+m/M}
\eneq
By defining $\gamma$ as the ratio of memories $m/M$, we recover the same effective frequency as before. 

This means that the strength $\gamma$ of the cognitive link can be interpreted as a ratio between memory sizes. If all sites were connected to each other, the occurrences expressing the contexts whose associated memory is the greatest would spread all over the network. However, not all sites lead to all others: There are pathways in the conceptual organization, which constrain possible semantic changes and allow for low-memory contexts to invade higher-memory ones. 

The main difference brought forth by this interpretation is that it allows for $\gamma$'s greater than one. In general, there would be no critical behavior and thus no latency, except if the conquering occurrence type comes from a very low memory context. This would suggest that, as grammaticalizations are well-characterized by the latency-growth pattern with sigmoidal increase, lexical meanings are allocated a much smaller memory than grammatical ones. However, it would also be the case within the lexicon, when a word goes from a concrete meaning to an abstract one. 

It is not clear why functional and abstract meanings should be allocated a greater memory than concrete meanings. There could be for instance some advantage in making the more abstract and structural part of the conceptual realm more stable in their linguistic expression than  other parts of speech, especially because they serve to constrain the processing of utterances and provide structure to the flow of speech. Were it the case, then we could understand the strong asymmetry evidenced by grammaticalization --- the fact that lexical forms are recruited to express grammatical meanings overwhelmingly more frequently than the reverse. Indeed, if the links were from the stable (i.e. supported by a large memory size) to the unstable parts of the language, then all those links would be associated to a very high $\gamma$ parameter, so that all parts of language would soon come to be expressed by the grammatical forms. This would right away lead to a complete communicative failure. There would thus be an obvious advantage in preventing the links from grammatical concepts to lexical ones, hence in the unidirectionality exhibited by grammaticalization. 

\subsubsection{Different probabilities of use}

We now introduce different calling probabilities for $C_0$ and $C_1$ in the hearer perspective. Let's say that the probability to call $C_0$ is $\alpha$. Here again $\gamma$ is set to $1$ (i.e. $C_0$ automatically entails $C_1$). The jump probabilities becomes thus:
\beq
R^H(x) = \left[ \alpha + (1 - \alpha) P_0(x) \right] (1 -x)
\eneq
and:
\beq
L^H(x) = (1 - \alpha) (1 - P_0(x) ) x \, . 
\eneq
We can factorize $R^H(x)$ by $1 - \alpha$. Then we recover the same computation as before, with the ratio of calling probabilities $\alpha / (1- \alpha)$ playing the role of $\gamma$. Furthermore, if we set the call probability to be proportional to memory size, then we recover the same $\gamma $ as in the preceding subsection. This assumption seems natural, since greater memory sizes would help stabilizing the linguistic expressions of widely used meanings. 

In such a case, the near-criticality associated to the latency-growth pattern is recovered only if the links in the conceptual network are from the seldom called contexts to the often called contexts (so as to insure low enough values of $\gamma$). This seems a natural assumption for grammaticalization phenomena, since functional meanings are much more frequently called than lexical ones. Such assumption remains of course to be carefully investigated. 

These two interpretations of the cognitive link point in the same direction: In short, the links of the conceptual network would be distributed so as to prevent highly frequent forms from invading the less frequent ones, i.e., to ensure linguistic diversity. The asymmetry evidenced by grammaticalization would thus be a consequence of the fact that the highly pervasive functional forms must be kept away from the lexical, referential, more context-specific forms. This puzzling unidirectionality could thus have been selected as a cognitive structure able to guarantee a wide spectrum of possibilities in linguistic expression. 

\subsection{Sociolinguistic interpretation \label{socio_interpretation}} 

We can give our model a completely different interpretation, taking a sociolinguistic view point. Instead of sites $C_0$ and $C_1$, we consider two separate communities of speakers, $C_0$ and $C_1$. Different tokens represent now different individuals, who make binary choices between either variant $X$ or variant $Y$. The different community sizes, $m$ and $M$, are then the analogous of the different memory sizes. The fact that $C_0$ influences unilaterally $C_1$ may be understood as the fact that community $C_0$ has some prestige compared to $C_1$, so that $C_1$ members listen to $C_0$ members while the reverse does not hold. Similarly, different call frequencies may represent different representations in society --- people from prestige communities being given media visibility to the exclusion of the other communities. With this purely sociolinguistic interpretation, the model formalism thus remains exactly the same. Note that this point of view is akin to the one in~\cite{blythe2012s}. 

In this interpretation, however, the model does not explain why the prestige community $C_0$ adopted $X$ in the first place; nor does it explain the regularities in semantic change. Another point in which this interpretation weakens is the timescale. Linguistic change can be very slow, taking up to several centuries, as shown in our corpus study. Is it reasonable to presume that the social structure holds and remains the same throughout centuries? On the contrary, some aspects of conceptual structure happen to be extremely stable, as they are both deeply constitutive of a culture, e.g. through entrenched metaphors~\cite{lakoff2008metaphors}, and due to the generic cognitive features of the mind (expressing time relations through spatial ones~\cite{heine1997cognitive}, for instance). As it happens, metaphors prove to be very stable, even if the reasons for this stability are still unclear. The astonishing persistence of myths schemata through the ages~\cite{da2016comparative} is another hint of the remarkable resilience of human cultural features. 

A last remark is in order. Sociolinguistic explanation describes change as happening through two successive steps~\cite{weinreich1968empirical}: `actuation' of the change (the seemingly sudden appearence of a new variant in the speech of an individual), and propagation of the innovation through social ties. Though Labov deemed actuation as irrelevant for the understanding of language change, numerous efforts have been devoted to make sense of it~\cite{mcmahon1994understanding,nettle1999using}. Recent modeling attemps, following Labov claim, have eluded the difficulty, positing a non-zero initial frequency of the new variant, or assuming that an influent agent is already making use of the variant exclusively~\cite{ke2008language,blythe2012s}. Latency, in particular, cannot fit within this framework. 

The actuation step, on the contrary, has received much attention in Cognitive Linguisitcs and more specifically in the literature on grammaticalization. Indeed, in grammaticalization phenomena, it appears that the actuation process is tightly constrain: not all innovations are equally likely, and changes appear to follow a limited number semantic chains. Several mechanisms of actuation have thus been proposed: invited inference~\cite{traugott1989rise}, conventionalization of an implicature~\cite{nicolle1998relevance}, subjectivation~\cite{marchello2006grammaticalisation}. They all bring forth the idea that a novel variant is always rooted in language use, so that a new form, or a new meaning, always arises out of a contigency from an existing speech practice. Actuation of the change is then an expected result of a particular cognitive organization of language. 

We showed that this process of cognitive actuation is sufficient to explain the S-curve. In a sense, the cognitive interpretation is more economic, as it explains the S-curve (and the latency) by the mechanism of actuation alone, instead of positing a prerequisite actuation, and then explaining the S-curve (but not the latency) as social propagation, which is the case in the sociolinguistic framework. Occam's razor inclines therefore towards the cognitive interpretation of our model and of language change in general. 

\section{Corpus data}

\subsection{Raw data \label{data}}

Raw data has been made available as part of a Supplementary Material on the \href{https://doi.org/10.6084/m9.figshare.c.3910621.v2}{Open Science website},where the folder \texttt{full\_data.zip} can be downloaded. To each studied linguistic form corresponds a file in this folder, named \texttt{form.csv}. This file contains a 70 rows table specifying, for each decade starting with 1321-1330, the number of occurrences of the form found in the corpus, the associated frequency, and the associated averaged frequency (over five decades, as described in Materials and Methods). Two additional files, respectively named \texttt{corpus\_stats.csv} and \texttt{corpus\_complet.csv}, encode all needed information on our corpus. The former is a 70 rows table listing all decades, and giving the number of occurrences associated with each (required to compute the frequency in the individual forms files). The latter is a list of all documents included in the corpus, identified by their Frantext ID. The corresponding date, the corresponding decade, and the associated number of occurrences are also specified. 

\subsection{Frantext textual database \label{A1}}

\begin{figure*}[!bp]
  \centering
  \includegraphics[scale=0.33]{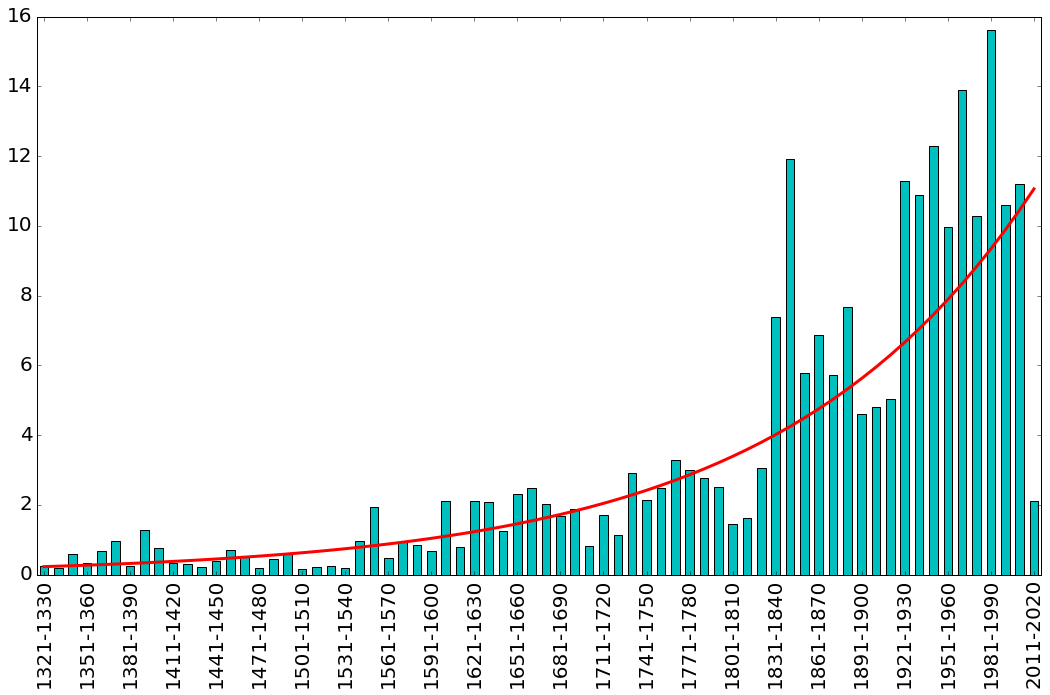}
  \caption{Number of millions of occurrences per decade in the Frantext database. Exponential fit is shown by a red line.}
  \label{fig:fran_stats}
\end{figure*}

The data we collected for the present study have been extracted from the {\it Frantext} database~\cite{frantext}, one of the most extensive databases available in French, to which one has access under subscription by the ATILF-CNRS laboratory. Frantext is an ever-expanding gathering of 4,746 texts to this day (8th december 2016), updated every year. This corpus presents various literary genres (epistolary, drama, poetry, essays, scientific books), but mainly novels, almost exclusively from French literature (with a few translated works). The publication year of the texts range from 950 to 2013. The allotment of the texts between the different time periods is however far from being homogeneous, and most of them belong to the twentieth century: Indeed, the number of texts by decade roughly follows an exponential increase (Fig.~\ref{fig:fran_stats}). 

Frantext, while being much smaller than Google Ngram, provides much cleaner and more controlled results (see~\ref{A6}).  We decided to start from the decade 1321-1330, as from this date all decades are associated with at least seven texts. In our corpus, we retained most of the texts, with a few exceptions, e.g. when the date provided by Frantext was unsatisfying (for instance, the text referred to as 6205, \tit{Le Canarien, pi\`eces justificatives} is dated `between 1327 and 1470'), or when we knew that the text has been written over too long a time period, as is the case for the text \tit{Chartes et documents de l'abbaye de Saint-Magloire} (ref 8203), whose publication year (1330) is far from covering the time span during which the document was compiled. Most interestingly, Frantext also provides the surrounding text on which a token is to be found, so that it is possible to check if the different occurrences make sense and truly correspond to the request.

Frantext is not flawless. Some parts of the scanned texts have been appended through posterior editing. This is clearly the case for the text A017,  \tit{Chroniques de Mor\'ee}, where some page notes from a contemporaneous edition of this medieval chronicle have been included, so that the request for `dans' may return an occurrence such as `Erreur dans la num\'erotation de l'\'edition' (`error {\it in} the edition numbering').  Some decades are also strongly unbalanced in the available texts. For instance, among the 2.7 million words of decade 1551-1560, more than one third of them comes from the works of a single author, Jean Calvin (references E198, B022, R849 to R852). Another bias comes from the fact that drama pieces, up to the end of the Modern Era, were making use of represented orality~\cite{marchello2012oral} much more than literary texts, so that many new constructions appear in them before spreading among the other texts. This would not be a problem if the proportion of dramas were more or less constant across the decades, which is not the case. This problem vanishes in more recent times, when represented orality appears also frequently in novels, while drama becomes itself more sophisticated and shifts further away from daily language. 

Frantext is not only a database. It comes also with built-in text-mining algorithms which allow to submit very refined queries to the database. Such queries can make use of booleans and a given number of blank words. For instance, the query \textbf{(\`a$|$a) \&q(1,2) (ins\c{c}u$|$insu$|$insceu)} (\textbf{\&q(1,2)} is a blank slot for any one or two words) will retrieve occurrences such as \textit{\`a l'insu}, \textit{\`a leur insu}, but also \textit{\`a son propre insu}. This kind of flexible requests are especially relevant when one is looking for specific constructions with a filling slot, as the corresponding possibilities cannot be exhaustively predicted. We studied for instance the construction \textit{d'une voix} + ADJ. If we cannot list all adjectives, we can rule out all the parasite occurrences with an elaborated request such as \textbf{\^{}(tous$|$receus) d'une voix \^{}(que$|$qui$|$qu'$|$et$|$ensemble$|$trestous$|$de$|$d'$|$vous$|$le$ |$la$|  $les$|$par$|$pour$|$dont$|$-$|$.$|$;$|$,$|$:)}, where \textbf{\^{}} and $\mathbf{|}$ respectively stands for the booleans `not' and `or'. Such a request makes it possible to capture unexpected adjectival constructs such as \tit{toute chang\'ee}, \tit{si peu effroy\'ee} or \tit{extraordinairement rauque et rouill\'ee}, while discarding all spurious occurrences. Frantext also allows for special requests, for instance if one wishes to encompass several orthographic variations in a single query, for instance \textbf{souventes?f*} captures all possible variants of  \textit{souventesfois}, such as \textit{souventeffoiz}, \textit{souvente fois}, \tit{souventez fois}, \tit{souventefoys}, etc. This kind of elaborations prove to be all the more useful in the first stages of the evolution, where a functional construction has not yet become entrenched into an idiomatic form and can still be found in a large diversity of variants. 

Once a request is submitted to the database, Frantext returns a datafile whose contents may vary according to the needs of the user. Depending on the options one chooses, the file displays, for each text, the text reference, the publication year, and the total number of occurrences of the query in that text. Next to this automatized procedure, we can also look across all individual occurrences in their context, as a sanity check. This was used frequently to help refining our queries. Unfortunately, it was impossible to ask Frantext for a file providing the statistics of the corpus itself, listing the number of occurrences per text reference. We extracted this information from an HTML page which does display this information (Corpus de travail $>$ Visualiser). The data file provided by Frantext was then directly treated by our own algorithm to compute average frequencies for each decade. 

\subsubsection*{A note on French}

We acknowledge that we restricted ourselves to instances of semantic expansions in French, a choice which may appear to restrict the scope of our findings. As we argue in the main text, we believe this is not the case. In the following, we stress, 1 - the necessity to conduct the analysis on a long timescale (i.e. long enough so that we can consider the language to have changed during that period, just as contemporary French has drifted sufficiently away from Middle French (XIV$^{\text{th}}$ century) so that, without specific training, the latter is only partially intelligible to speakers of the former), 2 - that few corpora are as efficient as Frantext to achieve such a goal. 

Given the issues addressed in this paper, it appears important to consider instances taken from a large time period (seven centuries in our case). Indeed, a frequently asked question is whether or not recent technological advances (radio, TV, the Internet) have had an influence on the way language changes. Sociologically, this influence is obvious: Languages tend to homogenize over greater geographical areas and dialects have constantly declined throughout the twentieth century. Yet, the pattern of change of an established language is something entirely different. Our statistical survey shows that the pattern of change is the same, no matter in which century it may happen. It is furthermore consistent with recent findings establishing that the rate of change did not increase in the most recent decades~\cite{dubossarsky2016verbs}. It also goes along our claim that the pattern we exhibit is cognitively driven by memory retrieval and conceptual organization, two cognitive mechanisms that the most recent technological evolutions could not have significantly altered. 

Alas, finding appropriate corpora covering a long time period in a given language is not obvious. As discussed in~\ref{A6}, Google Ngram cannot be used for texts earlier than the nineteenth centuries, since the scanning procedure does not lead to reliable digital data. For the English language, the reputed British National Corpus restricts itself to the twentieth century. The Helsinki Corpus spans a time period suited for our purposes, but the texts are too sparse (450 in total) for the corpus to be fitted for a statistical survey. The CORDE corpus, in Spanish, spans several centuries (XIII$^{\text{th}}$ to XX$^{\text{th}}$), and gathers an impressive amount of data as well (250$\,$M words), but it covers different variants of Spanish (Argentinian, Colombian, Castillan, etc.) which cannot be blended together when it comes to investigate semantic expansions (note that CORDE dutifully offers to treat them apart, but then the database is not extensive enough for each of the variant separately). The querying system also suffers from serious limitations, and it is not possible to submit complex queries as is the case with Frantext. This latter database is therefore truly remarkable in many aspects and has to be considered an exception. We thus leave to further studies the case of other languages.

A last remark is in order: We deliberately do not provide any translation of the studied forms (\ref{A7}), however obscure they may appear to the reader. Indeed, these forms have all undergone a semantic expansion, so that a translation would be most mistaking as it would concern only one among several meanings adopted by the form. The only satisfying way of glossing the items we studied would have been to find forms which not only have the same meaning, but have also undergone (at least roughly) the same meaning shifts, as in the case of \tit{anyway} and \tit{de toute fa\c{c}on} for the later stages of their respective semantic evolutions. Obviously, this would have been possible only for a handful of cases, and we chose to leave the items without translation.

\subsection{Why not using Google Ngram? \label{A6}}
Google Ngram ({\url{https:/books.google.com/ngrams}) gathers an impressive quantity of digitalized books from about the sixteenth century. It hosts about 800,000 texts in French (about two hundred times more than Frantext). Nevertheless, it presents some major limitations which make this database inappropriate for the present study, as we discuss in this section.

Some biases of the Google Ngram database have already been stressed in the recent past~\cite{pechenick2015characterizing}. However, these concerns are specifically relevant for lexical changes, most subject to socio-historical contingencies, and they do not straightforwardly apply to our aims. Functional words, unlike proper names like `Frodo' or items like `computer', are not that sensitive to cultural shifts. However, there are other serious limitations, more crucially relevant for our study, that we point out here. 
In the following, we use Google Books ({\url{https://books.google.com/}) as a probe to the contents of Google Ngram, though the two algorithms are different (e.g. the former does not regognize ponctuation while the latter does), and the exact overlapping between the contents of the two databases is unknown. 

The first concern about Google Ngram regards quality of digitalization. Texts older than the nineteenth century have been printed in fonts for which the character recognition algorithm has clearly not been optimized. For instance, the following sentence from \tit{The royal dictionary abridged, in two parts}, by Abel Boyer, 1715: `Parler avantageusement de quelqu'un, to speak well of one, to speak much to his advantage, to give a good character of him, to speak honourably of him.' has been transcribed as: `Parler avantageusement e quelqu'un, 1$^{\circ}$ speak well of one, te steak much to his advant 1ge, to ive a gead characier of him, to steak h2nourably of him.'  Some words, such as `steak' and `rince', consequently appear much more frequently than they should, as they are mistaken for `speak' and `Prince'. Another example of this poor scanning quality can be seen in the comparison between: `I found that the New-modelling of this Story, would force me sometimes on the difficult Task of making the chiefest Persons speak something like their Characters, on Matter whereof I had no Ground in my Author.' and `I faura that the Ne: -we kling of this Story, troi'i fr e ve \{ ctives on the di ili 7 k of making ti e li fist Perffns steak \{ like their Carefiers, en -i/attro sviereof. I had no Gr\'eard in » , Author.', to be found in \tit{The History of King Lear, A Tragedy. Acted as the King's-Theatre.} by Nahum Tate, 1736. The original text is admittedly hard to decipher, yet any posterior check on the scan would immediately detect such nonsensical concatenations of characters. By comparison, every text in the Frantext database has been digitalized with great care and such blatant errors are not to be found. 

The second point is the kind of available data. Google Ngram provides statistics on n-grams, which are strings of n successive items (the so-called `grams'), with n ranging from 1 to 5. For each n-gram, it is provided, per year, the number of times it appears and in how many texts. Thus one cannot identify in which texts it appears most; nor can one have access to its context of use. The only way to probe the contents of Google Ngram is through Google Books (which we used here for all the discussed examples), yet it seems impossible to know the exact overlap between the two databases. This data structure based on n-grams is furthermore limiting when it comes to slot constructions. For instance, the French construction `\`a X reprises', with X being a quantity, can hardly be tracked using Google Ngram, as it corresponds to far too many n-grams, which would need to be listed one by one: `\`a deux reprises', `\`a deux ou trois reprises', `\`a plusieurs reprises', `\`a de nombreuses reprises', etc. This search is made all the more difficult by the fact that `\`a' did not always take an accent in older texts. In contrast, with Frantext, as we have seen in~\ref{A1}, one can work out an elaborate request using booleans and blank words to capture the diverse uses of this construction and overcome the orthographic difficulties. 

The third and final point we want to stress here is the choice of texts and their dating. In Frantext, a text may appear in several editions, as is the case for \textit{Le Cid}, by Pierre Corneille, which appear thrice in the database, associated to the years 1637, 1637 and 1682. These dates usually correspond to the first edition of a book, rather than to the edition which is actually digitalized (such information being also provided). Google Ngram displays about thirty versions of \textit{Le Cid}, with publication ranging from 1775 to 2013, some of them being ascribed to Jean Racine (as they are found in several editions of a book called \tit{Oeuvres de J. Racine et de P. et T. Corneille}). The case of \tit{Le Cid} is, in Frantext, quite an exception, while in Google Ngram, most famous classical novels from past centuries are found in a dozen versions at least. 

The contents of the database is problematic as well. As highlighted in~\cite{pechenick2015characterizing}, Google Ngram over-represent academic literature. This also tends to bias the data. For instance, among the fourteen results of the request `par ma barbe' on the French Google Books subdatabase, for the years 1950-2000, only three of them are relevant, two being modern translations of older texts (Don Quixote and a nineteenth century German play by T\"opffer). The third one comes from an anthology of French folktales. All other occurrences are academic quotes and glosses of past works, or reprints of such works. In such a case, it means than only one fifth of the occurrences would be reliable as a reflect of language use in this time period (two of them being borderline cases). Frantext, on the other hand, has two occurrences of `par ma barbe', one of them from the song lyrics of singer Georges Brassens, the other from a 1988 translation of a Shakespeare play (and so more debatable). There is thus almost as many relevant occurrences in Frantext and Google Ngram (two versus three), while none in Frantext are completely irrelevant. 

This being said, Google Ngram is a formidable tool, which can lead to interesting insights and be of great use. It is not, however, fitted for the work that we performed, where we need an accuracy and a reliability that this database is unable to provide. 

\subsection{Studied forms \label{A7}}

Making use of the study of Frantext database and its retrieving tools, we looked at the frequency of use of about 400 hundred semantic expansions in the functional realm (with the exception of \tit{libert\'e} and some lexical constructions such as \textit{\'a court terme}, which we have shown to suggest the further generality of the pattern). We selected these forms according to several criteria: They must have undergone at least one semantic expansion towards a functional use during the time period under consideration; they must be easily distinguished from compositional uses (e.g. \tit{entre deux}, in the meaning of `in between', can be confused with occurrences of literal meaning `between two'). The set of chosen forms is far from exhausting the pool of possible examples. 

On the table below, we provide the full list of studied forms. For each of those, we display:
\begin{itemize} 
\item the length (in decades) of the latency part;
\item the length (in decades) of the growth part;
\item the slope of the logit transform of the growth part;
\item the $r^2$ parameter associated to the linear fit of this logit transform;
\item the $\chi^2$ of the sigmoidal fit of the data (including the boundaries $x_{min}$ and $x_{max}$);
\item the associated Cram\'er's V (which is the square root of the ratio between the $\chi^2$ and the width, or the square root of the mean $\chi^2$ per data point; the smaller the Cram\'er's V, the better the fit)
\item the result of the consistency check (either \checkmark if successful or $\times$ if failed), as described in section~\ref{section:robustness};
\item the total number of occurrences of the form in our corpus. 
\end{itemize}

 Some forms are listed several times; it corresponds to the case where a form underwent several semantic expansion processes, each associated with the latency-growth pattern. `BUG' corresponds to a flaw of Frantext, sometimes unable to build up the output file of the query. This bug cannot be overriden through a manual manoeuvre, for it is caused by a faulty encoding of some parts of the texts. The data thus exist, but could not be retrieved. An upper-case `NO' indicates that no such pattern has been found in the time-evolution of the frequency of that form. The fact that a form does not follow an S-curve during its semantic expansion may spread doubt on the genericity of this pattern. In many cases however, the pattern was rejected because the data was too spurious, but its overall behavior would not be incompatible with an S-curve. 

It is nonetheless interesting to note that the robustness of the pattern does not depend excessively on the scarcity of data. Indeed, instances associated to a very low number of occurrences can lead to a very clean pattern (e.g. \tit{\`a plus d'un titre}, whose growth lasts for 8 decades in total, scores as low as 59 occurrences, and still brings out a remarkable $r^2$ of $0.995$). What seems to be crucial is thus not the question of how much data we can get, but of whether or not the change is isolated. Indeed, some changes are not independent from one another. Many constructions beginning with the preposition \tit{par}, for instance, follow their own course of evolution, while the meaning of \tit{par} itself also expands. Several constructions can also compete for the same paradigm (e.g. \tit{il me semble}, \tit{je pense}, \tit{je suppose}). Their individual frequency pattern not following an S-curve of growth may thus be seen as resulting from interferences between the different semantic expansion processes. In these cases, only the refinement of linguistic queries can lead to better results. It thus confirms, once again, the necessity to rely on a clean and easily manipulable database rather than on giant databases where the sheer amount of data is of no help.

\newpage

\setcounter{table}{0}
\tablehead{\multicolumn{7}{c}%
{{\captionsize\bfseries \ \thetable{LIST OF FORMS}}} \\
\hline  \multicolumn{1}{|c|}{\textbf{Form}} &
\multicolumn{1}{c|}{\textbf{Lat.}} &
\multicolumn{1}{c|}{\textbf{Growth}} &
\multicolumn{1}{c|}{\textbf{Slope}} &
\multicolumn{1}{c|}{\textbf{r$^2$}} &
\multicolumn{1}{c|}{\textbf{$\chi^2$}} &
\multicolumn{1}{c|}{\textbf{C.'s V}} &
\multicolumn{1}{c|}{\textbf{Check}}  &
\multicolumn{1}{c|}{\textbf{\# occ.}} \\ \hline }

\tabletail{ \hline}
\tablelasttail{\hline}

\begin{center}
\begin{xtabular}{|p{4cm}|c|c|c|c|c|c|c|c|}
\hline
\`a base de & 7 & 10 & 0.57 & 0.994 & 0.1097 & 0.1047 & \checkmark & 607 \\ \hline
\`a bien des \'egards (i) & 0 & 8 & 0.79 & 0.983 & 0.0602 & 0.0867 & \checkmark & 147 \\ \hline
\`a bien des \'egards (ii) & 2 & 7 & 1.27 & 0.984 & 0.0567 & 0.0900 & \checkmark & 147 \\ \hline
\`a bord de & NO & NO & NO & NO & NO & NO & NO & 1728 \\ \hline
acabit & 0 & 7 & 1.20 & 0.992 & 0.0345 & 0.0702 & \checkmark & 148 \\ \hline
\`a cause de & NO & NO & NO & NO & NO & NO & NO & 24840 \\ \hline
\`a cause que & NO & NO & NO & NO & NO & NO & NO & 2516 \\ \hline
\`a ce moment (i) & 0 & 10 & 0.61 & 0.989 & 0.0777 & 0.0881 & \checkmark & 8861 \\ \hline
\`a ce moment (ii) & 7 & 7 & 0.69 & 0.992 & 0.0710 & 0.1007 & $\times$ & 8861 \\ \hline
\`a ce propos (i) & 2 & 7 & 2.22 & 0.988 & 0.0004 & 0.0076 & \checkmark & 1711 \\ \hline
\`a ce propos (ii) & 3 & 8 & 0.88 & 0.983 & 0.0623 & 0.0882 & \checkmark & 1711 \\ \hline
\`a ce sujet & 10 & 7 & 1.95 & 0.984 & 0.0217 & 0.0557 & \checkmark & 4001 \\ \hline
\`a cet \'egard & 0 & 8 & 1.56 & 0.992 & 0.1118 & 0.1182 & \checkmark & 4974 \\ \hline
\`a cet instant & 2 & 13 & 0.55 & 0.982 & 0.1123 & 0.0929 & \checkmark & 1198 \\ \hline
\`a condition de & 5 & 9 & 0.79 & 0.991 & 0.0427 & 0.0689 & \checkmark & 1151 \\ \hline
\`a condition que (i) & 11 & 6 & 1.19 & 0.997 & 0.0470 & 0.0885 & \checkmark & 1653 \\ \hline
\`a condition que (ii) & 6 & 8 & 0.83 & 0.971 & 0.0909 & 0.1066 & \checkmark & 1653 \\ \hline
\`a contre-courant & 0 & 16 & 0.59 & 0.971 & 0.0797 & 0.0706 & \checkmark & 171 \\ \hline
\`a côt\'e de & 22 & 14 & 0.57 & 0.965 & 0.0603 & 0.0656 & \checkmark & 18065 \\ \hline
\`a coup s\^ur (i) & 0 & 14 & 0.63 & 0.971 & 0.0916 & 0.0809 & \checkmark & 2546 \\ \hline
\`a coup s\^ur (ii) & 7 & 7 & 1.70 & 0.996 & 0.0088 & 0.0355 & \checkmark & 2546 \\ \hline
\`a court terme & 13 & 7 & 2.19 & 0.997 & 0.0072 & 0.0321 & \checkmark & 751 \\ \hline
\`a couvert & NO & NO & NO & NO & NO & NO & NO & 1144 \\ \hline
actuellement & 9 & 24 & 0.35 & 0.977 & 0.1300 & 0.0736 & $\times$ & 6618 \\ \hline
\`a d\'ecouvert & 1 & 7 & 1.33 & 0.981 & 0.0390 & 0.0746 & \checkmark & 930 \\ \hline
\`a d\'efaut de & NO & NO & NO & NO & NO & NO & NO & 1725 \\ \hline
afin de & 4 & 6 & 0.81 & 0.995 & 0.1155 & 0.1387 & $\times$ & 21833 \\ \hline
afin que & BUG & BUG & BUG & BUG & BUG & BUG & BUG & 19850 \\ \hline
\`a fond de & 0 & 6 & 1.31 & 0.994 & 0.0302 & 0.0709 & \checkmark & 486 \\ \hline
\`a fond de train & BUG & BUG & BUG & BUG & BUG & BUG & BUG & 180 \\ \hline
\`a force & NO & NO & NO & NO & NO & NO & NO & 294 \\ \hline
\`a force de & NO & NO & NO & NO & NO & NO & NO & 8178 \\ \hline
\`a grand renfort de & NO & NO & NO & NO & NO & NO & NO & 230 \\ \hline
ainsi donc & NO & NO & NO & NO & NO & NO & NO & 1247 \\ \hline
\`a la base & NO & NO & NO & NO & NO & NO & NO & 574 \\ \hline
\`a l'accoutum\'ee & 0 & 8 & 0.99 & 0.988 & 0.0327 & 0.0639 & \checkmark & 196 \\ \hline
\`a l'aide de & 13 & 13 & 0.42 & 0.966 & 0.1952 & 0.1225 & $\times$ & 5247 \\ \hline
\`a la limite & 7 & 11 & 0.76 & 0.983 & 0.0448 & 0.0638 & \checkmark & 603 \\ \hline
\`a la lisi\`ere de & 5 & 13 & 0.57 & 0.976 & 0.0638 & 0.0701 & \checkmark & 527 \\ \hline
\`a la longue & 9 & 7 & 0.57 & 0.987 & 0.2092 & 0.1729 & \checkmark & 1245 \\ \hline
\`a la lumi\`ere de & 2 & 9 & 0.87 & 0.975 & 0.0823 & 0.0956 & \checkmark & 1141 \\ \hline
\`a la mesure de (i) & 0 & 7 & 0.75 & 0.994 & 0.0925 & 0.1150 & \checkmark & 819 \\ \hline
\`a la mesure de (ii) & 24 & 9 & 1.14 & 0.988 & 0.0376 & 0.0646 & \checkmark & 819 \\ \hline
\`a la place & 22 & 27 & 0.31 & 0.983 & 0.0918 & 0.0583 & \checkmark & 5638 \\ \hline
\`a la rigueur & 9 & 8 & 1.14 & 0.983 & 0.0697 & 0.0933 & \checkmark & 1717 \\ \hline
\`a l'\'ecart & NO & NO & NO & NO & NO & NO & NO & 2517 \\ \hline
\`a l'\'ecart de & 24 & 19 & 0.30 & 0.970 & 0.1545 & 0.0902 & \checkmark & 854 \\ \hline
\`a l'\'egard de (i) & 4 & 13 & 1.01 & 0.968 & 0.1068 & 0.0906 & $\times$ & 13395 \\ \hline
\`a l'\'egard de (ii) & 2 & 7 & 0.98 & 0.978 & 0.0658 & 0.0970 & \checkmark & 13396 \\ \hline
\`a l'encontre de & 10 & 18 & 0.39 & 0.977 & 0.0946 & 0.0725 & \checkmark & 1272 \\ \hline
\`a l'envi & 0 & 9 & 0.88 & 0.991 & 0.0456 & 0.0712 & \checkmark & 817 \\ \hline
\`a l'exception de & 1 & 16 & 0.47 & 0.968 & 0.1093 & 0.0827 & \checkmark & 1883 \\ \hline
\`a l'heure actuelle & 0 & 11 & 0.95 & 0.981 & 0.0698 & 0.0797 & \checkmark & 858 \\ \hline
\`a l'heure dite & 3 & 9 & 0.83 & 0.969 & 0.0587 & 0.0808 & \checkmark & 234 \\ \hline
\`a l'heure o\`u & 10 & 11 & 0.58 & 0.963 & 0.1015 & 0.0961 & \checkmark & 1779 \\ \hline
\`a l'improviste & 4 & 10 & 0.65 & 0.996 & 0.0549 & 0.0741 & \checkmark & 1024 \\ \hline
\`a l'instant & 0 & 6 & 1.02 & 0.993 & 0.1038 & 0.1315 & \checkmark & 1550 \\ \hline
\`a l'instar de & 7 & 19 & 0.36 & 0.969 & 0.1502 & 0.0889 & \checkmark & 663 \\ \hline
\`a l'insu & 0 & 22 & 0.36 & 0.982 & 0.1347 & 0.0782 & \checkmark & 2776 \\ \hline
\`a l'inverse & 8 & 10 & 1.06 & 0.988 & 0.023 & 0.0476 & \checkmark & 764 \\ \hline
\`a l'occasion de & 6 & 8 & 1.52 & 0.983 & 0.0469 & 0.0766 & \checkmark & 2032 \\ \hline
\`a l'or\'ee de & 5 & 7 & 1.09 & 0.979 & 0.0793 & 0.1064 & \checkmark & 311 \\ \hline
alors que (i) & 3 & 7 & 1.01 & 0.983 & 0.0474 & 0.0823 & \checkmark & 28016 \\ \hline
alors que (ii) & 4 & 13 & 0.50 & 0.983 & 0.0700 & 0.0734 & $\times$ & 28016 \\ \hline
\`a mesure de & 4 & 8 & 0.71 & 0.990 & 0.1122 & 0.1184 & $\times$ & 774 \\ \hline
\`a mesure que (i) & 10 & 10 & 0.80 & 0.967 & 0.0850 & 0.0922 & \checkmark & 10183 \\ \hline
\`a mesure que (ii) & 1 & 11 & 0.50 & 0.965 & 0.1214 & 0.1051 & \checkmark & 10183 \\ \hline
\`a moins que & 0 & 13 & 0.96 & 0.964 & 0.0959 & 0.0859 & \checkmark & 5924 \\ \hline
\`a mon avis & NO & NO & NO & NO & NO & NO & NO & 1989 \\ \hline
\`a nouveau & 7 & 14 & 0.51 & 0.977 & 0.0801 & 0.0756 & $\times$ & 6039 \\ \hline
\`a outrance & NO & NO & NO & NO & NO & NO & NO & 552 \\ \hline
\`a part & 0 & 28 & 0.27 & 0.986 & 0.1034 & 0.0608 & \checkmark & 12506 \\ \hline
\`a part enti\`ere & 0 & 8 & 1.33 & 0.983 & 0.0467 & 0.0764 & \checkmark & 180 \\ \hline
\`a partir de & 13 & 12 & 0.56 & 0.965 & 0.1212 & 0.1005 & \checkmark & 10996 \\ \hline
\`a peine (i) & 0 & 6 & 1.73 & 0.994 & 0.0175 & 0.0540 & \checkmark & 40230 \\ \hline
\`a peu de chose pr\`es & 0 & 7 & 0.94 & 0.987 & 0.0693 & 0.0995 & \checkmark & 320 \\ \hline
\`a plus d'un titre & 1 & 7 & 1.02 & 0.995 & 0.0524 & 0.0865 & \checkmark & 59 \\ \hline
\`a plusieurs reprises (i) & 0 & 19 & 0.36 & 0.967 & 0.1261 & 0.0815 & \checkmark & 3873 \\ \hline
\`a plusieurs reprises (ii) & 9 & 7 & 1.22 & 0.994 & 0.0315 & 0.0671 & \checkmark & 3873 \\ \hline
apr\`es ce & NO & NO & NO & NO & NO & NO & NO & 101 \\ \hline
apr\`es que & 6 & 6 & 2.34 & 0.997 & 0.0078 & 0.0361 & \checkmark & 8487 \\ \hline
apr\`es quoi & 10 & 16 & 0.53 & 0.982 & 0.0742 & 0.0681 & \checkmark & 3468 \\ \hline
apr\`es tout & NO & NO & NO & NO & NO & NO & NO & 7741 \\ \hline
a priori & 3 & 9 & 1.21 & 0.985 & 0.0641 & 0.0844 & \checkmark & 1565 \\ \hline
\`a propos & NO & NO & NO & NO & NO & NO & NO & 1255 \\ \hline
\`a propos de & 1 & 19 & 0.39 & 0.972 & 0.0848 & 0.0668 & \checkmark & 9414 \\ \hline
\`a proprement parler & 10 & 11 & 0.49 & 0.965 & 0.1065 & 0.0984 & \checkmark & 1204 \\ \hline
\`a rebours (i) & 2 & 6 & 1.09 & 0.994 & 0.0627 & 0.1022 & \checkmark & 640 \\ \hline
\`a rebours (ii) & 6 & 12 & 0.66 & 0.979 & 0.0767 & 0.0799 & \checkmark & 640 \\ \hline
\`a qui mieux mieux & NO & NO & NO & NO & NO & NO & NO & 247 \\ \hline
\`a sa guise & 0 & 6 & 1.40 & 0.992 & 0.0503 & 0.0916 & \checkmark & 1079 \\ \hline
\`a son terme & 1 & 11 & 0.75 & 0.991 & 0.0518 & 0.0686 & \checkmark & 359 \\ \hline
\`a tel point que (i) & 0 & 7 & 0.75 & 0.996 & 0.1262 & 0.1343 & \checkmark & 555 \\ \hline
\`a tel point que (ii) & 4 & 9 & 0.73 & 0.975 & 0.1025 & 0.1067 & \checkmark & 555 \\ \hline
\`a terme & NO & NO & NO & NO & NO & NO & NO & 470 \\ \hline
\`a titre de & 5 & 14 & 0.47 & 0.964 & 0.1234 & 0.0939 & \checkmark & 1481 \\ \hline
\`a tous \'egards & 0 & 6 & 1.91 & 0.998 & 0.0138 & 0.0480 & \checkmark & 556 \\ \hline
\`a tout \`a l'heure & 0 & 10 & 0.93 & 0.983 & 0.0655 & 0.0809 & \checkmark & 280 \\ \hline
\`a tout instant & 5 & 11 & 0.94 & 0.969 & 0.0846 & 0.0877 & \checkmark & 903 \\ \hline
\`a tout moment & 0 & 17 & 0.42 & 0.968 & 0.1608 & 0.0973 & $\times$ & 2262 \\ \hline
\`a tout prendre & NO & NO & NO & NO & NO & NO & NO & 480 \\ \hline
au bord de & NO & NO & NO & NO & NO & NO & NO & 11850 \\ \hline
au bout de & NO & NO & NO & NO & NO & NO & NO & 23173 \\ \hline
au bout du compte & NO & NO & NO & NO & NO & NO & NO & 469 \\ \hline
au contraire & 3 & 9 & 0.93 & 0.978 & 0.0606 & 0.0821 & \checkmark & 29571 \\ \hline
au contraire de (i) & 0 & 8 & 1.09 & 0.977 & 0.0475 & 0.0771 & \checkmark & 1429 \\ \hline
au contraire de (ii) & 1 & 8 & 1.14 & 0.989 & 0.0341 & 0.0653 & \checkmark & 1429 \\ \hline
aucunefois & NO & NO & NO & NO & NO & NO & NO & 1248 \\ \hline
au demeurant & 0 & 12 & 0.68 & 0.983 & 0.0685 & 0.0756& \checkmark  & 1344 \\ \hline
au d\'epourvu & NO & NO & NO & NO & NO & NO & NO & 402 \\ \hline
au d\'etriment de & NO & NO & NO & NO & NO & NO & NO & 798 \\ \hline
au dernier moment & NO & NO & NO & NO & NO & NO & NO & 1370 \\ \hline
au final & NO & NO & NO & NO & NO & NO & NO & 38 \\ \hline
au fur et \`a mesure & 6 & 12 & 0.72 & 0.987 & 0.0340 & 0.0532 & $\times$ & 1908 \\ \hline
au jour d'aujourd'hui & NO & NO & NO & NO & NO & NO & NO & 87 \\ \hline
au m\^eme moment & 5 & 7 & 0.73 & 0.979 & 0.1091 & 0.1248 & \checkmark & 1437 \\ \hline
au moment o\`u & 6 & 19 & 0.49 & 0.984 & 0.0403 & 0.0461 & \checkmark & 12729 \\ \hline
\`a un moment donn\'e & 1 & 12 & 0.48 & 0.980 & 0.1249 & 0.1020 & \checkmark & 659 \\ \hline
au passage & 0 & 7 & 1.43 & 0.990 & 0.0492 & 0.0838 & \checkmark & 1754 \\ \hline
au pire (i) & 0 & 12 & 0.46 & 0.965 & 0.1424 & 0.1089 & \checkmark & 401 \\ \hline
au pire (ii) & 0 & 6 & 1.63 & 0.994 & 0.0315 & 0.0725 & \checkmark & 401 \\ \hline
au reste & 0 & 7 & 1.39 & 0.987 & 0.0350 & 0.0707 & \checkmark & 4375 \\ \hline
au sujet de & 1 & 12 & 0.75 & 0.981 & 0.0565 & 0.0686 & \checkmark & 4945 \\ \hline
au terme de (i) & 7 & 12 & 0.47 & 0.971 & 0.1214 & 0.1006 & \checkmark & 1492 \\ \hline
au terme de (ii) & 1 & 11 & 0.86 & 0.967 & 0.0984 & 0.0946 & \checkmark & 1492 \\ \hline
aux trousses & NO & NO & NO & NO & NO & NO & NO & 419 \\ \hline
avant tout & 27 & 10 & 0.91 & 0.986 & 0.0565 & 0.0752 & \checkmark & 5342 \\ \hline
avec force & NO & NO & NO & NO & NO & NO & NO & 324 \\ \hline
bah & 7 & 11 & 1.03 & 0.964 & 0.0579 & 0.0726 & \checkmark & 2681 \\ \hline
bien entendu (i) & 5 & 10 & 0.76 & 0.985 & 0.0555 & 0.0745 & \checkmark & 4476 \\ \hline
bien entendu (ii) & 2 & 19 & 0.40 & 0.979 & 0.1410 & 0.0861 & $\times$ & 4476 \\ \hline
bien s\^ur & 9 & 9 & 0.92 & 0.968 & 0.0839 & 0.0966 & \checkmark & 7997 \\ \hline
bref & 12 & 7 & 1.07 & 0.993 & 0.0353 & 0.0710 & \checkmark & 5536 \\ \hline
brusquement & 11 & 9 & 1.49 & 0.979 & 0.1103 & 0.1107 & $\times$ & 1783 \\ \hline
carr\'ement (i) & 0 & 10 & 1.02 & 0.964 & 0.0640 & 0.0800 & \checkmark & 1207 \\ \hline
carr\'ement (ii) & 1 & 7 & 1.55 & 0.982 & 0.0433 & 0.0786 & \checkmark & 1207 \\ \hline
ce faisant (i) & 0 & 6 & 1.88 & 0.992 & 0.0180 & 0.0548 & \checkmark & 781 \\ \hline
ce faisant (ii) & 19 & 8 & 0.64 & 0.994 & 0.0918 & 0.1071 & \checkmark & 781 \\ \hline
ce par quoi & 0 & 10 & 0.61 & 0.984 & 0.0707 & 0.0841 & \checkmark & 163 \\ \hline
c'est alors que & 5 & 11 & 0.69 & 0.967 & 0.0732 & 0.0816 & \checkmark & 3223 \\ \hline
c'est pour le coup que & 0 & 6 & 0.77 & 0.990 & 0.1309 & 0.1477 & \checkmark & 64 \\ \hline
c'est pourquoi (i) & 0 & 13 & 0.56 & 0.984 & 0.0380 & 0.0541 & $\times$ & 10994 \\ \hline
c'est pourquoi (ii) & 6 & 15 & 0.59 & 0.968 & 0.1755 & 0.1082 & $\times$ & 10994 \\ \hline
chemin faisant & 0 & 13 & 0.40 & 0.965 & 0.1220 & 0.0969 & \checkmark & 641 \\ \hline
compl\`etement & NO & NO & NO & NO & NO & NO & NO & 11560 \\ \hline
compte tenu de & 0 & 8 & 1.26 & 0.985 & 0.0291 & 0.0603 & \checkmark & 928 \\ \hline
concernant & 9 & 10 & 1.10 & 0.984 & 0.0615 & 0.0784 & $\times$ & 3477 \\ \hline
consid\'erant que & NO & NO & NO & NO & NO & NO & NO & 191 \\ \hline
contre mon attente & NO & NO & NO & NO & NO & NO & NO & 102 \\ \hline
contre toute attente (i) & 0 & 6 & 0.84 & 0.991 & 0.1103 & 0.1356 & \checkmark & 167 \\ \hline
contre toute attente (ii) & 8 & 8 & 0.95 & 0.971 & 0.0979 & 0.1106 & \checkmark & 167 \\ \hline
d'abord et avant tout & NO & NO & NO & NO & NO & NO & NO & 62 \\ \hline
d'ann\'ee en ann\'ee & NO & NO & NO & NO & NO & NO & NO & 505 \\ \hline
dans ce cas & 0 & 18 & 0.57 & 0.974 & 0.1264 & 0.0838 & $\times$ & 4289 \\ \hline
dans la mesure de & 6 & 12 & 0.49 & 0.983 & 0.0723 & 0.0776 & \checkmark & 480 \\ \hline
dans la mesure du possible & 0 & 12 & 0.64 & 0.980 & 0.0465 & 0.0622 & \checkmark & 188 \\ \hline
dans la mesure o\`u & 0 & 11 & 0.91 & 0.965 & 0.1322 & 0.1096 & $\times$ & 2753 \\ \hline
dans le cadre de & 11 & 8 & 1.16 & 0.971 & 0.0330 & 0.0642 & \checkmark & 1145 \\ \hline
dans le m\^eme temps (i) & 0 & 9 & 1.02 & 0.986 & 0.0333 & 0.0608 & \checkmark & 1217 \\ \hline
dans le m\^eme temps (ii) & 3 & 7 & 0.90 & 0.983 & 0.1137 & 0.1274 & \checkmark & 1217 \\ \hline
dans l'ensemble (i) & 0 & 9 & 0.74 & 0.967 & 0.0816 & 0.0952 & \checkmark & 1809 \\ \hline
dans l'ensemble (ii) & 10 & 9 & 0.93 & 0.969 & 0.0661 & 0.0857 & \checkmark & 1809 \\ \hline
dans l'imm\'ediat & 10 & 9 & 1.10 & 0.984 & 0.0303 & 0.0580 & \checkmark & 329 \\ \hline
dans quelque temps & 0 & 6 & 0.76 & 0.991 & 0.1579 & 0.1622 & \checkmark & 234 \\ \hline
dans son ensemble & 1 & 9 & 0.67 & 0.979 & 0.0725 & 0.0898 & \checkmark & 835 \\ \hline
dans un autre temps & NO & NO & NO & NO & NO & NO & NO & 143 \\ \hline
dans un cas comme dans l'autre & 1 & 8 & 1.04 & 0.983 & 0.0373 & 0.0683 & \checkmark & 111 \\ \hline
dans une large mesure & 5 & 8 & 0.66 & 0.994 & 0.0870 & 0.1043 & \checkmark & 381 \\ \hline
dans un instant & 1 & 11 & 0.67 & 0.969 & 0.0690 & 0.0792 & \checkmark & 661 \\ \hline
dans un moment & 0 & 15 & 0.47 & 0.984 & 0.0673 & 0.0670 & \checkmark & 1473 \\ \hline
dans un premier temps & NO & NO & NO & NO & NO & NO & NO & 229 \\ \hline
dans tous les cas & 5 & 14 & 0.71 & 0.978 & 0.1111 & 0.0891 & \checkmark & 1609 \\ \hline
d'autant plus & 1 & 10 & 0.74 & 0.976 & 0.0721 & 0.0849 & \checkmark & 11584 \\ \hline
d'autant plus que & 4 & 7 & 1.57 & 0.979 & 0.0446 & 0.0798 & \checkmark & 3339 \\ \hline
d'autre part (i) & 0 & 7 & 0.99 & 0.988 & 0.0697 & 0.0998 & \checkmark & 11012 \\ \hline
d'autre part (ii) & 12 & 12 & 0.64 & 0.982 & 0.0730 & 0.0780 & $\times$ & 11012 \\ \hline
de ce côt\'e & NO & NO & NO & NO & NO & NO & NO & 3665 \\ \hline
d\'ecid\'ement & 0 & 15 & 0.44 & 0.976 & 0.0931 & 0.0788 & $\times$ & 4795 \\ \hline
de ce fait & 2 & 9 & 0.69 & 0.974 & 0.1142 & 0.1126 & $\times$ & 628 \\ \hline
de fa\c{c}on que & NO & NO & NO & NO & NO & NO & NO & 1473 \\ \hline
de fait & 0 & 9 & 0.97 & 0.978 & 0.0569 & 0.0795 & \checkmark & 5018 \\ \hline
de jour en jour & NO & NO & NO & NO & NO & NO & NO & 2217 \\ \hline
de la part de & NO & NO & NO & NO & NO & NO & NO & 16400 \\ \hline
de la sorte & 8 & 13 & 0.61 & 0.965 & 0.0855 & 0.0811 & \checkmark & 3752 \\ \hline
de l'aveu de & 0 & 17 & 0.34 & 0.986 & 0.1072 & 0.0794 & \checkmark & 196 \\ \hline
de l'avis de & NO & NO & NO & NO & NO & NO & NO & 146 \\ \hline
de loin & 16 & 10 & 0.73 & 0.994 & 0.0314 & 0.0560 & \checkmark & 1262 \\ \hline
de loin en loin & 0 & 20 & 0.35 & 0.975 & 0.1541 & 0.0878 & \checkmark & 1348 \\ \hline
de long en large & 3 & 9 & 0.76 & 0.983 & 0.0550 & 0.0782 & \checkmark & 734 \\ \hline
d'embl\'ee & 3 & 10 & 0.72 & 0.986 & 0.0465 & 0.0682 & \checkmark & 1451 \\ \hline
de m\`eche & NO & NO & NO & NO & NO & NO & NO & 98 \\ \hline
de mieux en mieux & 0 & 6 & 1.32 & 0.996 & 0.0519 & 0.0930 & \checkmark & 445 \\ \hline
de moins en moins & 6 & 21 & 0.28 & 0.980 & 0.1005 & 0.0692 & \checkmark & 1536 \\ \hline
de mon côt\'e & 0 & 14 & 0.71 & 0.981 & 0.0836 & 0.0773 & $\times$ & 8788 \\ \hline
de mon fait & NO & NO & NO & NO & NO & NO & NO & 467 \\ \hline
de nulle part & 24 & 13 & 0.32 & 0.978 & 0.1708 & 0.1146 & \checkmark & 289 \\ \hline
de pair & 12 & 7 & 1.04 & 0.983 & 0.0575 & 0.0906 & \checkmark & 578 \\ \hline
de place en place & 15 & 9 & 0.89 & 0.974 & 0.0463 & 0.0717 & \checkmark & 376 \\ \hline
de point en point & 0 & 6 & 0.75 & 0.988 & 0.1511 & 0.1587 & \checkmark & 247 \\ \hline
de part en part (i) & 0 & 7 & 1.49 & 0.995 & 0.0218 & 0.0558 & \checkmark & 498 \\ \hline
de part en part (ii) & 7 & 6 & 1.42 & 0.989 & 0.0386 & 0.0802 & \checkmark & 498 \\ \hline
de part et d'autre & NO & NO & NO & NO & NO & NO & NO & 2505 \\ \hline
de plus en plus & 0 & 6 & 1.08 & 0.999 & 0.0587 & 0.0989 & \checkmark & 18226 \\ \hline
de pr\`es ou de loin & NO & NO & NO & NO & NO & NO & NO & 221 \\ \hline
de proche en proche & 3 & 9 & 1.00 & 0.987 & 0.0273 & 0.0551 & \checkmark & 702 \\ \hline
de quelque part & NO & NO & NO & NO & NO & NO & NO & 166 \\ \hline
des fois & 5 & 12 & 0.75 & 0.968 & 0.0770 & 0.0801 & \checkmark & 1423 \\ \hline
des fois que & 0 & 10 & 0.75 & 0.977 & 0.0673 & 0.0820 & \checkmark & 182 \\ \hline
d\`es l'instant (i) & 9 & 8 & 0.51 & 0.976 & 0.1873 & 0.1530 & \checkmark & 769 \\ \hline
d\`es l'instant (ii) & 1 & 8 & 0.96 & 0.988 & 0.0622 & 0.0882 & \checkmark & 769 \\ \hline
d\`es lors que & NO & NO & NO & NO & NO & NO & NO & 994 \\ \hline
de sorte que & 6 & 6 & 0.85 & 0.997 & 0.1307 & 0.1476 & \checkmark & 11320 \\ \hline
de surcro\^it & 36 & 11 & 1.08 & 0.918 & 0.0811 & 0.0859 & \checkmark & 720 \\ \hline
de temps \`a autre & 15 & 15 & 0.52 & 0.969 & 0.1061 & 0.0841 & \checkmark & 3547 \\ \hline
de temps en temps & 1 & 14 & 0.40 & 0.977 & 0.1176 & 0.0917 & \checkmark & 8916 \\ \hline
de toute fa\c{c}on & 35 & 11 & 0.76 & 0.996 & 0.0224 & 0.0451 & \checkmark & 3595 \\ \hline
de toute mani\`ere & 25 & 9 & 0.66 & 0.973 & 0.0858 & 0.0976 & \checkmark & 727 \\ \hline
de toutes fa\c{c}ons & 4 & 6 & 2.04 & 0.989 & 0.0306 & 0.0714 & \checkmark & 715 \\ \hline
de toutes parts & 13 & 10 & 0.81 & 0.972 & 0.0780 & 0.0883 & \checkmark & 4792 \\ \hline
d'heure en heure & NO & NO & NO & NO & NO & NO & NO & 573 \\ \hline
d'ici l\`a & 16 & 9 & 0.60 & 0.985 & 0.1554 & 0.1314 & $\times$ & 904 \\ \hline
dor\'enavant & NO & NO & NO & NO & NO & NO & NO & 256 \\ \hline
d'outre en outre & NO & NO & NO & NO & NO & NO & NO & 47 \\ \hline
du fait de & 24 & 8 & 0.73 & 0.986 & 0.0702 & 0.0937 & \checkmark & 1423 \\ \hline
du m\^eme coup & 14 & 17 & 0.47 & 0.991 & 0.0506 & 0.0546 & \checkmark & 1502 \\ \hline
du moment que & 2 & 8 & 0.84 & 0.979 & 0.0646 & 0.0899 & \checkmark & 1765 \\ \hline
d'une mani\`ere ou d'une autre & NO & NO & NO & NO & NO & NO & NO & 320 \\ \hline
d'une part (i) & 0 & 8 & 1.13 & 0.985 & 0.0354 & 0.0665 & \checkmark & 5671 \\ \hline
d'une part (ii) & 3 & 7 & 0.80 & 0.982 & 0.0844 & 0.1098 & $\times$ & 5671 \\ \hline
d'une voix claire (i) & 6 & 10 & 0.66 & 0.993 & 0.0542 & 0.0736 & \checkmark & 13511 \\ \hline
d'une voix claire (ii) & 4 & 9 & 0.71 & 0.985 & 0.0594 & 0.0812 & \checkmark & 13511 \\ \hline
du pareil au m\^eme & NO & NO & NO & NO & NO & NO & NO & 92 \\ \hline
du point de vue de & 7 & 8 & 0.80 & 0.995 & 0.0485 & 0.0779 & \checkmark & 899 \\ \hline
du reste & 1 & 6 & 1.82 & 0.994 & 0.0178 & 0.0545 & \checkmark & 5510 \\ \hline
en attendant & NO & NO & NO & NO & NO & NO & NO & 3351 \\ \hline
en attendant de & 4 & 8 & 1.09 & 0.979 & 0.0641 & 0.0895 & \checkmark & 510 \\ \hline
en attendant que & NO & NO & NO & NO & NO & NO & NO & 2270 \\ \hline
en bordure de & 0 & 11 & 0.83 & 0.983 & 0.0403 & 0.0605 & \checkmark & 434 \\ \hline
en bref & 1 & 6 & 1.06 & 0.996 & 0.0724 & 0.1098 & \checkmark & 339 \\ \hline
en ce moment (i) & 5 & 8 & 1.09 & 0.985 & 0.0624 & 0.0883 & \checkmark & 12751 \\ \hline
en ce moment (ii) & 1 & 11 & 0.53 & 0.970 & 0.1012 & 0.0959 & $\times$ & 12751 \\ \hline
en ce que & 0 & 8 & 1.24 & 0.972 & 0.0544 & 0.0825 & \checkmark & 3971 \\ \hline
en ce qui concerne & 28 & 14 & 0.47 & 0.973 & 0.1243 & 0.0942 & $\times$ & 3950 \\ \hline
en ce qui me concerne & 0 & 13 & 0.57 & 0.970 & 0.0922 & 0.0842 & \checkmark & 682 \\ \hline
en consid\'eration de & 0 & 10 & 0.74 & 0.967 & 0.0960 & 0.0980 & \checkmark & 409 \\ \hline
en cours de & 0 & 14 & 0.54 & 0.984 & 0.1289 & 0.0960 & $\times$ & 1110 \\ \hline
en cours de route & 0 & 8 & 0.91 & 0.986 & 0.0595 & 0.0862 & \checkmark & 301 \\ \hline
en d'autres termes & 0 & 10 & 0.70 & 0.971 & 0.0615 & 0.0784 & \checkmark & 1228 \\ \hline
en d\'efinitive & NO & NO & NO & NO & NO & NO & NO & 1538 \\ \hline
en d\'epit de & NO & NO & NO & NO & NO & NO & NO & 4016 \\ \hline
en face de & 18 & 17 & 0.39 & 0.982 & 0.1044 & 0.0784 & $\times$ & 10956 \\ \hline
en fa\c{c}on que & NO & NO & NO & NO & NO & NO & NO & 48 \\ \hline
en fait (i) & 0 & 6 & 0.91 & 0.990 & 0.1372 & 0.1512 & \checkmark & 8871 \\ \hline
en fait (ii) & 44 & 12 & 0.61 & 0.968 & 0.0726 & 0.0778 & $\times$ & 8871 \\ \hline
en fin de compte & 27 & 14 & 0.43 & 0.972 & 0.0955 & 0.0826 & \checkmark & 1417 \\ \hline
en gros & 23 & 10 & 0.86 & 0.975 & 0.0461 & 0.0679 & \checkmark & 320 \\ \hline
en guise de & 7 & 10 & 0.93 & 0.982 & 0.0390 & 0.0624 & \checkmark & 1598 \\ \hline
en instance de & 1 & 9 & 0.80 & 0.967 & 0.1036 & 0.1073 & \checkmark & 77 \\ \hline
en l'occurrence & 0 & 11 & 0.58 & 0.993 & 0.0476 & 0.0658 & \checkmark & 525 \\ \hline
en long et en large & NO & NO & NO & NO & NO & NO & NO & 108 \\ \hline
en m\^eme temps & 3 & 10 & 0.88 & 0.996 & 0.0242 & 0.0492 & \checkmark & 18370 \\ \hline
en m\^eme temps que & 0 & 18 & 0.47 & 0.968 & 0.1421 & 0.0889 & \checkmark & 8241 \\ \hline
en mesure de & NO & NO & NO & NO & NO & NO & NO & 1470 \\ \hline
en particulier (i) & 2 & 6 & 1.26 & 0.993 & 0.0352 & 0.0766 & \checkmark & 8949 \\ \hline
en particulier (ii) & 17 & 15 & 0.39 & 0.984 & 0.1001 & 0.0817& \checkmark  & 8949 \\ \hline
en particulier (iii) & 4 & 7 & 0.80 & 0.987 & 0.0853 & 0.1104 & $\times$ & 8949 \\ \hline
en partie & NO & NO & NO & NO & NO & NO & NO & 5645 \\ \hline
en passe de & NO & NO & NO & NO & NO & NO & NO & 46 \\ \hline
en plein & NO & NO & NO & NO & NO & NO & NO & 183 \\ \hline
en plein qqch & 31 & 11 & 0.94 & 0.968 & 0.0720 & 0.0809 & $\times$ & 15939 \\ \hline
en quelque sorte & NO & NO & NO & NO & NO & NO & NO & 6422 \\ \hline
en sorte que & NO & NO & NO & NO & NO & NO & NO & 4786 \\ \hline
en suspens & BUG & BUG & BUG & BUG & BUG & BUG & BUG & 961 \\ \hline
en tant que tel & 8 & 6 & 0.91 & 0.997 & 0.0836 & 0.1180 & \checkmark & 314 \\ \hline
entre autres & NO & NO & NO & NO & NO & NO & NO & 4402 \\ \hline
en v\'erit\'e (i) & 0 & 6 & 0.93 & 0.988 & 0.1352 & 0.1501 & \checkmark & 8194 \\ \hline
en v\'erit\'e (ii) & 5 & 8 & 0.65 & 0.971 & 0.1104 & 0.1175 & \checkmark & 8194 \\ \hline
en voie de & 0 & 24 & 0.36 & 0.970 & 0.1239 & 0.0719 & \checkmark & 1027 \\ \hline
en vue de (i) & 5 & 6 & 1.31 & 0.995 & 0.0348 & 0.0762 & \checkmark & 3625 \\ \hline
en vue de (ii) & 11 & 14 & 0.34 & 0.964 & 0.1586 & 0.1064 & $\times$ & 3625 \\ \hline
\'epoque & 7 & 14 & 0.75 & 0.993 & 0.0340 & 0.0493 & $\times$ & 32290 \\ \hline
essentiellement & 3 & 7 & 0.80 & 0.976 & 0.0892 & 0.1129 & $\times$ & 5471 \\ \hline
\'etant donn\'e que & 2 & 13 & 0.62 & 0.984 & 0.0474 & 0.0604 & \checkmark & 341 \\ \hline
et apr\`es & NO & NO & NO & NO & NO & NO & NO & 7562 \\ \hline
except\'e & 5 & 7 & 0.66 & 0.990 & 0.1250 & 0.1336 & \checkmark & 5042 \\ \hline
faute de (i) & 5 & 7 & 1.76 & 0.990 & 0.0133 & 0.0436 & \checkmark & 6725 \\ \hline
faute de (ii) & 7 & 12 & 0.78 & 0.978 & 0.0642 & 0.0731 & \checkmark & 6725 \\ \hline
faute de quoi & NO & NO & NO & NO & NO & NO & NO & 262 \\ \hline
force est de & NO & NO & NO & NO & NO & NO & NO & 84 \\ \hline
fors & BUG & BUG & BUG & BUG & BUG & BUG & BUG & 4451 \\ \hline
graduellement & NO & NO & NO & NO & NO & NO & NO & 827 \\ \hline
hormis & 4 & 11 & 0.83 & 0.964 & 0.0934 & 0.0921 & \checkmark & 1464 \\ \hline
il me semble & NO & NO & NO & NO & NO & NO & NO & 1822 \\ \hline
il s'agit de & 3 & 17 & 0.33 & 0.978 & 0.1358 & 0.0894 & \checkmark & 11558 \\ \hline
il y a moyen & 8 & 8 & 1.31 & 0.979 & 0.0307 & 0.0619 & \checkmark & 1295 \\ \hline
j'ai l'impression & 0 & 9 & 0.57 & 0.983 & 0.1066 & 0.1088 & \checkmark & 74 \\ \hline
ja soit ce que & NO & NO & NO & NO & NO & NO & NO & 268 \\ \hline
je pense & 5 & 6 & 1.48 & 0.989 & 0.0258 & 0.0656 & \checkmark & 4033 \\ \hline
je suppose & 0 & 8 & 1.00 & 0.995 & 0.0258 & 0.0568 & \checkmark & 1110 \\ \hline
j'imagine & NO & NO & NO & NO & NO & NO & NO & 824 \\ \hline
jusque l\`a & NO & NO & NO & NO & NO & NO & NO & 6908 \\ \hline
juste un & 26 & 11 & 0.59 & 0.973 & 0.1148 & 0.1022 & $\times$ & 1366 \\ \hline
l'autre jour & NO & NO & NO & NO & NO & NO & NO & 4438 \\ \hline
lendemain & NO & NO & NO & NO & NO & NO & NO & 28780 \\ \hline
le temps de & 20 & 13 & 0.46 & 0.972 & 0.1007 & 0.0880 & \checkmark & 1195 \\ \hline
libert\'e & 2 & 9 & 0.87 & 0.990 & 0.0500 & 0.0745 & \checkmark & 46705 \\ \hline
l'un dans l'autre & NO & NO & NO & NO & NO & NO & NO & 69 \\ \hline
l'un apr\`es l'autre & NO & NO & NO & NO & NO & NO & NO & 2010 \\ \hline
m'est avis & NO & NO & NO & NO & NO & NO & NO & 797 \\ \hline
nettement & 0 & 10 & 0.45 & 0.972 & 0.1682 & 0.1297 & \checkmark & 6109 \\ \hline
nomm\'ement & NO & NO & NO & NO & NO & NO & NO & 453 \\ \hline
non pas tant & 3 & 6 & 1.25 & 1.000 & 0.0312 & 0.0721 & \checkmark & 855 \\ \hline
non seulement & 11 & 18 & 0.57 & 0.966 & 0.1195 & 0.0815 & \checkmark & 22599 \\ \hline
non pas seulement & 2 & 10 & 0.94 & 0.987 & 0.1166 & 0.1080 & \checkmark & 1605 \\ \hline
notamment & 28 & 9 & 0.52 & 0.976 & 0.1690 & 0.1370 & $\times$ & 7508 \\ \hline
nulle part & 5 & 12 & 0.57 & 0.980 & 0.0679 & 0.0752 & \checkmark & 5006 \\ \hline
or donc & NO & NO & NO & NO & NO & NO & NO & 237 \\ \hline
ouille & 0 & 10 & 0.98 & 0.977 & 0.0399 & 0.0632 & \checkmark & 106 \\ \hline
outre mesure & 3 & 11 & 0.52 & 0.970 & 0.1227 & 0.1056 & \checkmark & 664 \\ \hline
par \`a-coups & 0 & 13 & 0.46 & 0.974 & 0.1224 & 0.0970 & \checkmark & 212 \\ \hline
par ailleurs & 27 & 13 & 0.99 & 0.971 & 0.0602 & 0.0680 & $\times$ & 2676 \\ \hline
par amour & 5 & 8 & 0.87 & 0.981 & 0.0718 & 0.0947 & \checkmark & 303 \\ \hline
par avance & 2 & 13 & 0.68 & 0.969 & 0.0555 & 0.0653 & \checkmark & 1265 \\ \hline
par besoin de & 7 & 9 & 1.05 & 0.990 & 0.0357 & 0.0630 & \checkmark & 156 \\ \hline
par ce fait & 0 & 9 & 1.05 & 0.990 & 0.0320 & 0.0596 & \checkmark & 101 \\ \hline
par cons\'equent & 6 & 6 & 1.30 & 0.998 & 0.0525 & 0.0935 & \checkmark & 12234 \\ \hline
par contre & 18 & 12 & 0.72 & 0.989 & 0.0501 & 0.0646 & $\times$ & 3014 \\ \hline
par crainte de & 19 & 12 & 0.53 & 0.981 & 0.0796 & 0.0814 & \checkmark & 534 \\ \hline
par degr\'es & 37 & 9 & 0.61 & 0.973 & 0.0960 & 0.1033 & \checkmark & 1447 \\ \hline
par dessus tout & 2 & 10 & 0.73 & 0.991 & 0.0463 & 0.0680 & \checkmark & 1433 \\ \hline
pareil \`a  & 11 & 12 & 0.51 & 0.973 & 0.0964 & 0.0896 & $\times$ & 6787 \\ \hline
par excellence & NO & NO & NO & NO & NO & NO & NO & 1749 \\ \hline
par faute de & 4 & 7 & 0.91 & 0.983 & 0.1526 & 0.1476 & \checkmark & 353 \\ \hline
parfois & 12 & 21 & 0.46 & 0.972 & 0.2241 & 0.1033 & $\times$ & 39445 \\ \hline
par go\^ut de & 15 & 11 & 0.63 & 0.973 & 0.1203 & 0.1046 & \checkmark & 143 \\ \hline
par hasard & 3 & 10 & 0.98 & 0.971 & 0.0675 & 0.0822 & \checkmark & 7071 \\ \hline
par instants & 0 & 9 & 0.73 & 0.993 & 0.0462 & 0.0716 & \checkmark & 1357 \\ \hline
par manque de & 0 & 18 & 0.23 & 0.970 & 0.1972 & 0.1047 & \checkmark & 268 \\ \hline
par m\'egarde & NO & NO & NO & NO & NO & NO & NO & 578 \\ \hline
parmi d'autres & 25 & 19 & 0.52 & 0.971 & 0.2387 & 0.1121 & \checkmark & 620 \\ \hline
par moments & 0 & 12 & 0.88 & 0.984 & 0.0467 & 0.0624 & \checkmark & 2774 \\ \hline
par ordre de & 8 & 12 & 0.69 & 0.968 & 0.0725 & 0.0777 & \checkmark & 877 \\ \hline
par peur de & 0 & 11 & 0.59 & 0.987 & 0.0564 & 0.0716 & \checkmark & 268 \\ \hline
par pr\'ecaution & NO & NO & NO & NO & NO & NO & NO & 241 \\ \hline
par rapport \`a (i) & 0 & 10 & 0.97 & 0.978 & 0.0422 & 0.0650 & \checkmark & 5290 \\ \hline
par rapport \`a (ii) & 2 & 9 & 0.81 & 0.970 & 0.0594 & 0.0812 & \checkmark & 5290 \\ \hline
par souci de & 4 & 10 & 1.12 & 0.967 & 0.0789 & 0.0888 & \checkmark & 186 \\ \hline
par surcro\^it & 19 & 12 & 0.59 & 0.982 & 0.0586 & 0.0699 & \checkmark & 498 \\ \hline
particuli\`erement (i) & 14 & 18 & 0.51 & 0.982 & 0.0753 & 0.0647 & \checkmark & 12784 \\ \hline
particuli\`erement (ii) & 3 & 7 & 1.18 & 0.978 & 0.0486 & 0.0833 & \checkmark & 12784 \\ \hline
par voie de & NO & NO & NO & NO & NO & NO & NO & 976 \\ \hline
par voie de cons\'equence & 0 & 10 & 0.47 & 0.968 & 0.1316 & 0.1147 & \checkmark & 130 \\ \hline
petit \`a petit & 3 & 10 & 0.77 & 0.985 & 0.0857 & 0.0926 & \checkmark & 1547 \\ \hline
peu \`a peu (i) & 11 & 8 & 0.94 & 0.970 & 0.0638 & 0.0893 & \checkmark & 16450 \\ \hline
peu \`a peu (ii) & 0 & 10 & 0.80 & 0.970 & 0.0736 & 0.0858 & \checkmark & 16450 \\ \hline
peu s'en faut & 0 & 9 & 0.92 & 0.982 & 0.0498 & 0.0744 & \checkmark & 221 \\ \hline
pour ainsi dire & 0 & 15 & 0.86 & 0.973 & 0.2276 & 0.1232 & $\times$ & 7704 \\ \hline
pour autant & 0 & 13 & 0.79 & 0.994 & 0.0116 & 0.0299& \checkmark  & 457 \\ \hline
pour finir & 23 & 15 & 0.63 & 0.982 & 0.0759 & 0.0711 & \checkmark & 838 \\ \hline
pour le coup & NO & NO & NO & NO & NO & NO & NO & 464 \\ \hline
pour l'essentiel & 0 & 9 & 0.97 & 0.985 & 0.0282 & 0.0560 & \checkmark & 284 \\ \hline
pour le moment & 0 & 13 & 0.39 & 0.971 & 0.1255 & 0.0983 & \checkmark & 2986 \\ \hline
pour l'heure & NO & NO & NO & NO & NO & NO & NO & 546 \\ \hline
pour l'instant & 11 & 14 & 0.63 & 0.988 & 0.0450 & 0.0567 & \checkmark & 1859 \\ \hline
pour ma part & 5 & 6 & 1.12 & 0.997 & 0.1018 & 0.1303 & \checkmark & 2744 \\ \hline
pour peu que & 0 & 10 & 0.74 & 0.977 & 0.0645 & 0.0803 & \checkmark & 2479 \\ \hline
pour surcro\^it de & NO & NO & NO & NO & NO & NO & NO & 90 \\ \hline
pourtant que (i) & 0 & 6 & 1.05 & 0.989 & 0.1257 & 0.1447 & \checkmark & 4220 \\ \hline
pourtant que (ii) & 0 & 6 & 1.70 & 0.989 & 0.0385 & 0.0801 & \checkmark & 4220 \\ \hline
pour tout dire & 2 & 7 & 1.14 & 0.986 & 0.0781 & 0.1056 & \checkmark & 655 \\ \hline
pour un temps & NO & NO & NO & NO & NO & NO & NO & 1333 \\ \hline
pr\'esentement & NO & NO & NO & NO & NO & NO & NO & 2683 \\ \hline
probablement (i) & 2 & 8 & 1.22 & 0.981 & 0.0568 & 0.0843 & \checkmark & 8497 \\ \hline
probablement (ii) & 2 & 10 & 0.70 & 0.981 & 0.0700 & 0.0837 & \checkmark & 8497 \\ \hline
proprement & 4 & 6 & 1.05 & 0.989 & 0.0825 & 0.1173 & \checkmark & 9817 \\ \hline
principalement & 17 & 9 & 1.03 & 0.970 & 0.0497 & 0.0743 & \checkmark & 6695 \\ \hline
progressivement & 3 & 9 & 0.59 & 0.978 & 0.1325 & 0.1213 & $\times$ & 2235 \\ \hline
quand m\^eme & 0 & 18 & 0.43 & 0.972 & 0.1269 & 0.0840 & \checkmark & 12171 \\ \hline
quant \`a & 5 & 15 & 0.37 & 0.966 & 0.1417 & 0.0972 & $\times$ & 20878 \\ \hline
quant \`a cela & NO & NO & NO & NO & NO & NO & NO & 91 \\ \hline
quant \`a moi & NO & NO & NO & NO & NO & NO & NO & 4875 \\ \hline
que dalle & NO & NO & NO & NO & NO & NO & NO & 163 \\ \hline
quelquefois & 11 & 9 & 1.11 & 0.967 & 0.0589 & 0.0809 & \checkmark & 34408 \\ \hline
quelque part & NO & NO & NO & NO & NO & NO & NO & 6454 \\ \hline
relatif \`a  & 12 & 10 & 0.57 & 0.980 & 0.1375 & 0.1173 & $\times$ & 2850 \\ \hline
relativement \`a & NO & NO & NO & NO & NO & NO & NO & 1469 \\ \hline
rien de plus & NO & NO & NO & NO & NO & NO & NO & 1537 \\ \hline
sans ambages & 1 & 13 & 0.46 & 0.969 & 0.1474 & 0.1065 & \checkmark & 130 \\ \hline
sans commune mesure & 2 & 9 & 1.12 & 0.986 & 0.0433 & 0.0694 & \checkmark & 112 \\ \hline
sans crier gare & 0 & 13 & 0.55 & 0.972 & 0.0651 & 0.0708 & \checkmark & 211 \\ \hline
sans d\'etour & 6 & 15 & 0.51 & 0.970 & 0.1423 & 0.0974 & \checkmark & 467 \\ \hline
sans fa\c{c}on & NO & NO & NO & NO & NO & NO & NO & 650 \\ \hline
sans tenir compte de & 5 & 8 & 0.85 & 0.983 & 0.0655 & 0.0905 & \checkmark & 143 \\ \hline
sauf & 5 & 10 & 0.82 & 0.965 & 0.0660 & 0.0812 & \checkmark & 11138 \\ \hline
sauf si & 3 & 12 & 0.74 & 0.994 & 0.0339 & 0.0532 & \checkmark & 247 \\ \hline
sauf que & 0 & 14 & 0.41 & 0.968 & 0.1141 & 0.0903 & \checkmark & 910 \\ \hline
selon moi & 1 & 13 & 0.39 & 0.982 & 0.1292 & 0.0997 & \checkmark & 1055 \\ \hline
si besoin est (i) & 5 & 6 & 2.30 & 0.992 & 0.0231 & 0.0620 & \checkmark & 106 \\ \hline
si besoin est (ii) & 3 & 6 & 2.27 & 0.995 & 0.0239 & 0.0631 & \checkmark & 106 \\ \hline
si bien que & 4 & 10 & 0.60 & 0.970 & 0.0905 & 0.0951 & \checkmark & 4831 \\ \hline
si \c{c}a se trouve & 0 & 8 & 0.94 & 0.974 & 0.0640 & 0.0894 & \checkmark & 144 \\ \hline
s'il en est & NO & NO & NO & NO & NO & NO & NO & 88 \\ \hline
si possible & 22 & 13 & 0.72 & 0.965 & 0.0801 & 0.0785 & \checkmark & 760 \\ \hline
soit dit en passant & NO & NO & NO & NO & NO & NO & NO & 276 \\ \hline
soudain & 22 & 20 & 0.47 & 0.970 & 0.1051 & 0.0725 & $\times$ & 3498 \\ \hline
soudainement & 1 & 9 & 1.08 & 0.968 & 0.0438 & 0.0698 & \checkmark & 94 \\ \hline
sous peu & 0 & 6 & 1.83 & 0.993 & 0.0165 & 0.0524 & \checkmark & 291 \\ \hline
sous pr\'etexte de & 0 & 9 & 1.03 & 0.976 & 0.0486 & 0.0735 & \checkmark & 2341 \\ \hline
sous pr\'etexte que & 6 & 10 & 0.64 & 0.977 & 0.0928 & 0.0963 & \checkmark & 1364 \\ \hline
sous r\'eserve que & NO & NO & NO & NO & NO & NO & NO & 89 \\ \hline
souventes fois & NO & NO & NO & NO & NO & NO & NO & 530 \\ \hline
sp\'ecialement & NO & NO & NO & NO & NO & NO & NO & 3764 \\ \hline
sur ce th\`eme & NO & NO & NO & NO & NO & NO & NO & 130 \\ \hline
sur le champ & 9 & 10 & 1.00 & 0.975 & 0.0834 & 0.0913 & \checkmark & 5152 \\ \hline
sur le moment & 8 & 16 & 0.38 & 0.992 & 0.0621 & 0.0623 & \checkmark & 715 \\ \hline
sur le sujet de  & 0 & 10 & 0.98 & 0.969 & 0.0834 & 0.0913 & \checkmark & 292 \\ \hline
sur le point de  & 13 & 18 & 0.28 & 0.967 & 0.1737 & 0.0982 & \checkmark & 3321 \\ \hline
sur l'heure & NO & NO & NO & NO & NO & NO & NO & 720 \\ \hline
sur l'instant & 0 & 15 & 0.44 & 0.971 & 0.1206 & 0.0897 & \checkmark & 162 \\ \hline
tandis que & BUG & BUG & BUG & BUG & BUG & BUG & BUG & 39303 \\ \hline
tant et plus & NO & NO & NO & NO & NO & NO & NO & 155 \\ \hline
tel quel & 4 & 8 & 1.07 & 0.983 & 0.0405 & 0.0712 & \checkmark & 985 \\ \hline
tour \`a tour & 11 & 22 & 0.37 & 0.982 & 0.1513 & 0.0829 & \checkmark & 4480 \\ \hline
tout \`a coup & 3 & 9 & 1.11 & 0.983 & 0.0424 & 0.0686 & \checkmark & 20468 \\ \hline
tout \`a fait & 20 & 15 & 0.73 & 0.971 & 0.1452 & 0.0984 & $\times$ & 25611 \\ \hline
tout \`a l'heure (i) & 6 & 12 & 0.57 & 0.967 & 0.0822 & 0.0828 & \checkmark & 12853 \\ \hline
tout \`a l'heure (ii) & 4 & 13 & 0.71 & 0.977 & 0.0717 & 0.0743 & \checkmark & 12853 \\ \hline
tout au long de & 12 & 13 & 0.71 & 0.974 & 0.0842 & 0.0805 & \checkmark & 1363 \\ \hline
tout au plus & 6 & 23 & 0.33 & 0.971 & 0.1487 & 0.0804 & \checkmark & 2954 \\ \hline
tout bien consid\'er\'e & 2 & 6 & 1.16 & 0.994 & 0.0453 & 0.0869 & \checkmark & 152 \\ \hline
tout compte fait & NO & NO & NO & NO & NO & NO & NO & 390 \\ \hline
tout court & NO & NO & NO & NO & NO & NO & NO & 1149 \\ \hline
tout de m\^eme (i) & 8 & 8 & 0.85 & 0.979 & 0.0758 & 0.0973 & \checkmark & 13315 \\ \hline
tout de m\^eme (ii) & 26 & 13 & 0.74 & 0.991 & 0.0166 & 0.0357 & \checkmark & 13315 \\ \hline
tout du long & NO & NO & NO & NO & NO & NO & NO & 302 \\ \hline
toutefois & NO & NO & NO & NO & NO & NO & NO & 20576 \\ \hline
tout juste & 17 & 14 & 0.46 & 0.968 & 0.0894 & 0.0799 & \checkmark & 2055 \\ \hline
tout juste de & NO & NO & NO & NO & NO & NO & NO & 197 \\ \hline
tout plein de & NO & NO & NO & NO & NO & NO & NO & 1014 \\ \hline
tout sauf & 3 & 10 & 0.69 & 0.968 & 0.1026 & 0.1013 & \checkmark & 158 \\ \hline
tout sp\'ecialement & 0 & 12 & 0.63 & 0.969 & 0.0948 & 0.0889 & \checkmark & 164 \\ \hline
tout un chacun & 0 & 11 & 0.70 & 0.975 & 0.0812 & 0.0859 & \checkmark & 260 \\ \hline
tr\`es tr\`es & NO & NO & NO & NO & NO & NO & NO & 356 \\ \hline
un de ces jours & 5 & 6 & 1.31 & 0.996 & 0.0442 & 0.0858 & \checkmark & 983 \\ \hline
une esp\`ece de & 16 & 8 & 1.17 & 0.976 & 0.0501 & 0.0791 & \checkmark & 12365 \\ \hline
une sorte de (i) & 7 & 13 & 0.51 & 0.969 & 0.1208 & 0.0964 & \checkmark & 31306 \\ \hline
une sorte de (ii) & 2 & 8 & 1.29 & 0.991 & 0.0396 & 0.0704 & \checkmark & 31306 \\ \hline
un lendemain & 8 & 8 & 0.56 & 0.993 & 0.1277 & 0.1263 & \checkmark & 505 \\ \hline
un petit peu & 0 & 15 & 0.52 & 0.973 & 0.1000 & 0.0816 & \checkmark & 692 \\ \hline
un surcro\^it de & NO & NO & NO & NO & NO & NO & NO & 454 \\ \hline
un tas de & 23 & 9 & 1.30 & 0.990 & 0.0243 & 0.0520 & \checkmark & 4352 \\ \hline
venir de & 20 & 20 & 0.38 & 0.973 & 0.1164 & 0.0763 & $\times$ & 35884 \\ \hline
vis \`a vis de & 4 & 17 & 0.37 & 0.978 & 0.0966 & 0.0754 & \checkmark & 3384 \\ \hline
voil\`a & 0 & 17 & 0.77 & 0.975 & 0.0470 & 0.0526 & \checkmark & 90090 \\ \hline
vu que & 0 & 11 & 1.00 & 0.973 & 0.0463 & 0.0649 & \checkmark & 1230 \\ \hline
zut & 0 & 10 & 0.99 & 0.987 & 0.0422 & 0.0650 & \checkmark & 525 \\ \hline

\end{xtabular}
\end{center}

\newpage

\begin{center}\textbf{REFERENCES}\end{center}

\bibliography{Sigmobiblio.bib}

\end{document}